\documentclass[a4paper,11pt]{article}
\usepackage{jheppub_modified} % for details on the use of the package, please
\usepackage{graphicx}   % auto-detectable
\usepackage[T1]{fontenc} % if needed
\usepackage{slashed}
\usepackage{epsfig}
\usepackage{tikz}
\usepackage{enumerate}
\newcounter{exno}

\newcommand{\mpl}{m_{\mbox{\tiny{Pl}}}}
\newcommand{\Beq}{\begin{equation}\begin{aligned}}
\newcommand{\Eeq}{\end{aligned}\end{equation}}

\newcommand{\bk}{{\bf{k}}}
\newcommand{\bx}{{\bf{x}}}

\newcommand{\bq}{{\bf{q}}}

\title{Lectures on Reheating after Inflation}
\subheader{2018 MPA Lecture Series on Cosmology}

\author{Kaloian Lozanov}
\affiliation{Max Planck Institute for Astrophysics,\\Karl-Schwarzschild-Str. 1,
\\ 85748 Garching,
Germany.\\ \\ \today}

\emailAdd{klozanov@mpa-garching.mpg.de}

\begin{document} 
\maketitle
\flushbottom

\acknowledgments

I would like to thank my PhD supervisors, Mustafa Amin, who has taught me everything I know about reheating, and Anthony Challinor, who introduced me to the field of Cosmology. Without them, the writing of these notes would have not been possible. I am also immensely grateful to Eiichiro Komatsu for giving me the opportunity to continue my research in the area of reheating and to give short lecture series on it.

%\newpage

\section{Avant propos}

Over the past decades, the understanding of our cosmic history has improved spectacularly. Precise measurements of the temperature anisotropies in the cosmic microwave background (CMB) \cite{Ade:2015xua,Ade:2015lrj} have revealed a homogeneous and isotropic universe on large scales with tiny (if any) spatial curvature, $\Omega_K$, and nearly scale-invariant primordial curvature perturbations. The observations provide compelling evidence for an inflationary phase in the early universe \cite{PhysRevD.23.347,LINDE1982389,PhysRevLett.48.1220,Starobinsky:1980te,Mukhanov:1981xt,Senatore:2016aui}, implying a connection between gravity and quantum mechanics. Measurements of the relative abundances of light-elements are also in excellent agreement with our expectations \cite{Steigman:2007xt}. The predictions of big-bang nucleosynthesis (BBN), based on the well-understood physics of nuclear reactions, point towards a hot and dense universe, in local thermal equilibrium at late times. They tell us that the first light-elements started forming at cosmic time $t_{\text{BBN}}\gtrsim 1\,\text{s}$ at equilibrium temperature $T_{\text{BBN}}\sim E_{\text{BBN}}\sim1\,\text{MeV}$. %= \mathcal{O}(1)\,\text{MeV}$. 

Connecting these two remarkable epochs, however, could be challenging. Since the energy scale at the end of inflation can be as high as $E_{\text{inf}} \sim 10^{16}\,\text{GeV}$, with the duration of inflation corresponding to $\Delta t_{\text{inf}}\gtrsim 10^{-36}\,\text{s}$, there is a huge range of energy (and time) scales which is poorly understood and poorly observationally constrained. Current cosmological experiments cannot probe the period between inflation and BBN. This is because typically the effects from the interesting post-inflationary dynamics are on subhorizon scales due to causality and are washed out by the later non-linear evolution of structure.  The thermal state of the universe, required for BBN, also hides information about earlier times. Collider experiments can shed light on some intermediate-energy phenomena, e.g., the electro-weak symmetry breaking and QCD phase transitions, $E_\text{EW} \sim 10^2 \,\text{GeV}$ and $E_{\text{QCD}} \sim 10^2 \,\text{MeV}$, respectively, but they will not be able to cover the entire energy range in the foreseeable future.

Nevertheless, it is crucial that we try to understand the period between inflation and BBN better, for both theoretical and observational reasons. In the standard lore, the universe at the end of inflation is cold and dark, virtually empty of particles and dominated by the approximately homogeneous inflaton field. The energy of the inflaton that drove inflation, must be somehow transferred to other species of matter, eventually populating all relevant degrees of freedom of the Standard Model, leaving a hot, thermal, radiation-dominated universe, setting the scene for BBN. Importantly, this reheating process explains not only the cosmic origin of the matter that we are made of, but it also accounts for the production of cosmic relics such as photons and neutrinos, and perhaps dark matter and gravitational waves, as well as the generation of the observed matter-antimatter asymmetry in our universe (baryogenesis). Any unified theory of high-energy physics must include a complete understanding of inflation, reheating and the later evolution of the universe. 

Given the current and planned advances in observational cosmology and the improvement of constraints on inflation, reheating will be an integral part of research in the coming years. Arguably, one of the most important observational implications of the post-inflationary dynamics is its effect on the expansion history of the universe between inflation and BBN. It determines how we map perturbation modes from their exiting the Hubble horizon during inflation to horizon re-entry at late times. Thereby, the poorly constrained and understood post-inflationary expansion history influences directly the predictions for cosmological observables of specific models of inflation. For instance, it leads to significant uncertainties in the predictions for the spectral index, $n_s$, and the tensor-to-scalar ratio, $r$, in different models. It is critical that we understand these uncertainties better, if we wish to narrow the range of the observationally-allowed models of inflation. Works on reheating have also shown the possibility of formation of relics such as solitons and cosmic defects, helping us further constrain the variety of scenarios. 

The initial stage of reheating, also known as preheating, can involve highly non-perturbative processes, during which the universe gets populated via parametric resonances. They cannot be described with the usual perturbative expansions in coupling constants, even in cases with weak couplings \cite{Dolgov:1989us,Traschen:1990sw,Kofman:1994rk,Kofman:1996mv,Kofman1997,Bassett:2005xm,Allahverdi:2010xz,Amin2014}. Such resonances arise as the inflaton condensate (or generally any light scalar, that has attained a non-zero vacuum expectation value during inflation) begins to oscillate about the minimum of its potential, soon after inflation. The oscillations induce an effective time-dependence in the couplings of the inflaton to the other species of matter. While the background inflaton field dominates the energy budget of the system, the evolution of the remaining fields it is coupled to can be linearised. As the effective frequencies of the individual Fourier modes of the daughter fields change non-adiabatically every time the inflaton crosses the origin, we observe `explosive' (or resonant) particle production. This can be quite efficient, since it involves the collective decay of many inflatons from the condensate. When the energy of the newly-populated degrees of freedom becomes comparable to the background, back-reaction effects become important. Typically, the condensate fragments and the subsequent evolution is non-linear. It can be studied in the classical approximation, using classical lattice simulations, since all relevant modes have large occupancies and hence quantum effects are negligible.\footnote{However, since the universe at the beginning of BBN is in local thermal equilibrium, late-stage reheating analysis should eventually include a full quantum mechanical computation of the approach of all relevant degrees of freedom to states with maximal local entropy such as Bose-Einstein or Fermi-Dirac distributions \cite{Micha:2002ey,Micha:2004bv}.} The non-linear dynamics can lead to many interesting phenomena, e.g., the production of solitons that can delay the thermalisation required as an initial condition for BBN, field configurations evolving self-similarly in a turbulent manner \cite{Micha:2002ey,Micha:2004bv}, non-thermal phase transitions and the production of cosmic defects \cite{Tkachev:1998dc,Rajantie:2000fd,Dufaux2010}.

Research in the field of reheating has been divided into three main areas. On the theoretical side, there is a need to consider more realistic high-energy physics models \cite{GarciaBellido:2008ab,Bezrukov2008,Repond:2016sol,Figueroa2015,Figueroa:2016ojl,Kari,Enqvist:2015sua,Adshead:2015kza,Adshead:2015pva,Lozanov:2016pac}, including fermions and gauge bosons, in addition to the more traditional scalars, in the quest for a unified description of our cosmic history. Another direction for future work concerns the phenomenology of the many stages of reheating: from the non-perturbative particle production during preheating and the following non-linear classical evolution, to the late-time approach to a radiation-dominated period of expansion in local thermal equilibrium \cite{Figueroa:2016wxr,Lozanov:2016hid,Podolsky:2005bw,Amin:2011hj,Deskins2013,Hertzberg:2013jba,Hertzberg2013,Lozanov:2016pac,Hertzberg:2014jza,Hertzberg:2014iza}. A lot of effort has been dedicated to observational signatures of reheating \cite{Giblin:2014gra,Bond:2009xx,Dufaux2010,Antusch:2016con,Figueroa:2016ojl}, as well as their important implications for inflationary observables \cite{Munoz:2014eqa,Lozanov:2016hid,Dai:2014jja,Martin:2014nya,Martin:2016oyk,Hardwick:2016whe,Martin:2013tda,Liddle:2003as}. These three areas have also formed the common thread of our lecture notes.

These notes are meant to serve as a generic introduction to the field of reheating after inflation, starting with a brief summary of the inflationary paradigm (Section \ref{sec:InflIntroThesis}), followed by a review of the different preheating mechanisms (Section \ref{sec:PreheatIntro}) and the ensuing non-linear evolution (Section \ref{sec:NonLinRehThesis}), and finally considering some high-energy physics (Section \ref{sec:RehHEPmodels}) and observational (Section \ref{sec:ObsImplReh}) aspects of reheating.

\newpage

\section{Bedtime reading}

\begin{itemize}
  \item {\it Towards the Theory of Reheating After Inflation}, Lev Kofman, Andrei Linde, Alexei Starobinsky, \textit{\underline{arXiv:hep-ph/9704452}}.

The seminal paper that helped launch the modern understanding of (p)reheating.
  \item {\it Inflation Dynamics and Reheating}, Bruce A. Bassett, Shinji Tsujikawa, David Wands, \textit{\underline{arXiv:astro-ph/0507632}}.

A comprehensive review on reheating after inflation, covering a broad range of theoretical, phenomenological and observational aspects, many of which are still relevant, as well as providing a pedagogical introduction to inflation.
  \item {\it Reheating in Inflationary Cosmology: Theory and Applications}, Rouzbeh Allahverdi, Robert Brandenberger, Francis-Yan Cyr-Racine, Anupam Mazumdar, \textit{\underline{arXiv:1001.2600}}.

An excellent concise introduction to reheating.
  \item {\it Non-linear Dynamics and Primordial Curvature Perturbations from Preheating}, Andrei Frolov, \textit{\underline{arXiv:1004.3559}}.

A review on the non-linear dynamics of reheating and dedicated state-of-the-art numerical techniques.
  \item {\it Nonperturbative Dynamics Of Reheating After Inflation: A Review}, Mustafa A. Amin, Mark P. Hertzberg, David I. Kaiser, Johanna Karouby, \textit{\underline{arXiv:1410.3808}}.

The most recent review, including a pedagogical treatment of the linear stage of reheating in multi-field models.
\end{itemize}

\newpage

\section{Notation and conventions}

We use natural units in which $\hbar=c=k_{\rm{B}}=\epsilon_0=1$. In these units, the reduced Planck mass is given by $\mpl=1/\sqrt{8\pi G}$.

\vspace{2.5mm}

\noindent Greek indices $\mu$, $\nu$ and so on go over the four space-time coordinates $x^{\mu}=[x^{0}\!,\,x^{1}\!,\,x^{2}\!,\,x^{3}]^{T}$ with $x^0$ for the time coordinate.

\vspace{2.5mm}

\noindent Minkowski metric is given by $\eta_{\mu\nu}=\text{diag}[1,-1,-1,-1]$.

\vspace{2.5mm}

\noindent Latin labels $i$, $j$, $k$ and so on go over the three spatial coordinates.

\vspace{2.5mm}

\noindent Spatial vectors are written in boldface.

\vspace{2.5mm}

\noindent Summation over repeated indeces is assumed unless otherwise stated.

\vspace{2.5mm}

\noindent The Ricci tensor, defined in terms of the Christoffel symbols, is

\Beq
R_{\mu\nu}\equiv \partial_{\lambda}\Gamma^{\lambda}_{\mu\nu}-\partial_{\nu}\Gamma^{\lambda}_{\mu\lambda}+\Gamma^{\lambda}_{\lambda\rho}\Gamma^{\rho}_{\mu\nu}-\Gamma^{\rho}_{\mu\lambda}\Gamma^{\lambda}_{\nu\rho}\,,
\Eeq
and the Ricci scalar is $R=g^{\mu\nu}R_{\mu\nu}$.

\vspace{2.5mm}

\noindent The spatial Fourier transform of a field $f({\bx})$ is $f_{\bk}=\int f({\bx})\text{e}^{-i{\bk}\cdot{\bx}} \text{d}^\text{3}\bx/(2\pi)^{3}$ and the inverse transform is $f({\bx})=\int f_{\bk}\text{e}^{i{\bk}\cdot{\bx}} \text{d}^\text{3}\bk$.

\newpage

%-----------------------------------------------------------------------------------------------------------------------------
%				Inflation and initial conditions for reheating
%-----------------------------------------------------------------------------------------------------------------------------

\section{Inflation and initial conditions for reheating}
\label{sec:InflIntroThesis}

%\begin{adjustwidth}{5.0em}{0pt}
\hfill\begin{minipage}{\dimexpr\textwidth-3.7cm}
`{\it With the new cosmology the universe must have been started off in some very simple way. What, then, becomes of the initial conditions required by dynamical theory? Plainly there cannot be any, or they must be trivial. We are left in a situation which would be untenable with the old mechanics. If the universe were simply the motion which follows from a given scheme of equations of motion with trivial initial conditions, it could not contain the complexity we observe. Quantum mechanics provides an escape from the difficulty. It enables us to ascribe the complexity to the quantum jumps, lying outside the scheme of equations of motion. The quantum jumps now form the uncalculable part of natural phenomena, to replace the initial conditions of the old mechanistic view.}'
%\xdef\tpd{\the\prevdepth}
\end{minipage}

%\prevdepth\tpd
%\end{adjustwidth}

\hfill {\it P. A. M. Dirac (1939)}

\subsection{Standard cosmology, its puzzles and why we need inflation}
\label{sec:InflationIntro}

Standard cosmology is based on the empirical observation that the universe is homogeneous and isotropic on large scales \cite{Ade:2015xua}. In the context of General Relativity, it means that the space-time metric takes the Friedmann-Robertson-Walker (FRW) form
\Beq
\label{eq:FRWmetricIntro}
ds^2=dt^2-a(t)^2\left(\frac{dr^2}{1-Kr^2}+r^2d\theta^2+r^2\sin^2\theta d\phi^2\right)\,,
\Eeq
where $t$ is the cosmic time and $a(t)$ is the Robertson-Walker scale factor. The term in brackets represents the line element of the three-dimensional homogeneous and isotropic space. For positive, zero and negative $K$  this hypersurface can be considered as a 3-dimensional sphere embedded in a 4-dimensional Euclidean space, a 3-dimensional Euclidean space and a 3-dimensional hypersphere embedded in a 4-dimensional pseudo-Euclidean space, respectively \cite{Weinberg:2008zzc}. The positive, zero and negative cases are better known as the closed, flat and open universes, respectively.

The evolution of $a(t)$ is determined by the Einstein equations
\Beq
\label{eq:Einst}
R_{\mu\nu}-\frac{1}{2}g_{\mu\nu}R=\frac{T_{\mu\nu}}{\mpl^2}+g_{\mu\nu}\Lambda\,.
\Eeq
Here $\Lambda$ is the cosmological constant, introduced by Einstein to make the universe static. Henceforth, we set $\Lambda=0$\footnote{Observations favor a small, but non-zero value of $\Lambda$, unjustifiable by Quantum Field Theory (QFT) if interpreted as the energy of the vacuum. This drawback of standard cosmology and QFT is known as the {\it Cosmological constant problem}. The contribution of the $\Lambda$ term to the energy budget of the universe (known as dark energy) becomes significant only at very late times, at a redshift of about $1$, so it is safe to ignore $\Lambda$ at earlier epochs.}. Near the origin of locally Cartesian co-moving coordinates, the components of the energy-momentum tensor, $T^{\mu\nu}$, in a homogeneous universe are functions of $t$ only. Isotropy also imposes the additional constraints $T^{i0}=0$ and $T^{ij}\propto\delta^{ij}$. Conventionally, the energy density, $\rho$, and the pressure, $p$, of the perfect fluid filling the homogeneous and isotropic universe are defined locally as
\Beq
\label{eq:TmunuMeanFRW}
T^{00}=\rho(t)\,,\qquad T^{ij}=-a(t)^{-2}\delta^{ij}p(t)\,.
\Eeq
Then only the $00$ and the $ii$ Einstein equations, eq. \eqref{eq:Einst}, do not vanish
\Beq
\label{eq:FRWeqs}
H^2&=\frac{\rho}{3\mpl^2}-\frac{K}{a^2}\,,\\
\frac{\ddot{a}}{a}&=-\frac{\rho+3p}{6\mpl^2}\,,
\Eeq
and are known as the Friedmann and Raychaudhuri equations, respectively. The Einstein equations, eq. \eqref{eq:Einst}, also imply the conservation of the energy-momentum tensor $T^{\nu\mu}{}_{;\mu}=0$. Due to isotropy the momentum-conservation law $T^{i\mu}{}_{;\mu}=0$ is automatically satisfied. The energy-conservation $T^{0\mu}{}_{;\mu}=0$ yields
\Beq
\label{eq:EnergyConservationFRW}
\dot{\rho}+3H(\rho+p)=0\,.
\Eeq
This expression could be derived from a combination of the Friedmann and Raychaudhuri equations, eq. \eqref{eq:FRWeqs}. For a constant equation of state of the form $w=p/\rho$, the energy-conservation law implies $\rho\propto a^{-3-3w}$. Using this result in eq. \eqref{eq:FRWeqs} we find the power-law solution
\Beq
\label{eq:FRWPowerLaw}
a(t)\propto t^{2/(3+3w)}\,.
\Eeq
In standard cosmology the typical sources of gravity are non-relativistic matter (dust) and relativistic matter (radiation). If one of these components is dominant, then
\Beq
{\rm{for\,\,\, dust}}\!:&\qquad a\propto t^{2/3}\,,\qquad\rho\propto a^{-3}\,,\qquad w=0\,,\\
{\rm{for\,\,\, radiation}}\!:&\qquad a\propto t^{1/2}\,,\qquad\rho\propto a^{-4}\,,\qquad w=1/3\,.
\Eeq
The second column also applies for individual species, even if subdominant, provided they are self-interacting only. Note that the universe always decelerates, $\ddot{a}<0$.

The Friedmann equation, eq. \eqref{eq:FRWeqs}, can be rewritten as 
\Beq
\label{eq:OmegaK}
\Omega-1=\frac{K}{a^2H^2}\,,
\Eeq
where the energy density makes up a fraction $\Omega=\rho/\rho_{\rm{c}}$ of the critical energy density, $\rho_{\rm{c}}=3\mpl^2H^2$. Similarly, one often writes the spatial curvature term as $\Omega_{K}=-K/(a^2H^2)$. Cosmological observations give very tight constraints on this quantity, consistent with zero. The $95\%$ limit, from the most recent measurement of the anisotropies in the CMB \cite{Ade:2015xua}, on the spatial curvature today is $|\Omega_{K,0}|<0.005$. This small value leads to one of the fine-tuning problems in standard cosmology. Since $\ddot{a}<0$, $aH=\dot{a}$ increases when going backwards in time. Hence, $\Omega_K$ becomes even smaller at earlier times, or in other words, the energy density tends to the critical one, $\Omega\rightarrow 1$, with unnaturally high precision. To get some rough idea about the degree of fine-tuning in the initial condition for $\Omega$ at some early time, $t_{\rm{early}}$, required by current measurements, consider the ratio
\Beq
\label{eq:OmegaEarly}
\left|\frac{\Omega_{\rm{early}}-1}{\Omega_{0}-1}\right|=\left(\frac{a_0H_0}{a_{\rm{early}}H_{\rm{early}}}\right)^2=\left(\frac{a_0H_0}{a_{\rm{eq}}H_{\rm{eq}}}\right)^2\left(\frac{a_{\rm{eq}}H_{\rm{eq}}}{a_{\rm{early}}H_{\rm{early}}}\right)^2\sim \frac{1+z_{\rm{eq}}}{(1+z_{\rm{early}})^2}\,.
\Eeq
We assume the universe to be radiation dominated at early times, $t_{\rm{early}}<t<t_{\rm{eq}}$, and matter dominated after the epoch of radiation-matter equality $t_{\rm{eq}}<t<t_0$. For $t_{\rm{early}}\sim t_{\rm{BBN}}$, $|\Omega_{\rm{BBN}}-1|\ll 10^{-17}$, assuming $T_{\rm{BBN}}\sim1\,\rm{MeV}$ and $z_{\rm{eq}}\sim 10^3$. At the GUT epoch, $T_{\rm{GUT}}\sim10^{16}\,\rm{GeV}$, $|\Omega_{\rm{GUT}}-1|\ll 10^{-55}$. Unless the initial conditions are set very precisely, the universe either collapses too quickly or expands too fast before large-scale structure can form. This is known as the {\it flatness problem}.

The deceleration of the scale factor in standard cosmology also leads to contradictions with measurements of the causal structure of the observable universe. The physical length, $l_{\rm{phys}}=al$, of a given co-moving length-scale $l$, increases as the universe expands. On the other hand, its ratio with the Hubble radius $l_{\rm{phys}}/H^{-1}=\dot{a}l$ decreases with time, since $\ddot{a}<0$. Hence, any co-moving length-scale becomes much greater than the Hubble scale at sufficiently early times. This implies that the causally-connected region in the universe today should lie deep inside the Hubble volume. We can easily calculate the expected size of this region at the epoch of recombination. Ignoring spatial curvature, the physical size of the region at recombination is equal to the particle horizon $d_{\rm{rec}}=a_{\rm{rec}}\int_{t_{\rm{early}}}^{t_{\rm{rec}}}dt/a(t)$. For $a\sim t^{n}$, $0<n<1$ and $t_{\rm{rec}}\gg t_{\rm{early}}$ the upper bound of the integral determines its value. For dust or radiation or any mixture of the two, $d_{\rm{rec}}\sim H_{\rm{rec}}^{-1}$. We also have $d_{\rm{rec}}=a_{\rm{rec}}r_{\rm{rec}}\Delta\theta$, where the co-moving distance to the CMB is $r_{\rm{rec}}=\int_{t_{\rm{rec}}}^{t_{0}}dt/a(t)\sim1/(a_0H_0)$ and $\Delta\theta$ is the angular size of the causal region. Assuming matter domination between recombination and today, $\Delta\theta\sim(1+z_{\rm{rec}})^{-1/2}$, which corresponds to about $1^{\rm{o}}$ for $z_{\rm{rec}}\sim10^3$. This is in conflict with observations of the microwave sky, showing the same temperature to high accuracy in all directions \cite{Ade:2015xua}. Within standard cosmology this isotropy of the CMB cannot be accounted for, since there is no way for points separated by more than a degree to be in thermal equilibrium (and in causal contact) before the epoch of last-scattering. This constitutes the {\it horizon problem}.

\begin{figure*}[t] %  figure placement: here, top, bottom, or page
   \centering
   \includegraphics[width=4.22in]{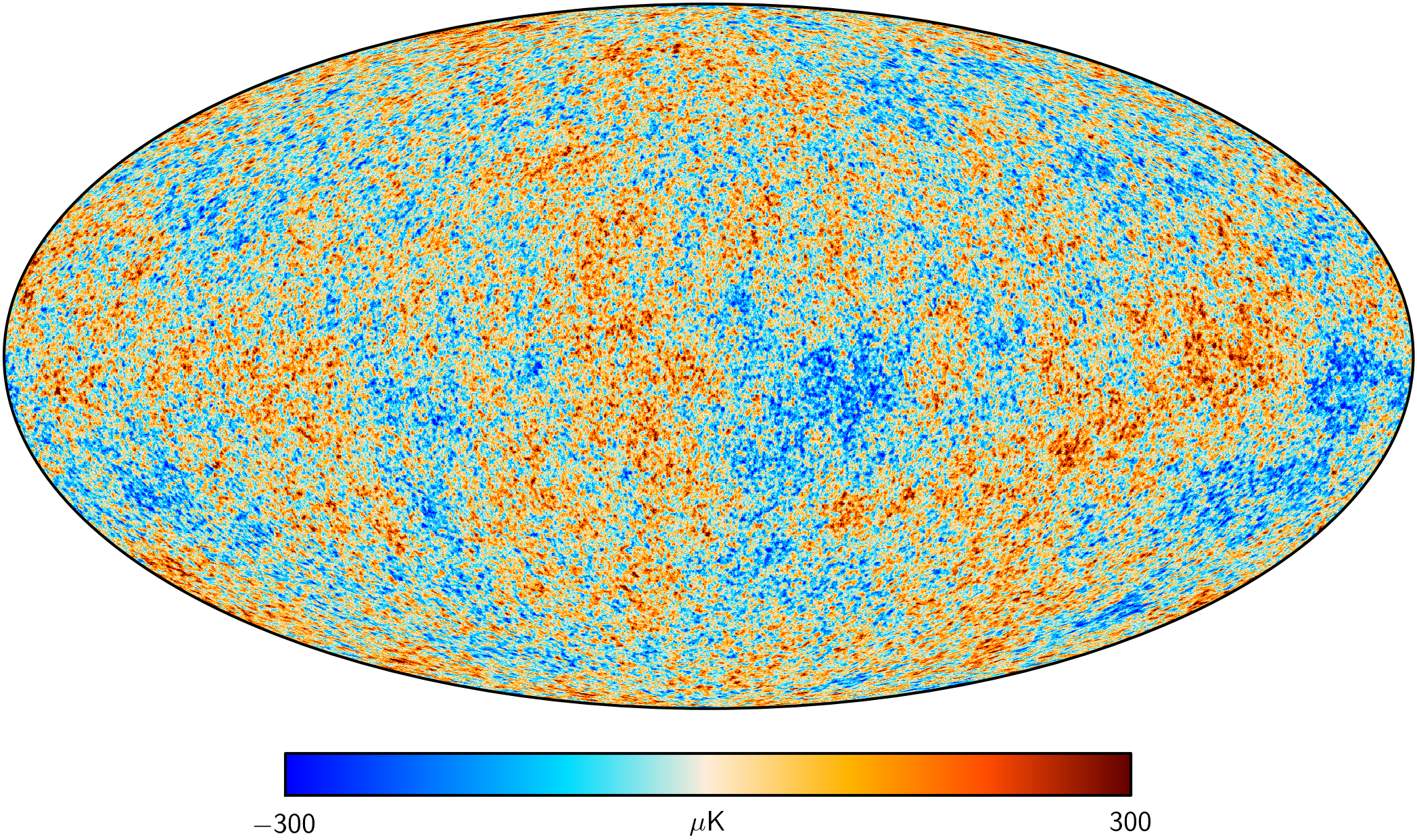} \\
\hspace{2.2in}\\
   \includegraphics[width=4.22in]{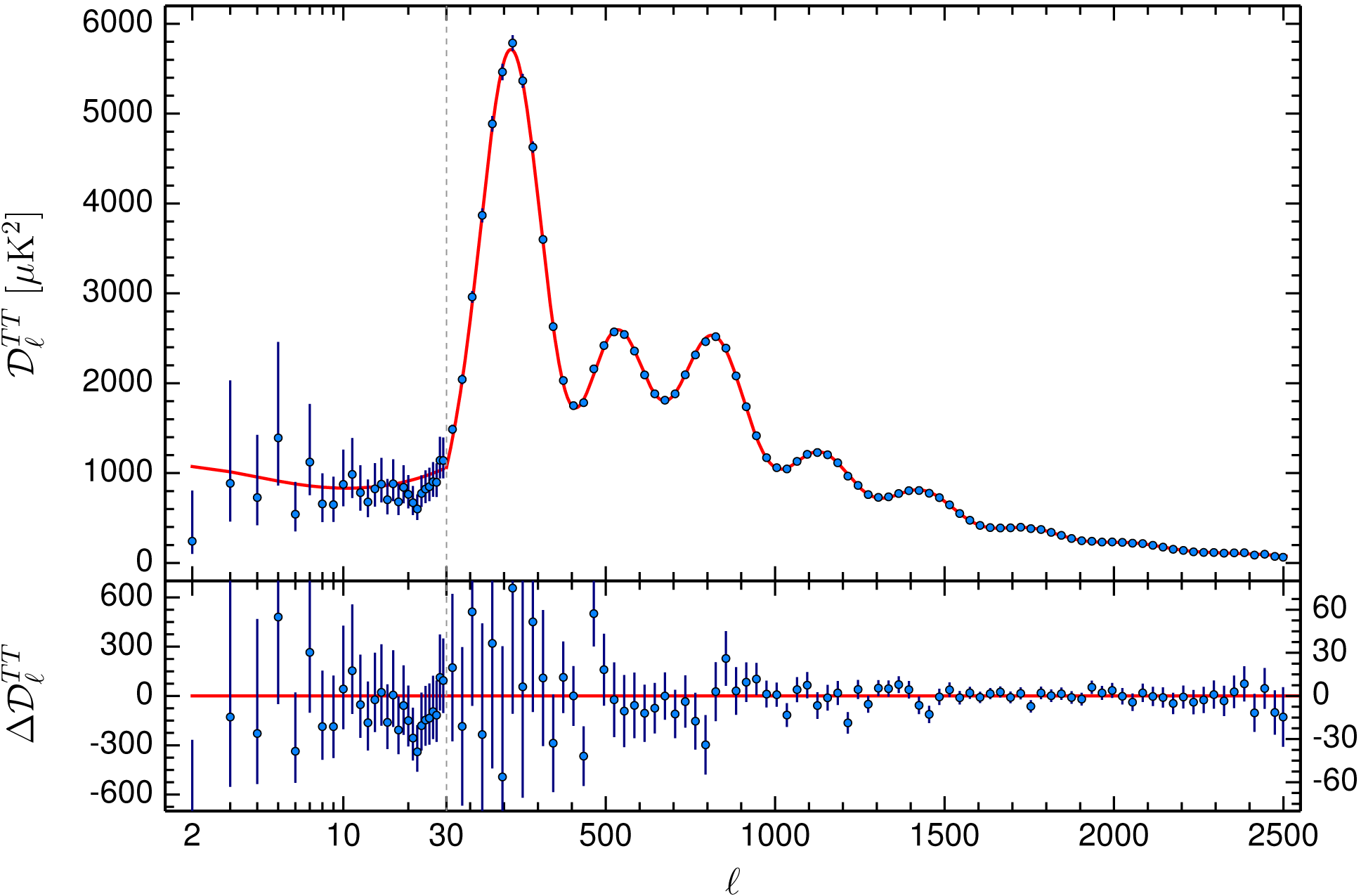}\\ 
\hspace{-2.2in}
   \caption{A full-sky map of the CMB temperature (top) and its power spectrum (bottom) \cite{Ade:2015xua} with the mean temperature $\bar{T}_{\rm{CMB}}=2.7255\,\rm{K}$ subtracted. The angular size of the particle horizon at recombination, $\Delta\theta\sim1^{\rm{o}}$, corresponds roughly to $l\approx200$.}
   \label{fig:PlanckMap}   
\end{figure*}

The tiny anisotropies measured in the CMB have a nearly scale-invariant power-spectrum, even on large, causally disconnected at the time of last-scattering, scales \cite{Ade:2015xua}, see Fig. \ref{fig:PlanckMap}. They reflect small variations in the matter density at the epoch of recombination. Later on, these density fluctuations act as seeds for the formation of large-scale structure. Hence, standard cosmology also fails to explain the deviations from the FRW metric.\footnote{The appearance of well-defined acoustic peaks on small, causally connected at the time of last-scattering scales, is also a compelling evidence for deviations from standard cosmology. Only if all temperature perturbation modes of a given $k$, irrespective of the direction of the wavevector $\bk$, oscillate in phase can we observe these prominent acoustic peaks. This phase coherence suggests that fluctuations were seeded in the early universe, before horizon entry (the moment they started oscillating). Alternative models in which fluctuations are sourced continuously during the radiation and matter domination eras, e.g., by a network of cosmic defects, do not give rise to phase coherence. The role of such sources for structure formation can be only subdominant.}

%~~~~~~~~~~~~~~~~~
%~~~~~~~~~~~~~~~~~

High-energy physics theories take the view that the physical laws look simpler at higher energies. This implies that gauge symmetries inevitably get broken during the evolution of the early universe, leaving behind them topological defects such as monopoles. The density of monopoles is bounded from below, due to the existence of a maximal correlation length determined by the causal length, i.e., the particle horizon, during the phase transition. This presents a serious problem for standard cosmology. If these relics do not annihilate efficiently, their abundance on sub-horizon scales can become large after the phase transition. Furthermore, these massive relics behave as dust. Their energy density can become the dominant component at dangerously early times, e.g., before or around BBN, since it decays more slowly with $a(t)$ than that of radiation. This is known as the {\it monopole problem}\footnote{This issue plagues all theories featuring massive relics, e.g., gravitinos, Kaluza-Klein particles and moduli fields.}.

All of the above problems can be shown to have the same origin -- the expansion with time of the co-moving Hubble sphere, $(aH)^{-1}=\dot{a}^{-1}$, following from $\ddot{a}<0$. One can resolve all of these puzzles by postulating the existence of an earlier stage of inflation, during which the universe undergoes accelerated expansion and the co-moving Hubble sphere shrinks \cite{PhysRevD.23.347,LINDE1982389,PhysRevLett.48.1220}. The most common expansion history of inflation is the quasi-exponential one, i.e., $a\sim \exp(Ht)$, with $H$ varying very slowly with time. Another possibility is power-law inflation, $a\sim t^n$, $n>1$. For $n\gg1$, $H$ again varies very slowly with time.

The measured tiny value of $|\Omega_{K,0}|$ is in fact a prediction of inflation. If we assume that $|\Omega_{K,\rm{init}}|$ at the beginning of inflation is of order unity then from eq. \eqref{eq:OmegaK} follows
\Beq
\left|\frac{\Omega_{\rm{init}}-1}{\Omega_{0}-1}\right|>1\,.
\Eeq
For a constant expansion rate during inflation, $H_{\rm{inf}}$, the number of {\it e}-folds of expansion until the end of inflation, $N=\ln(a_{\rm{end}}/a_{\rm{init}})$, is bounded to be
\Beq
\label{eq:NefoldsPuzzles}
e^{N}>\frac{a_{\rm{end}}H_{\rm{inf}}}{a_0H_0}.
\Eeq
Making the tentative assumption of $t_{\rm{end}}\sim t_{\rm{early}}$, see eq. \eqref{eq:OmegaEarly}, i.e., of the universe becoming radiation dominated immediately after the end of inflation, we find that for $t_{\rm{early}}\sim t_{\rm{BBN}}$ we need at least about $20$ {\it e}-folds to resolve the flatness problem, whereas for $t_{\rm{early}}\sim t_{\rm{GUT}}$, $N$ has to be over $60$.

To account for the isotropy of the CMB, we need to make sure that the co-moving particle horizon at recombination is greater than the co-moving distance photons travel after recombination until today, i.e.,
\Beq
\label{eq:HorizonProblem}
\int_{t_{\rm{init}}}^{t_{\rm{rec}}}\frac{dt}{a(t)}>\int_{t_{\rm{rec}}}^{t_{0}}\frac{dt}{a(t)}\,. %r_{\rm{rec}}=\int_{t_{\rm{rec}}}^{t_{0}}\frac{dt}{a(t)}\sim\frac{1}{a_0H_0}\,.
\Eeq
Note that $\int dt/a(t)=\int d\ln a/\dot{a}$. Hence, each integral is dominated by the smallest $\dot{a}$, i.e., the moment when the co-moving Hubble sphere is the largest. It implies for the right-hand side of eq. \eqref{eq:HorizonProblem} a value $\sim 1/(a_{0}H_{0})$ as already shown above and $\sim 1/(a_{\rm{init}}H_{\rm{inf}})$ for the left-hand side. Thus, the condition in eq. \eqref{eq:NefoldsPuzzles} also applies to the horizon problem.

The scale-invariance of the power-spectrum of the small density fluctuations, imprinted on the CMB as tiny anisotropies, is also a consequence of the accelerated expansion during inflation. At the beginning of inflation, small-scale perturbations, lying deep inside the Hubble radius, are generated by Minkowski space-time quantum fluctuations (since the space-time curvature can be neglected). As the universe undergoes accelerated expansion, perturbations of fixed co-moving wavelength cross outside the shrinking co-moving Hubble sphere. As they become superhorizon, the Hubble friction term starts to dominate and they become over-damped. Since $H$ is approximately constant during inflation and is the only scale determining the evolution of perturbations after Hubble exit, the perturbations which leave the Hubble sphere during inflation have an almost scale-invariant power-spectrum. As the co-moving Hubble radius begins to increase after inflation, perturbations of a given co-moving wavelength start to re-enter the horizon, accounting for the observed approximate scale-invariance of density perturbations over a range of different scales. By letting the Hubble sphere shrink during inflation and then begin to expand as the universe becomes radiation and then matter dominated we provide a causal mechanism for producing seemingly-acausal correlations in the density perturbations.\footnote{This also explains the well-defined acoustic peaks in the CMB on shorter lengthscales.} Basically, inflation opens up the past light cones for fundamental observers (those who are stationary with respect to the cosmic grid).

Inflation also provides a straightforward solution to the monopole problem. If the phase transition occurs during or before inflation, we need to make sure that there is sufficient number of {\it e}-folds of accelerated expansion after the transition to dilute the concentration of the relics. If the gauge symmetry is broken after inflation, the correlation length at the phase transition, equal to the particle horizon, is substantially increased in comparison to standard cosmology and we can again put a lower bound on the duration of inflation. For instance, let us consider the generation of monopoles after spontaneous symmetry breaking at the GUT scale, $E_{\rm{GUT}}=10^{16}\,\rm{GeV}$, immediately after the end of inflation. The number density of monopoles is determined by the correlation length of the scalar fields, which is set by the particle horizon and is $\sim a_{\rm{end}}/(a_{\rm{init}}H_{\rm{inf}})=e^N/H_{\rm{inf}}$. On the other hand, $H_{\rm{inf}}\sim E_{\rm{GUT}}^2/\mpl$. Hence, there is roughly one monopole per $E_{\rm{GUT}}^{-6}\mpl^{3}e^{3N}$. Assuming the universe becomes radiation dominated and reaches local thermal equilibrium soon after the formation of monopoles, the number density of photons at that time is $\sim E_{\rm{GUT}}^3$. Ignoring subsequent processes that can change the number of photons and annihilation of monopoles, the ratio of the number densities of monopoles and photons remains constant, since each $\propto a^{-3}$. Thus, $n^{\rm{mon}}_0/n^{\gamma}_0\sim e^{3N}E_{\rm{GUT}}^{3}/\mpl^{3}=10^{-9}e^{3N}$. For less than $10^{-39}$ monopoles per photon \cite{Weinberg:2008zzc}, as suggested by terrestrial experiments, $N>23$. An identical bound is obtained if GUT scale monopoles are generated during or before inflation. A possibility that cannot be resolved by inflation is the production of massive particles during the reheating process after inflation. If the reheating temperature is low enough, thermal particle production of dangerous massive relics, that can ruin the successful BBN in standard cosmology, can be evaded. On the other hand, non-thermal production has to be dealt with on a case-by-case basis.

\subsection{Dynamics of inflation and setting the scene for reheating}

%Inflation is introduced to solve the most serious issues undermining the successful standard cosmology. 
The semi-classical theory of inflation provides not only a solution to the horizon, flatness and monopole problems, but more importantly predicts the generation of density perturbations exhibiting a nearly scale-invariant power-spectrum. These act as seeds for the large-scale structure in the late universe. 

The inflationary paradigm can be interpreted in at least two different ways. We could think of it as an approximate description (some sort of a parametrisation)% of the early universe
, which does not capture the actual physical laws, due to its semi-classical nature. We could also argue that the universe genuinely underwent a stage of accelerated expansion, driven by a scalar condensate whose origin can be traced back to any of the high-energy models going beyond the Standard Model of Particle Physics. This way of thinking makes inflation the link between quantum gravity or/and extensions of the Standard Model, and the well-understood physics of BBN in standard cosmology.

In this note, we consider the most common (and consistent with observations \cite{Ade:2015lrj}) models of inflation, in which a single scalar field $\phi$, called the inflaton, sources the accelerated expansion of the universe, with action
\Beq
\label{eq:ActionInflaton}
S=\int d^4x \sqrt{-g}\left[-\frac{\mpl^2}{2}R+\frac{1}{2}\nabla_{\mu}\phi\nabla^{\mu}\phi-V(\phi)\right]+S_{\rm{matter}}\,.
\Eeq
We limit ourselves to models, minimally coupled to gravity, with canonical kinetic terms\footnote{There are examples of non-minimal and non-canonical models in which a conformal transformation or a field redefinition can reduce the action to the form given in eq. \eqref{eq:ActionInflaton}.}. The matter action term, $S_{\rm{matter}}$, contains the entire information regarding the other constituents of the matter sector, including the Standard Model Lagrangian as well as the terms describing the couplings of the inflaton to other fields. 

\subsubsection{Homogeneous dynamics of inflation}
Isotropy and homogeneity require that the dominant component of the scalar field depends on $t$ only, $\bar{\phi}(t)$. This scalar condensate provides the classical background configuration during inflation (and the initial stages of reheating). The energy density and the pressure of the isotropic and homogeneous scalar fluid are simply
\Beq
\label{eq:RhoPInfl}
\rho_{\bar{\phi}}=\frac{1}{2}{\dot{\bar{\phi}}}^{\,2}+V(\bar{\phi})\,,\qquad p_{\bar{\phi}}=\frac{1}{2}{\dot{\bar{\phi}}}^{\,2}-V(\bar{\phi})\,.
\Eeq
To have acceleration, $\ddot{a}>0$, the Raychaudhuri equation, eq. \eqref{eq:FRWeqs}, demands $\rho_{\bar{\phi}}+3p_{\bar{\phi}}<0$. This means that inflation occurs as long as ${\dot{\bar{\phi}}}^{\,2}<V(\bar{\phi})$. The Friedmann equation, eq. \eqref{eq:FRWeqs}, and the Euler-Lagrange equation for $\bar{\phi}$ following from the action in eq. \eqref{eq:ActionInflaton} are
\Beq
\label{eq:FriedKG}
H^2=\frac{1}{3\mpl^2}\left[\frac{1}{2}{\dot{\bar{\phi}}}^{\,2}+V(\bar{\phi})\right]\,,\qquad\ddot{\bar{\phi}}+3H\dot{\bar{\phi}}+\partial_{\bar{\phi}}V(\bar{\phi})=0\,.
\Eeq

\subsubsection{Slow-roll inflation}

Having derived the equations of motion and shown that accelerated expansion is possible, we need to find what conditions $V$ has to satisfy to have enough number of {\it e}-folds of inflation to solve the horizon and flatness problems. Note that $\ddot{a}=a(\dot{H}+H^2)$, hence
\Beq
\epsilon_{H}\equiv -\frac{\dot{H}}{H^2}<1\,,
\Eeq
has to hold to have $\ddot{a}>0$. Inflation ends when $\epsilon_{H}=1$, corresponding to $\ddot{a}=0$. Since $H^{-1}$ is the characteristic time-scale for one {\it e}-fold of expansion (recall $dN=d\ln a=Hdt$), known as the Hubble time, $\epsilon_{H}=-d\ln H/dN<1$ implies that the time-scale over which the fractional decrease in $H$ is significant is greater than a Hubble time. Or in other words the rate of decrease of $H$ must be slower than the rate of expansion of the universe in order to have inflation. To achieve sufficiently many {\it e}-folds (at least $40$ up to $60$, see Section \ref{sec:InflationIntro}) of inflation we need $\epsilon_{H}$ to be much less than $1$ for a long enough period, implying that for most of the time $\epsilon_{H}\ll1$. The parameter quantifying the rate of change of $\epsilon_{H}$ is
\Beq
\eta_{H}\equiv \frac{\dot{\epsilon}_H}{H\epsilon_H}=\frac{d\ln \epsilon_H}{dN}\,.
\Eeq
For $\epsilon_{H}$ to increase slowly, over many Hubble times, $|\eta_{H}|$ has to be less than unity%\footnote{An absolute value is taken, because for $\eta_{H}<-1$ inflation never ends.}
. But because of the large number of {\it e}-folds required by observations, $|\eta_{H}|\ll1$ has to hold\footnote{If $\eta_{H}\sim\mathcal{O}(1)$, inflation still takes place for roughly $\Delta N\sim\ln(1/\epsilon_H)$. If $\epsilon_H$ is several orders of magnitude less than unity, the number of {\it e}-folds of inflation is much less than the one required by observations.}. All of this basically means that $H\approx \rm{const}$ for most of inflation, and that the scale factor increases quasi-exponentially. That is why this period is also called quasi-de Sitter expansion. Current observational constraints are roughly $\epsilon_H<0.01$ and $\eta_H\approx0.03\pm0.01$ \cite{Ade:2015lrj} and are in support of this picture.

Using the Friedmann equation, eq. \eqref{eq:FRWeqs}, the energy conservation equation, eq. \eqref{eq:EnergyConservationFRW}, and the expressions for the energy density and pressure of the inflaton condensate, eq. \eqref{eq:RhoPInfl}, we find that $\epsilon_{H}={3\dot{\bar{\phi}}}^{\,2}/(2\rho_{\bar{\phi}})$. Thus, $\epsilon_{H}\ll1$ implies a negligible contribution of the kinetic energy density to the total energy density of the condensate, which also means that ${\dot{\bar{\phi}}}^{\,2}/2\ll V(\bar{\phi})$ during inflation, consistent with our comments under eq. \eqref{eq:RhoPInfl}. Hence, $V$ has to be very flat, for $\bar{\phi}$ to roll sufficiently slowly. This is called {\it slow-roll inflation}. For slow-roll inflation to last long enough, we need the kinetic energy density to remain small. This means that the fractional change in $\dot{\bar{\phi}}$ during one expansion time $H^{-1}$, $|\ddot{\bar{\phi}}/\dot{\bar{\phi}}|H^{-1}$, has to be much less than $1$. It implies, given $\epsilon_{H}\ll1$, that $|\eta_{H}|\ll1$, since one can show that $\eta_H=2\epsilon_H+2\ddot{\bar{\phi}}/(\dot{\bar{\phi}}H)$.

Finally, we are in a position to put constraints on the form of $V$ that can support inflation for sufficiently long periods. Applying ${\dot{\bar{\phi}}}^{\,2}/2\ll V(\bar{\phi})$ to the first expression in eq. \eqref{eq:FriedKG}, yields $H^2\approx V/3\mpl^2$, whereas substituting $|\ddot{\bar{\phi}}/\dot{\bar{\phi}}|H^{-1}\ll1$ in the second expression in eq. \eqref{eq:FriedKG} gives $3H\dot{\bar{\phi}}\approx-\partial_{\bar{\phi}}V(\bar{\phi})$. Hence, $\epsilon_{H}\approx (\mpl^2/2)(\partial_{\bar{\phi}}V/V)^2$. Taking the time derivative of $3H\dot{\bar{\phi}}\approx-\partial_{\bar{\phi}}V(\bar{\phi})$ yields $\epsilon_H-\ddot{\bar{\phi}}/(\dot{\bar{\phi}}H)\approx\mpl^2\partial^2_{\bar{\phi}}V/V$. These two ratios of $V$ and its derivatives are conventionally denoted as
\Beq
\label{eq:SlowRollPotential}
\epsilon_{V}\equiv\frac{\mpl^2}{2}\left(\frac{\partial_{\bar{\phi}}V(\bar{\phi})}{V}\right)^2\,,\qquad \eta_{V}\equiv\mpl^2\frac{\partial_{\bar{\phi}}^2V(\bar{\phi})}{V}\,,
\Eeq
and are known as the potential slow-roll parameters (similarly, $\epsilon_H$ and $\eta_H$ are known as the Hubble slow-roll parameters). $\epsilon_H$ and $|\eta_H|$ $\ll1$ is equivalent to $\epsilon_V$ and $|\eta_V|$ $\ll1$.

To see that within the slow-roll approximation, the expansion during inflation can be exponentially large consider\footnote{The total number of {\it e}-folds of inflation is defined as the first integral with $\epsilon_H(t_{\rm{init}})=\epsilon_H(t_{\rm{end}})=1$ and $\epsilon_H(t)<1$ for $t_{\rm{init}}<t<t_{\rm{end}}$.}
\Beq
e^{N}\equiv\frac{a_{\rm{end}}}{a_{\rm{init}}}=\exp\left[\int_{t_{\rm{init}}}^{t_{\rm{end}}}H(t)dt\right]&=\exp\left[\int_{{\bar{\phi}}_{\rm{init}}}^{{\bar{\phi}}_{\rm{end}}}H(\bar{\phi})\frac{d\bar{\phi}}{\dot{\bar{\phi}}}\right]\\&\approx\exp\left[-\mpl^{-2}\int_{{\bar{\phi}}_{\rm{init}}}^{{\bar{\phi}}_{\rm{end}}}\frac{V(\bar{\phi})}{\partial_{\bar{\phi}}V(\bar{\phi})}d\bar{\phi}\right]\,,
\Eeq
where we assume $0<V({\bar{\phi}}_{\rm{end}})<V({\bar{\phi}}_{\rm{init}})$ and $\Delta\bar{\phi}\equiv{\bar{\phi}}_{\rm{init}}-{\bar{\phi}}_{\rm{end}}>0$, implying a positive argument in the last exponential. The slow-roll condition $\epsilon_V\ll1$ leads to $N\gg \Delta\bar{\phi}/\mpl$. If the value of the inflaton changes by $\sim\mpl$, we definitely get a huge number of {\it e}-foldings. Note that such large field values do not mean that the quantum nature of gravity becomes important. For this to happen the energy density of the condensate must be $\sim\mpl^4$. This can be easily avoided, even for $\bar{\phi}\gtrsim\mpl$, if $V(\bar{\phi})$ is proportional to a sufficiently small coupling constant. None of the potential slow-roll parameters, eq. \eqref{eq:SlowRollPotential}, depends on it.

We can find the approximate trajectory of the inflaton during inflation. Since, during slow-roll $\epsilon_H\approx\epsilon_V\ll1$, $\dot{\phi}\approx-\mpl\partial_{\bar{\phi}}V/\sqrt{3V}$. As a test case, we consider the simplest form for the inflaton potential, i.e., $V(\phi)=m^2\phi^2/2$. In general, all monomial potentials satisfy the slow-roll conditions, eq. \eqref{eq:SlowRollPotential}, for some $\Delta\bar{\phi}\gtrsim\mpl$. These models belong to the class of Chaotic inflation \cite{LINDE1983177,Linde:2005ht}. It encompasses all models having $V$ that supports slow-roll inflation for $\Delta\bar{\phi}\sim\mathcal{O}(\mpl)$ or smaller. 
%a large initial value for the inflaton which then slow rolls towards the minimum of $V$ at smaller field values. 
In these lecture notes, we will concentrate on Chaotic inflation models\footnote{Examples of single-field models that do not belong to Chaotic inflation include Small-field models in which necessarily $\Delta\bar{\phi}\ll\mpl$, models in which higher-order kinetic terms or higher-order curvature terms, instead of $V$, drive inflation and models in which phase transitions stop or trigger inflation, e.g., Old and New inflation, respectively.}. For the quadratic potential the slow-roll trajectory and the expansion law take the simple approximate analytic form ($t_{\rm{init}}=0$)
\Beq
\label{eq:AttractorSlowRoll}
\dot{\bar{\phi}}\approx-\sqrt{\frac{2}{3}}\mpl m\,,\qquad a\approx a_{\rm{init}}\exp\left[\frac{mt}{\sqrt{6\mpl}}\left(\bar{\phi}_{\rm{init}}-\frac{\mpl m}{\sqrt{6}}t\right)\right]\,.
\Eeq
It breaks down towards the end of inflation, $\bar{\phi}_{\rm{end}}\approx\mpl$ for which the slow-roll conditions in eq. \eqref{eq:SlowRollPotential} are violated. This solution is also known as the attractor solution, since one can show that for a broad range of ${\bar{\phi}}_{\rm{init}}$ \cite{GOLDWIRTH1992223,Handley:2014bqa}, even for large $\dot{\bar{\phi}}_{\rm{init}}$ such that $\epsilon_H(t_{\rm{init}})>1$, the field velocity decays very rapidly and $\bar{\phi}(t)$ and $a(t)$ quickly approach eq. \eqref{eq:AttractorSlowRoll}. This goes to show how broad the set of initial conditions is that can lead to an inflationary stage in chaotic models.

Speaking of initial conditions, the term `chaotic' derives from the possibility of having initially a scalar field varying randomly with position, i.e., having almost arbitrary initial conditions for the inflaton, and still getting slow-roll inflation after that. Even if the value of the inflaton varies from one spatial region to another, there should be a patch of space in which the inflaton looks uniform enough and has a value for which the slow-roll conditions in eq. \eqref{eq:SlowRollPotential} are satisfied, e.g., a super-Planckian value for monomial potentials \cite{GOLDWIRTH1992223}. One can easily show that the initial physical size of the homogeneous patch, $L_{\rm{init}}=a_{\rm{init}}l$, has to obey $H_{\rm{init}}L_{\rm{init}}\gg\bar{\phi}_{\rm{init}}/\mpl$ for the gradients to be negligible. For monomial potentials, this implies that a sufficiently large uniform patch has to be super-Hubble initially\footnote{Which is interpreted as a requirement for fine-tuning of the initial conditions for inflation by some authors \cite{NYAS:NYAS249,Carroll:2005it}.} (and super-Planckian after imposing the condition of sub-Planckian $V_{\rm{init}}$).

\subsubsection{End of slow-roll inflation}

For inflation to be successful it must feature a graceful exit into the deceleration stage of standard cosmology; otherwise the homogeneity and isotropy of the universe are destroyed. A famous example of non-graceful exit is Alan Guth's Old inflation \cite{PhysRevD.23.347} in which the inflaton is initially trapped in a false vacuum. As the inflaton leaks through the potential barrier and forms bubbles of true vacuum, the energy released in the transition ends up concentrated within the bubble walls. If the bubbles are able to merge, a homogeneous and isotropic universe emerges. However, the bubbles never collide, since the background false-vacuum space in which they formed, never stops inflating. Hence, for an observer located inside a bubble the universe would appear highly anisotropic and inhomogeneous, since structure has to grow out of the energy concentrated in the bubble walls. The graceful exit problem is naturally avoided in Chaotic inflation. For power-law potentials, the homogeneous inflaton background simply begins to oscillate about the potential minimum. One can easily determine the oscillatory attractor solution for a quadratic minimum. We put $\bar{\phi}=\sqrt{6}H(\mpl/m)\cos\theta$ and $\dot{\bar{\phi}}=\sqrt{6}H\mpl\sin\theta$ to satisfy the first expression in eq. \eqref{eq:FriedKG}. After differentiating it with respect to time and using the second expression in eq. \eqref{eq:FriedKG} we find $\dot{H}=-3H^2\sin^2\theta$. This implies that $H$ decays during the oscillatory stage as $t^{-1}$. Taking the time derivative of the new definition of $\bar{\phi}$ in terms of $\theta$ and using the definition of $\dot{\bar{\phi}}$ in terms of $\theta$ we find $\dot{\theta}=-m-(3/2)H\sin(2\theta)$. The second term on the right decays with time, so $\theta\approx mt$ up to a constant for $mt\gg1$. We can use this result in the expression for $\dot{H}$. After integration we find
\Beq
\label{eq:PrehBackAppr}
H\approx \frac{2}{3t}\left(1+\frac{\sin(2mt)}{2mt}\right)\,, \qquad {\rm{and}} \qquad \bar{\phi}\approx\frac{2\sqrt{2}\mpl\cos(mt)}{\sqrt{3}mt}\left(1+\frac{\sin(2mt)}{2mt}\right)\,.
\Eeq
This decaying scalar field condensate provides the classical background during the reheating phase. Note that $a\propto t^{2/3}$ up to subdominant decaying oscillating terms, which implies that the universe is in a dust-dominated state of expansion. Ultimately, the universe has to reheat itself to reach eventually a radiation-like state of expansion, with the inflaton energy transferred into radiation, baryons and leptons. Also note that even if the oscillating terms are very small, they can play an important role for the space-time curvature (neglecting the spatial curvature for simplicity)
\Beq
\label{eq:RicciOsc}
R=-6\left(\frac{\ddot{a}}{a^2}+\frac{\dot{a}^2}{a^2}\right)\approx -\frac{4}{3t^2}\left(1+3\cos(2mt)\right)\,.
\Eeq

Before moving forward, we should point out that the state of expansion of a universe dominated by a homogeneous oscillating scalar, about the minimum of its potential, depends on the form of $V$. For simple power-laws, $V\propto|\phi|^{2n}$, where $n$ need not be an integer, one can easily determine the temporal mean equation of state during the oscillatory phase. Ignoring expansion, since $H$ decays with time after inflation, and multiplying by $\bar{\phi}$ the second expression in eq. \eqref{eq:FriedKG}, we find $\langle\bar{\phi}\partial_{\bar{\phi}}V\rangle=-\langle\bar{\phi}\ddot{\bar{\phi}}\rangle=\langle{\dot{\bar{\phi}}}^2\rangle$. The angle brackets represent time averaging over many oscillations. The last equality follows from virialization, i.e., $0=\langle d(\bar{\phi}\dot{\bar{\phi}})/dt\rangle=\langle{\dot{\bar{\phi}}}^2\rangle+\langle\bar{\phi}\ddot{\bar{\phi}}\rangle$. Thus, assuming $\langle p_{\bar{\phi}}/\rho_{\bar{\phi}}\rangle\approx\langle p_{\bar{\phi}}\rangle/\langle\rho_{\bar{\phi}}\rangle$, we find \cite{Turner:1983he}
\Beq
\langle w\rangle\approx\frac{\langle{\dot{\bar{\phi}}}^{\,2}\rangle/2-\langle V\rangle}{\langle{\dot{\bar{\phi}}}^{\,2}\rangle/2+\langle V\rangle}=\frac{\langle\bar{\phi}\partial_{\bar{\phi}}V\rangle/2-\langle V\rangle}{\langle\bar{\phi}\partial_{\bar{\phi}}V\rangle/2+\langle V\rangle}=\frac{n-1}{n+1}\,.
\Eeq
For quadratic potentials, $n=1$, we have the expected $w=0$ matter-like equation of state, whereas for quartic potentials, $n=2$, we have a radiation-like equation of state, $w=1/3$. When $n\leq1/2$, $w\leq-1/3$, and eq. \eqref{eq:FRWPowerLaw} tells us that $\ddot{a}>0$ -- the universe inflates. However, consistency requires that the inflaton oscillates around a potential that has a non-singular first derivative at its minimum, for the equation of motion to be well-defined for all field values, implying the condition $n>1/2$, i.e., the potential has to be steeper than linear at the minimum. Oscillations about such minima always lead to a decelerating stage of expansion with $w>-1/3$.

\subsubsection{Cosmological perturbations from inflation}

Having described what the homogeneous and isotropic universe looks like at the end of inflation, we now consider the small deviations from the FRW approximation. After all, these small departures enable us to distinguish between different models. As mentioned above, the tiny anisotropies measured in the CMB, as well as the seeds for Large Scale Structure (LSS) can be explained within the inflationary paradigm, as being microscopic quantum fluctuations, stretched to cosmic sizes during inflation. As we will see in later sections, they also laid down the seeds for particle production during reheating. To understand the initial conditions for this process, we need to consider their evolution during inflation.

We expand the metric and the energy momentum tensor about their background values \cite{PhysRevD.22.1882}
\Beq
g_{\mu\nu}(x^\alpha)=\bar{g}_{\mu\nu}(t)+\delta g_{\mu\nu}(x^{\alpha})\,,\qquad T_{\mu\nu}(x^\alpha)=\bar{T}_{\mu\nu}(t)+\delta T_{\mu\nu}(x^{\alpha})\,,
\Eeq
where $\bar{g}_{\mu\nu}(t)$ is given in eq. \eqref{eq:FRWmetricIntro} and we set the spatial curvature, $K$, to zero, since during inflation it quickly becomes negligible. While we know the non-zero components of $\bar{T}_{\mu\nu}(t)$, see eq. \eqref{eq:TmunuMeanFRW}, it is easier to work with a co-variant form for the tensor. Since the FRW universe is filled with a perfect homogeneous fluid, i.e., a fluid that looks the same in all directions for all co-moving observers at equal cosmic times, the rank-2 tensor $\bar{T}_{\mu\nu}(t)$ has to be a linear combination of $\bar{\rho}(t)\bar{g}_{\mu\nu}(t)$, $\bar{p}(t)\bar{g}_{\mu\nu}(t)$, $\bar{\rho}(t)\bar{u}_{\mu}(t)\bar{u}_{\nu}(t)$ and $\bar{p}(t)\bar{u}_{\mu}(t)\bar{u}_{\nu}(t)$, where $\bar{u}_{\mu}(t)$ is the 4-velocity of a co-moving observer. The only linear combination that respects homogeneity and isotropy is $\bar{T}_{\mu\nu}(t)=(\bar{\rho}+\bar{p})\bar{u}_{\mu}\bar{u}_{\nu}-\bar{g}_{\mu\nu}\bar{p}$, where we used $[1,0,0,0]^{T}$ for $\bar{u}^{\mu}(t)$. More generally, in an arbitrary gravitational field, a perfect fluid is a medium% which appears isotropic for all observers in co-moving with the fluid, locally inertial Cartesian frames, i.e.,
, with energy momentum tensor
\Beq
T_{\mu\nu}(x^{\alpha})=\left[\rho(x^{\alpha})+p(x^{\alpha})\right]u_{\mu}(x^{\alpha})u_{\nu}(x^{\alpha})-g_{\mu\nu}(x^{\alpha})p(x^{\alpha})\,.
\Eeq

We now write the actual forms of the perturbations $\delta g_{\mu\nu}(x^{\alpha})$ and $\delta T_{\mu\nu}(x^{\alpha})$. We use the conformal time, $d\tau\equiv dt/a(t)$, which simplifies the background metric, $\bar{g}_{\mu\nu}(\tau)=a^{2}(\tau)\eta_{\mu\nu}$. The most general metric perturbations are
\Beq
\label{eq:PerturbedMetricIntro}
ds^2&=(\bar{g}_{\mu\nu}+\delta g_{\mu\nu})dx^{\mu}dx^{\nu}\\
    &=\left(1+2\varphi\right)a^2(\tau)d\tau^2+2\left(\partial_i\mathcal{B}+\mathcal{S}_{i}\right)a^2(\tau)dx^id\tau\\
    &\qquad-\big[\left(1-2\psi\right)\delta_{ij}-2\partial_i\partial_j\mathcal{E}-\partial_j\mathcal{K}_{i}-\partial_i\mathcal{K}_{j}-\tilde{h}_{ij}\big]a^2(\tau)dx^idx^j\,,
\Eeq
where $\varphi(x^\sigma)$, $\mathcal{B}(x^{\sigma})$, $\psi(x^{\sigma})$, $\mathcal{E}(x^{\sigma})$ are scalar perturbations, $\mathcal{S}_{i}(x^{\sigma})$, $\mathcal{K}_i(x^{\sigma})$ are divergence-free $3$-vector perturbations, and $\tilde{h}_{ij}(x^{\sigma})$ is a traceless transverse $3$-tensor perturbation. Consistency requires the energy density, pressure and 4-velocity fields to be also perturbed
\Beq
\rho(x^{\alpha})=\bar{\rho}(t)+\delta{\rho}(x^{\alpha})\,,\qquad p(x^{\alpha})=\bar{p}(t)+\delta{p}(x^{\alpha})\,,\qquad u_{\mu}(x^{\alpha})=\bar{u}_{\mu}(t)+\delta{u}_{\mu}(x^{\alpha})\,,
\Eeq
where $\bar{u}_\mu=[a,0,0,0]^T$, $\delta{u}_{\mu}\equiv[\delta u_0,\partial_i \delta u^{\parallel}+\delta u^{\perp}_i]^T$ and $\partial_i u^{\perp}_i=0$. Since the 4-velocity of an observer is normalized, i.e., $\bar{g}_{\mu\nu}\bar{u}^{\mu}\bar{u}^{\nu}=1$ and $g_{\mu\nu}u^{\mu}u^\nu=1$ one can show to linear order that $\delta u_0=a\varphi$. In deriving this expression, we have used $\delta g^{\mu\nu}=-\bar{g}^{\mu\alpha}\delta g_{\alpha\beta}\bar{g}^{\beta\nu}$, which also holds to linear order for $\bar{g}_{\mu\nu}\bar{g}^{\nu\alpha}=\delta^{\alpha}_{\mu}$ and $g_{\mu\nu}g^{\nu\alpha}=\delta^{\alpha}_{\mu}$. This also implies $u^{\mu}=a^{-1}[1-\varphi,-a^{-1}\partial_i \delta u^{\parallel}-a^{-1}\delta u^{\perp}_i-\partial_i\mathcal{B}-\mathcal{S}_i]^T$, to first order in perturbations. The perturbations in the energy momentum tensor then take the form
\Beq
\label{eq:TmunuPerfectFluid1stOrder}
\delta{T}^{\mu}{}_{\nu}&=(\delta\rho+\delta p)\bar{u}^{\mu}\bar{u}^{\nu}-\delta p\delta^{\mu}_{\nu}-(\bar{\rho}+\bar{p})\bar{u}^{\mu}\delta u_{\nu}-(\bar{\rho}+\bar{p})\delta u^{\mu}\bar{u}_{\nu}\,,\\
\delta{T}^{0}{}_{0}&=\delta{\rho}\,,\qquad \delta T^0{}_i=-(\bar{\rho}+\bar{p})a^{-1}(\partial_i \delta u^{\parallel}+\delta u^{\perp}_i)\,,\\
\delta T^{i}{}_j&=-\delta p\delta_{ij}\,,\qquad \delta T^i{}_0=(\bar{\rho}+\bar{p})a^{-1}(a^{-1}\partial_i \delta u^{\parallel}+a^{-1}\delta u^{\perp}_i+\partial_i\mathcal{B}+\mathcal{S}_i)\,.
\Eeq

Decomposing perturbations into scalars, divergence-free vectors and traceless transverse tensors, also known as the scalar-vector-tensor decomposition, is very useful since the Einstein equations decouple the three kinds of modes to linear order. This is a consequence of the symmetries of the FRW background. The Einstein equations are also invariant under diffeomorphisms, i.e., space-time coordinate transformations
\Beq
\label{eq:DiffDeff}
x^{\mu}\rightarrow x'^{\mu}=x^{\mu}+\xi^{\mu}(x^{\alpha}).
\Eeq
We assume $\xi^{\mu}=[\xi^{0},\partial_i\xi^{\parallel}+\xi^{\perp}_i]^{T}$ to be small, of the order of the metric and energy momentum tensor perturbations. Since the metric transforms as
\Beq
g'_{\mu\nu}(x'^{\alpha})=\frac{\partial x^{\beta}}{\partial x'^{\mu}}\frac{\partial x^{\gamma}}{\partial x'^{\nu}}g_{\beta\gamma}(x^{\alpha})\,,
\Eeq
the metric perturbations at $x^{\alpha}$ transform, to linear order, as\footnote{It is understood that $\bar{g}_{\mu\nu}(x^{\alpha})=\bar{g}_{\mu\nu}'(x^{\alpha})$, i.e., eq. \eqref{eq:DiffDeff} yields $\delta g_{\mu\nu}(x^{\alpha})\rightarrow \delta g_{\mu\nu}'(x^{\alpha})=\delta g_{\mu\nu}(x^{\alpha})+\Delta\delta g_{\mu\nu}(x^{\alpha})$. }
\Beq
\Delta\delta g_{\mu\nu}(x^{\alpha})=g'_{\mu\nu}(x^{\alpha})-g_{\mu\nu}(x^{\alpha})&\approx g'_{\mu\nu}(x'^{\alpha})-\frac{\partial g_{\mu\nu}}{\partial x^{\beta}}\xi^{\beta}-g_{\mu\nu}(x^{\alpha})\\
&\approx -\bar{g}_{\beta\nu}(x^{\alpha})\frac{\partial \xi^{\beta}}{\partial x^{\mu}}-\bar{g}_{\mu\beta}(x^{\alpha})\frac{\partial \xi^{\beta}}{\partial x^{\nu}}-\frac{\partial \bar{g}_{\mu\nu}(x^{\alpha})}{\partial x^{\beta}}\xi^{\beta}\,.
\Eeq
Similarly,
\Beq
\label{eq:DiffEnergyMomentumTensor}
\Delta\delta T^{\mu}{}_{\nu}(x^{\alpha})\approx \bar{T}^{\beta}{}_{\nu}(x^{\alpha})\frac{\partial \xi^{\mu}}{\partial x^{\beta}}-\bar{T}^{\mu}{}_{\beta}(x^{\alpha})\frac{\partial \xi^{\beta}}{\partial x^{\nu}}-\frac{\partial \bar{T}^{\mu}{}_{\nu}(x^{\alpha})}{\partial x^{\beta}}\xi^{\beta}\,.
\Eeq
From this follows that perturbations depend on our choice of space-time coordinates, e.g.,
\Beq
\label{eq:DiffTransfPsiVel}
\Delta\psi=\mathcal{H}\xi^0\,,\qquad\Delta\delta u^{\parallel}=-a\xi^{0}\,.
\Eeq
This issue can be resolved by working in diffeomorphism invariant quantities that take the same values for all choices of coordinates, i.e., in all gauges. For instance, the quantity
\Beq
\mathcal{R}=\psi+\frac{\mathcal{H}}{a}\delta u^{\parallel}\,,
\Eeq
is gauge invariant. It is known as the co-moving curvature perturbation. We can construct other gauge-invariant quantities, e.g., 
\Beq
\label{eq:BardeenIntro}
\Phi &= \varphi - \frac{1}{a}\partial_{\tau}\left[a\left(\mathcal{B}-\partial_{\tau}\mathcal{E}\right)\right]\,,\\
\Psi &= \psi + \mathcal{H}\left(\mathcal{B}-\partial_{\tau}\mathcal{E}\right)\,,
\Eeq
known as Bardeen variables. We can either calculate such quantities directly, by solving their equations of motion, or we can fix the gauge first by imposing conditions on the gauge dependent perturbations and then solve for the metric and energy momentum tensor perturbations. No matter what gauge we choose in the latter case, the gauge-invariant quantities always have the same values.

To make further progress, we need to specify the energy momentum tensor. In single-field inflation, see eq. \eqref{eq:ActionInflaton}, the contribution from the matter action term to the energy momentum tensor vanishes at linear order. Hence, the scalar metric perturbations are coupled to linear order only to the perturbation in the inflaton
\Beq
\phi(x^{\alpha})=\bar{\phi}(t)+\delta \phi(x^{\alpha})\,.
\Eeq
The unperturbed energy momentum tensor is
\Beq
\label{eq:EMTIntro}
T_{\mu\nu }=\nabla_{\mu } \phi \nabla_{\nu } \phi-g_{\mu \nu } \left[\frac{1}{2}\nabla^{\alpha } \phi  \nabla_{\alpha } \phi -V\right],
\Eeq
from which follows that
\Beq
\delta T^0{}_{i}=\frac{\partial_{\tau}\bar{\phi}}{a^2}\partial_{i}\delta{\phi}\,.
\Eeq
Given the expressions for the background pressure and energy density in eq. \eqref{eq:RhoPInfl}, the scalar velocity perturbation can be found directly by using eq. \eqref{eq:TmunuPerfectFluid1stOrder} to be
\Beq
\delta u^{\parallel}=\frac{a\delta \phi}{\partial_{\tau}\bar{\phi}}\,,
\Eeq
implying $\mathcal{R}=\psi+\mathcal{H}\delta \phi/\partial_{\tau}\bar{\phi}$. We can calculate this quantity during inflation, working in the slow-roll approximation. The most suitable gauge for analytical analysis is the spatially-flat gauge defined as $\psi=\mathcal{E}=0$ (fixing the two scalar perturbations removes the gauge freedom due to $\xi^0$ and $\xi^{\parallel}$). This implies that the scalar modes in $\delta g_{ij}$ vanish. The second order terms in the action in eq. \eqref{eq:ActionInflaton} which couple the non-zero scalar metric perturbations, $\delta g_{00}$ and $\delta g_{0i}$, to the inflaton perturbations, coming from the $g_{\mu\nu}\partial^{\mu}\phi\partial^{\nu}\phi/2$ term are proportional to $\dot{\bar{\phi}}$, whereas those coming from $\sqrt{-g}V$ are $\sim \partial_{\bar{\phi}}V$. This means that all couplings between metric perturbations and inflaton perturbations are slow-roll suppressed and can be ignored during inflation. The effective mass term due to the inflaton potential also vanishes in the slow-roll limit, $\partial^2_{\bar{\phi}}V\sim\eta_V H^2$. Thus, the second order action for the Fourier components of the inflaton perturbations, without the slow-roll suppressed terms, reduces to
\Beq
\label{eq:2ndOrderAction}
&S^{(2)}_{\rm{sr}}=\int \text{d}\tau L^{(2)}_{\rm{sr}}(\tau)=\int d\tau{\int d^3k\,\,b_{\rm{sr}}(k,\tau) \left[\frac{1}{2}|\partial_\tau \delta\phi_{\bk}|^2-\frac{1}{2}\omega^2_{\rm{sr}}(k,\tau)|\delta\phi_{\bk}|^2\right]}\,,\\
&b_{\rm{sr}}(k,\tau)=a^2\,,\qquad \omega_{\rm{sr}}^2(k,\tau)=k^2\,.
\Eeq
Note that due to the $b_{\rm{sr}}$ pre-factor the kinetic term in the action is not canonically-normalized. If one wishes, $b_{\rm{sr}}$ can be absorbed into a field re-definition which makes the kinetic term canonical. 

Having derived the second order action for the inflaton fluctuations, we can now quantize them. First we need the conjugate momentum density 
\Beq
\label{eq:piIntro}
\pi^{\rm{sr}}_{\bk}(\tau)=\frac{\delta\left(L^{(2)}_{\rm{sr}}(\tau)\right)}{\delta\left(\partial_\tau \delta\phi_{\bk}\right)}=b_{\rm{sr}}\partial_\tau \delta\phi_{-\bk}\,,
\Eeq
where we have taken the functional derivative of the Lagrangian and made use of the reality of the inflaton fluctuations, i.e., $\delta\phi^*_{\bk}=\delta\phi_{-\bk}$. In the Heisenberg picture, the field operators $\delta\hat{\phi}_\bk$ and their conjugate momenta operators $\hat{\pi}^{\rm{sr}}_\bk$  satisfy the equal time commutators:
\Beq
\label{eq:Introcommutators}
\left[\delta\hat{\phi}_\bk(\tau),\delta\hat{\phi}_{\bq}(\tau)\right]=0\,,\quad\left[\hat{\pi}^{\rm{sr}}_{\bk}(\tau),\hat{\pi}^{\rm{sr}}_{\bq}(\tau)\right]=0\,,\quad\left[\delta\hat{\phi}_\bk(\tau),\hat{\pi}^{\rm{sr}}_{\bq}(\tau)\right]=i\left(2\pi\right)^{-3}\delta(\bk-\bq)\,.
\Eeq
The only non-vanishing commutator and the expression for the conjugate momentum in eq. \eqref{eq:piIntro} yield
\Beq
\label{eq:CommIntro}
\left[\delta\hat{\phi}_\bk(\tau),\partial_\tau\delta\hat{\phi}_{\bq}(\tau)\right]=i\left(2\pi\right)^{-3}\left(b_{\rm{sr}}(k,\tau)\right)^{-1}\delta(\bk+\bq)\,.
\Eeq
The quantized perturbations $\delta\hat{\phi}_\bk$ can be written in terms of operators $\hat{a}_\bk$ and mode functions $u_k(\tau)$ as 
\Beq
\label{eq:ftouIntro}
\delta\hat{\phi}_\bk(\tau)=\hat{a}_{\bk}u_{k}(\tau)+\hat{a}^{\dagger}_{-\bk} u^{*}_{k}(\tau)\,.
\Eeq
The two mode functions, $u_{k}(\tau)$ and its complex conjugate, span the space of solutions to the classical equation of motion for $\delta\phi_{\bk}$ obtained by varying the action in eq. \eqref{eq:2ndOrderAction}, i.e.,
\Beq
\label{eq:uIEoMIntro}
\partial_\tau^2u_k+\left(\partial_\tau \ln b_{\rm{sr}}\right)\partial_\tau u_k+\omega_{\rm{sr}}^2u_k=0\,.
\Eeq
Given a set of initial conditions for the mode functions, we can evolve them forwards in time. To calculate $\delta\hat{\phi}$ and ultimately $\hat{\mathcal{R}}_{\bk}$ we need to know not only the initial conditions for the mode functions, but also the commutators for $\hat{a}_\bk$ and $\hat{a}_\bk^{\dagger}$. Since, during inflation the co-moving Hubble sphere shrinks, observationally-relevant co-moving modes lie inside the sphere at early times and cross out of it at some point before the end of inflation. Hence, at very early times, the physical wavelength of these modes is much shorter than the Hubble radius, and they are not affected by the space-time curvature. At these early times, the mode functions should resemble those of free fields in flat space-time, while the $\hat{a}_\bk$ and $\hat{a}_\bk^{\dagger}$ operators should be interpreted as creation and annihilation operators. Note that the latter are time-independent, meaning their commutation relations remain the same even after Hubble exit. We can show all of this rigorously.

Consider eq. \eqref{eq:uIEoMIntro}. Since we work in the slow-roll approximation, we can set $H=\partial_{\tau}a(\tau)/a^2=\rm{const}$, implying $a(\tau)=-1/(H\tau)$. We can then find the general solution exactly
\Beq
\label{eq:DeSitterSolution}
u_k=c_{k1}\tau\left(1-\frac{i}{k\tau}\right)e^{-ik\tau}+c_{k2}\tau\left(1+\frac{i}{k\tau}\right)e^{ik\tau}\,.
\Eeq
The two constant coefficients, $c_{k1}$ and $c_{k2}$, can be found by imposing that at early times, when modes are still sub-Hubble, $k/aH=-k\tau\gg1$, $\delta\hat{\phi}$ behaves as a free, massless\footnote{During slow-roll $\partial^2_{\bar{\phi}}V\ll H^2$.} scalar with creation and annihilation operators satisfying
\Beq
\label{eq:CommaadagIntro}
\left[\hat{a}_{\bk},\hat{a}_{\bq}\right]=0\,,\qquad[a_{\bf{k}},a_{\bf{q}}^{\dagger}]=\delta(\bk-\bq)\,.
\Eeq
Using eq. \eqref{eq:CommIntro}, we find $c_{k1}=-(2\pi)^{-3/2}H/\sqrt{2k}$ and $c_{k2}=0$. This means that mode functions of co-moving modes that are still sub-Hubble are of the form $e^{-ik\tau}/a$, i.e., plane waves with a decaying amplitude, scaling inversely with $a(\tau)$. The scale factor is simply a consequence of the non-canonical kinetic term in the action in eq. \eqref{eq:2ndOrderAction}. One can easily check that the rescaled field $a\delta\phi$ has a canonically-normalized action and the equation of motion for its mode functions is of the form given in eq. \eqref{eq:uIEoMIntro}, with $b_{\rm{sr}}=1$ and $\omega_{\rm{sr}}^2=k^2-\partial_{\tau}^2a/a$. The plane wave factor can also be checked to enforce the vacuum state that is annihilated by $\hat{a}_{\bk}$, $|0\rangle$, as the ground state of the second order Hamiltonian. This state is known as the Bunch-Davies vacuum. Conventionally, it is assumed that observationally-relevant modes started in the Bunch-Davies vacuum while lying deep inside the Hubble sphere during inflation. As the universe expands these modes become super-Hubble and according to eq. \eqref{eq:DeSitterSolution}, for $k/aH=-k\tau\rightarrow0$, $u_k\approx i(2\pi k)^{-3/2}H/\sqrt{2}$, i.e., the inflaton perturbations freeze. The canonically-normalized field, $a\delta\hat{\phi}_{\bk}$, then simply grows linearly with scale factor, whereas its conjugate momentum is equal to $a\mathcal{H}\delta\hat{\phi}_{-\bk}$, and therefore the two effectively commute. Hence, the superhorizon inflaton perturbations behave classically, and can be treated as a classical stochastic field. The quantum expectation value with respect to the Bunch-Davies vacuum translates into the classical ensemble average over field realizations drawn from a Gaussian probability distribution. Since we work in the spatially-flat gauge, the curvature perturbation, $\hat{\mathcal{R}}=H\delta\hat{\phi}/\dot{\bar{\phi}}$, on super-Hubble scales is Gaussian, too. Its power-spectrum, $\Delta_{\mathcal{R}}^2$, is defined as follows
\Beq
\langle0|\mathcal{R}(\tau,{\bf{x}})\mathcal{R}(\tau,{\bf{x}})|0\rangle&=\int 4\pi k^2dk \frac{H^2}{\dot{\bar{\phi}}^2}|u_{k}(\tau)|^2\,,\\
                                                                  &\equiv\int d\ln k \Delta_{\mathcal{R}}^2(k,\tau)\,.
\Eeq
Note that the expressions on the right do not depend on the arbitrary space coordinate $\bf{x}$. It can be understood in terms of the isotropy and homogeneity of the universe. On super-Hubble scales during inflation the power-spectrum is scale-invariant, i.e., independent of $k$, $\Delta_{\mathcal{R}}^2\approx H^2/(8\pi^2\mpl^2\epsilon_H)$ and constant in time if we assume $\eta_H=\dot{\epsilon}_H/H\epsilon_H=0$. This result was derived in the de Sitter approximation in which $H$ is approximated to be constant. The conservation of the co-moving curvature on super-Hubble scales can be proven to hold more generally, independently of the assumption of de Sitter expansion. We will discuss it further below. However, it is important to note that since $H$ and $\epsilon_H$ vary slowly with time during inflation, the conserved value for $\Delta_{\mathcal{R}}^2$ is slightly different for different $k$. Essentially, the value of the conserved power-spectrum is determined by the value of $H$ and $\epsilon_H$ at horizon exit, $k=aH$. This introduces a weak scale-dependence. The power-spectrum of the co-moving curvature perturbation generated during slow-roll inflation is
\Beq
\label{eq:Rslowroll}
\Delta_{\mathcal{R}}^2(k)\approx \frac{H^2}{8\pi^2\mpl^2\epsilon_H}\Big|_{k=aH}\,.
\Eeq
This result is obtained in the spatially-flat gauge, ignoring interactions of the inflaton perturbations with those in the metric due to slow-roll suppression. The approximation breaks down towards the end of inflation, when the slow-roll coefficients become of order unity. However, this does not affect eq. \eqref{eq:Rslowroll} for observationally-relevant modes, since the expression is evaluated at the time of Hubble horizon exit, when the slow-roll approximation still holds.

One can derive the above results by working in gauge-invariant variables only. Under the diffeomorphism given in eq. \eqref{eq:DiffDeff}, the inflaton perturbation transforms as $\Delta\delta\phi=-\partial_{\tau}\bar{\phi}\xi^0$. We then define the gauge-invariant inflaton perturbation
\Beq
\delta \tilde{\phi} =\delta \phi - \left(\partial_{\tau}\bar{\phi}\right)\left(\mathcal{B}-\partial_{\tau}\mathcal{E}\right)\,.
\Eeq
The gauge-invariant co-moving curvature perturbation can be then expressed only in terms of gauge-invariant quantities, $\mathcal{R}=\Psi+H\delta \tilde{\phi}/\dot{\bar{\phi}}$; see eq. \eqref{eq:BardeenIntro}. The linearised equation of motion for the gauge-invariant field perturbation is
\Beq
\label{eq:DiffInvEoM}
\partial_{\tau}^2\delta \tilde{\phi}_{\bk}&+2\mathcal{H}\partial_{\tau}\delta\tilde{\phi}_{\bk}+k^2 \delta\tilde{\phi}_{\bk}+a^2\frac{\partial^2V}{\partial\bar{\phi}^2}\delta\tilde{\phi}_{\bk}-\partial_{\tau}\bar{\phi}\left(3\partial_{\tau}\Psi_{\bk}+\partial_{\tau}\Phi_{\bk}\right)+2a^2\frac{\partial V}{\partial \bar{\phi}}\Phi_{\bk} = 0\,,
\Eeq
and the linearised Einstein equations yield
\Beq
\label{eq:BardeenEoM}
&\Phi_{\bk}=\Psi_{\bk}\,,\\
&\left(\partial_{\tau}\mathcal{H}-\mathcal{H}^2+k^2\right)\Psi_{\bk}=\frac{1}{2m_{\text{pl}}^2}\left[-\partial_{\tau}\bar{\phi}\left(\partial_{\tau}\delta\tilde{\phi}_{\bk}+\mathcal{H}\delta\tilde{\phi}_{\bk}\right)+\delta\tilde{\phi}_{\bk}\partial_{\tau}^2\bar{\phi}\right]\,,\\
&\partial_{\tau}\Psi_{\bk}+\mathcal{H}\Psi_{\bk}=\frac{1}{2m_{\text{pl}}^2}\delta\tilde{\phi}_{\bk}\partial_{\tau}\bar{\phi}\,.
\Eeq
These equations can be most easily derived in the Newtonian gauge, $\mathcal{B}=\mathcal{E}=0$, in which the only non-zero scalar metric perturbations $\varphi$ and $\psi$ are equal to $\Phi$ and $\Psi$, respectively, whereas $\delta\phi=\delta\tilde{\phi}$ and therefore, the gauge-invariant quantities should obey the same equations of motion as the perturbations in the Newtonian gauge. In the equation of motion for the gauge-invariant inflaton perturbation, eq. \eqref{eq:DiffInvEoM}, the couplings to the Bardeen variables are slow-roll suppressed. Similarly, the source terms involving the inflaton perturbation in the Einstein equations, eq. \eqref{eq:BardeenEoM}, are also slow-roll suppressed (they also vanish in the limit $k/aH\gg1$). Thus, the evolution of the gauge-invariant $\delta\tilde{\phi}$ is identical to the one of $\delta\phi$ in the spatially-flat gauge. Furthermore, since the source terms for $\Psi$ vanish during slow-roll inflation and the contribution of $\delta\tilde{\phi}$ to $\mathcal{R}$ dominates due to division by $\sqrt{\epsilon_H}$, we find the same value for $\Delta^2_{\mathcal{R}}(k)$ as in eq. \eqref{eq:Rslowroll}, but this time using gauge-invariant variables. We should also point out that $\Psi$ (as well as $\Phi$) plays the role of an auxiliary field. One can see this most easily by substituting for the scalar metric perturbations in eq. \eqref{eq:DiffInvEoM}, using eq. \eqref{eq:BardeenEoM}
\Beq
\label{eq:dPhi1QuantIntro}
&\partial_{\tau}^2\delta\tilde{\phi}_{\bk}+2\mathcal{H}\partial_{\tau}\delta\tilde{\phi}_{\bk}+k^2\delta\tilde{\phi}_{\bk}+a^2\frac{\partial^2 V}{\partial \bar{\phi}^2}\delta\tilde{\phi}_{\bk}\\
&\qquad\qquad\quad+\frac{2}{\mpl^2}\left[\left(\mathcal{H}\partial_{\tau}\bar{\phi}+\frac{a^2}{2}\frac{\partial V}{\partial \bar{\phi}}\right)\frac{\delta\tilde{\phi}_{\bk}\partial_{\tau}^2\bar{\phi}-\partial_{\tau}\bar{\phi}\left(\partial_{\tau}\delta\tilde{\phi}_{\bk}+\mathcal{H}\delta\tilde{\phi}_{\bk}\right)  }{\partial_{\tau}\mathcal{H}-\mathcal{H}^2+k^2} - \left(\partial_{\tau}\bar{\phi}\right)^{\!2}\delta\tilde{\phi}_{\bk} \right]=0\,.
\Eeq
This is a second-order ordinary differential equation for $\delta\tilde{\phi}$. Its quantized solution is of the form given in eq. \eqref{eq:ftouIntro}. This means that the scalar metric perturbations do not have their own creation and annihilation operators. They can be expressed in terms of $\delta\tilde{\phi}$ according to eq. \eqref{eq:BardeenEoM} and do not represent gravitational radiation. It is also obvious that during slow-roll, only the first three terms in eq. \eqref{eq:dPhi1QuantIntro} are important, as expected, so the same considerations as before apply to the initial conditions for the mode functions and ultimately the expression in eq. \eqref{eq:Rslowroll} can be shown to hold.

So far we have shown that $\mathcal{R}$ is conserved on super-Hubble scales during single-field slow-roll inflation. Using eqs. \eqref{eq:DiffInvEoM} and \eqref{eq:BardeenEoM}, we can obtain the equation of motion for the co-moving curvature perturbation
\Beq
\label{eq:Adiabat}
a^{-4}\epsilon_H^{-1}\partial_{\tau}\left(a^2\epsilon_H\partial_{\tau}\mathcal{R}_{\bk}\right)+\frac{k^2}{a^2}\mathcal{R}_{\bk}=0\,.
\Eeq
This equation, often rearranged in a different form, is referred to as the Mukhanov-Sasaki equation. In the limit of $k/aH\ll1$ it has a constant solution and a decaying solution going like $\int d\tau/(a^2\epsilon_H)$. The constant solution is the relevant one for observations. It remains constant even after the end of slow-roll of inflation. 

In fact, a theorem due to Weinberg \cite{Weinberg:2008zzc} states that no matter what the constituents of the universe are, for scalar and tensor perturbations about an FRW background, in the limit $k/aH\ll1$ there always exist two {\it adiabatic solutions}, one constant and one decaying. Adiabatic solutions have the same ratio $\delta s/\dot{\bar{s}}$ for any 4-scalar, $s$.\footnote{This can occur if the universe is in thermal equilibrium even when perturbed, so that $\delta p(T)=\bar{p}'(T)\delta T$ and $\delta \rho(T)=\bar{\rho}'(T)\delta T$ from which follows $\delta \rho/\dot{\bar{\rho}}=\delta p/\dot{\bar{p}}=\delta T/\dot{\bar{T}}$, hence the name.} In single-field inflation, there is only one degree of freedom, $\delta\tilde{\phi}$ (the scalar metric perturbations are auxiliary fields), which implies that there are two solutions to its second-order differential equation. Since there are only two solutions, they must approach the adiabatic limit for $k/aH\ll1$ according to Weinberg. One can check this by considering the gauge-invariant quantity known as the non-adiabatic pressure, $\delta p_{\rm{nad}}=\delta p-\delta \rho\dot{\bar{p}}/\dot{\bar{\rho}}$. In single-field models, it can be shown to vanish on super-Hubble scales. However, in multi-field models, the non-adiabatic pressure does not vanish necessarily. When it does not, the equation of motion for $\mathcal{R}$ has an additional source term, due to the non-adiabatic (entropy) perturbations. There are more than two solutions for $\mathcal{R}$, implying that $\mathcal{R}$ is not generally conserved in these cases.

The tensor metric perturbations given in eq. \eqref{eq:PerturbedMetricIntro}, $\tilde{h}_{ij}$, are gauge invariant and evolve independently of the matter instabilities. The $\tilde{h}_{ij}$ represent gravitational waves. There are no constraint equations on them and they represent the gravitational degrees of freedom. Since the 3-tensor $\tilde{h}_{ij}$ is traceless and transverse, it has two degrees of freedom only. They are frequently denoted as $h^{+}=\tilde{h}_{11}/\sqrt{2}=-\tilde{h}_{22}/\sqrt{2}$ and $h^{\times}=\tilde{h}_{21}/\sqrt{2}=\tilde{h}_{12}/\sqrt{2}$ and referred to as the $+$ and $\times$ polarizations, respectively. In this notation $h_{3i}=h_{i3}=0$ (the transverse plane waves are propagating along the {\it z}-direction). Perturbing the Einstein-Hilbert term in the action in eq. \eqref{eq:ActionInflaton}, one can show that the second order action governing each polarization state is of the form given in eq. \eqref{eq:2ndOrderAction} with $b=\mpl^2a^2$ and $\omega^2=k^2$. Note that in deriving the gravitational waves action we do not make the slow-roll assumption. We can then follow the standard quantization procedure, eqs. (\ref{eq:piIntro}--\ref{eq:uIEoMIntro}), separately for each polarization state. The equation of motion governing the mode functions reduces to
\Beq
\label{eq:GWsEoMIntro}
\partial_{\tau}^2u_k^{(+,\times)}+2\mathcal{H}\partial_{\tau}u_k^{(+,\times)}+k^2u_k^{(+,\times)}=0\,,
\Eeq
manifesting the free nature of the tensor perturbations. The calculation of the mode function evolution during inflation is then identical to the one in the spatially-flat gauge for the scalar perturbations after ignoring slow-roll suppressed terms. At early times, for modes lying deep inside the Hubble sphere, one can show that $u_k^{(+,\times)}=(2\pi)^{-3/2}e^{-ik\tau}/(a\sqrt{2k}\mpl)$, corresponding to the ground state of the Hamiltonian calculated in the Bunch-Davies vacuum $|0\rangle$, annihilated by $\hat{a}_{\bk}^{(+,\times)}$. Later on, as $k/aH=-k\tau\ll1$, the mode function freezes to a constant $u_k^{(+,\times)}\approx i(2\pi)^{-3/2}H/(\sqrt{2}\mpl)$. Like in the scalar perturbations case, one can again show that on superhorizon scales the canonically-normalized tensor perturbation operators $\hat{h}_c^{(+,\times)}=a\mpl \hat{h}^{(+,\times)}$ effectively commute with their conjugate momenta. Hence, the gravitational waves become classical and Gaussian. Their total power is conventionally given by
\Beq
\langle0|4h^{+}(\tau,{\bf{x}})h^{+}(\tau,{\bf{x}})+4h^{\times}(\tau,{\bf{x}})h^{\times}(\tau,{\bf{x}})|0\rangle&=\int 4\pi k^2dk \left(4|u_{k}^{+}(\tau)|^2+4|u_{k}^{\times}(\tau)|^2\right)\,,\\
                                                                  &\equiv\int d\ln k \Delta_{\rm{t}}^2(k,\tau)\,,
\Eeq
where in the last line we define the tensor power-spectrum. Similarly to the power-spectrum of the co-moving curvature perturbation, see eq. \eqref{eq:Rslowroll}, the tensor power-spectrum generated during slow-roll inflation is
\Beq
\label{eq:hslowroll}
\Delta_{\rm{t}}^2(k)\approx \frac{2H^2}{\pi^2\mpl^2}\Big|_{k=aH}\,.
\Eeq
The tensor perturbations are generally conserved in the limit $k/(aH)\ll1$, just like the co-moving curvature perturbation. One can see that most easily from eq. \eqref{eq:GWsEoMIntro}, which shows that the mode functions become overdamped in the super-Hubble limit. Hence, again there is a constant and a decaying solution.

The weak scale-dependences in $\Delta_{\mathcal{R}}^2(k)$ and $\Delta_{\rm{t}}^2(k)$ are characterised by their logarithmic derivatives 
\Beq
\label{eq:LogDer}
n_{\rm{s}}-1\equiv\frac{d\ln \Delta_{\mathcal{R}}^2}{d\ln k}\,,\qquad n_{\rm{t}}\equiv\frac{d\ln \Delta_{\rm{t}}^2}{d\ln k}\,.
\Eeq 
In other words, one can approximate the scale-dependences by simple power-laws
\Beq
\label{eq:InflPSDef}
\Delta_{\mathcal{R}}^2\approx A_{\rm{s}}\left(\frac{k}{k_{\star}}\right)^{n_{\rm{s}}-1}\,,\qquad \Delta_{\rm{t}}^2\approx A_{\rm{t}}\left(\frac{k}{k_{\star}}\right)^{n_{\rm{t}}}\,.
\Eeq
The quantities $n_{\rm{s}}$ and $n_{\rm{t}}$ are known as the scalar and tensor spectral indices, respectively, $k_{\star}$ is the pivot scale, and $A_{\rm{s}}$ and $A_{\rm{t}}$ are the amplitudes of the scalar and tensor power-spectra, respectively. Normally, the tensor amplitude is normalized by the scalar amplitude
\Beq
\label{eq:ttsr}
r=\frac{A_{\rm{t}}}{A_{\rm{s}}}\,.
\Eeq
The quantity is known as the tensor-to-scalar ratio.

Slow-roll inflation predicts small values for the logarithmic derivatives in eq. \eqref{eq:LogDer}, $n_{\rm{s}}-1=-2\epsilon_{H}-\eta_H$ and $n_{\rm{t}}=-2\epsilon_{H}$. All slow-roll parameters are evaluated at Hubble exit of the pivot scale, $k_{\star}=aH$, during inflation. In deriving these expressions, we have made use of the identity $d/d\ln k|_{k=aH}\approx H^{-1}d/dt|_{k=aH}$, which follows from the assumption that during slow-roll inflation $d\ln a/d\ln k|_{k=aH}\approx1$ as $H\approx \rm{const}$. The scalar and tensor amplitudes can be also written in terms of the Hubble slow-roll parameters during inflation, $A_{\rm{s}}=H^2/(8\pi^2\mpl^2\epsilon_H)$, $A_{\rm{t}}=2H^2/(\pi^2\mpl^2)$ and $r=16\epsilon_H$ with again all quantities evaluated at $k_{\star}=aH$. Note that $r=-8n_{\rm{t}}$ and is known as the consistency relation for slow-roll inflation. To connect with the shape of the inflaton potential in models of single-field slow-roll inflation, we recall that $\eta_H/2+\eta_V=2\epsilon_H\approx2\epsilon_V$ and $H^2\approx V/(3\mpl^2)$, implying
\Beq
\label{eq:SlowRollInflatonPot}
n_{\rm{s}}-1\approx-6\epsilon_{V}+2\eta_V\,, \qquad r=-8n_{\rm{t}}\approx16\epsilon_{V}\,,\qquad A_{\rm{s}}\approx \frac{V}{24\pi^2\mpl^4\epsilon_V}\,,
\Eeq
with all potential and potential derivative terms evaluated at $\bar{\phi}=\bar{\phi}_{\star}$, corresponding to the inflaton value at the Hubble exit of the pivot scale. CMB observations \cite{Ade:2015lrj} yield $A_{\rm{s}}=2.2\times10^{-9}$, $n_{\rm{s}}=0.968\pm0.006$ and the constraint $r<0.11\,{\rm{at}}\,95\,\%$ confidence level. They are consistent with adiabatic primordial fluctuations, as predicted by single-field inflation.

In the above analysis of cosmological perturbations, we made several approximations. We ignored the contribution to $\delta T^{\mu}{}_{\nu}$ due to anisotropic stresses, $\Pi^{\mu}{}_{\nu}$. They are a complimentary source of perturbations to the isotropic pressure term, i.e., $\Pi^{0}{}_{0}=\Pi^{0}{}_i=\Pi^{i}{}_i=0$, while $\Pi^{i}{}_j\neq0$ for $i\neq j$. The anisotropic stress in single-field inflation is zero at linear order. Multi-field models involving scalar fields only, also have $\Pi^{i}{}_j=0$ at the linear level. More complicated models with vector fields in some homogeneous and isotropic background configuration for instance, can feature a non-negligible $\Pi^{i}{}_j$. Even in the presence of anisotropic stresses, according to the Weinberg theorem, there always exist a constant and a decaying solution for the scalar and tensor perturbations on super-Hubble scales. We have also not talked about the vector metric perturbations. The reason is that according to Einstein equations, the vector metric perturbations are always redshifted away in the absence of sources. % for $k/(aH)\rightarrow0$.

\bigskip 
\bigskip 

The aim of this section was to show that inflation can make the universe homogeneous and isotropic at the level required by observations. However, this comes at a price. At the end of inflation, the universe is in a cold and non-thermal state. On the other hand the successful theory of big-bang nucleosynthesis calls for a universe very close to thermal equilibrium at temperatures at least around $1\,\rm{MeV}$. That is why reheating is an integral part of inflationary cosmology. Any successful theory of inflation must give an account of the production of Standard Model matter out of the energy stored overwhelmingly in the oscillating inflaton condensate at the end of the period of accelerated expansion. Reheating should also include baryogenesis and perhaps the production of dark matter. In the remaining sections of the lecture notes we review our current understanding of reheating. The early transfer of energy, from the inflaton condensate to the fields it is coupled to, is the subject of the next section. The main focus is on preheating -- the exponential particle production due to non-perturbative resonances and tachyonic instabilities. Section \ref{sec:NonLinRehThesis} discusses the non-linear dynamics ensuing after the fragmentation of the inflaton condensate, and the approach to thermalization. Sections \ref{sec:RehHEPmodels} and \ref{sec:ObsImplReh} connect phenomenological models of reheating with High-Energy Physics models and cosmological observations. We should point out that the details of the reheating process depend on the underlying particle physics theory beyond the Standard Model. Since there are so many possible extensions of the Standard Model, it makes more sense to begin by studying simple toy models inspired by High-Energy Physics to clarify the relative importance of different reheating mechanisms. Many toy models of reheating allow for a thermal universe at the epoch of big-bang nucleosynthesis. To some this is disappointing, since it shows that the current precision of observations does not let us distinguish between different models of inflation and reheating, but to others it is encouraging, since it advocates the inflationary scenario.

\newpage

%-----------------------------------------------------------------------------------------------------------------------------
%				Preheating: the decay of the inflaton condensate
%-----------------------------------------------------------------------------------------------------------------------------
\section{Preheating: the decay of the inflaton condensate}
\label{sec:PreheatIntro}

\hfill\begin{minipage}{\dimexpr\textwidth-3.7cm}
`{\it The career of a young theoretical physicist consists of treating the harmonic oscillator in ever-increasing levels of abstraction.}'
%\xdef\tpd{\the\prevdepth}
\end{minipage}

\bigskip

\hfill {\it Sidney Coleman}

\bigskip

Around the end of inflation, $\epsilon_H=1$, the homogeneous inflaton begins to oscillate about the minimum of its potential. The inflaton condensate must decay into other forms of matter and radiation, eventually giving the particle content of the Standard Model and perhaps dark matter. These more familiar forms of matter and radiation must eventually reach thermal equilibrium at temperatures greater than $1\,\rm{MeV}$ in order to recover the successful big-bang nucleosynthesis scenario. The transition of the universe from the supercooled state at the end of inflation to the hot, thermal, radiation dominated state required for big-bang nucleosynthesis is called reheating. The subject of this section is the early transfer of energy from the inflaton condensate to the fields it is coupled to. We begin with the perturbative theory of reheating -- historically, the process was first treated this way. We then show the importance of non-perturbative effects arising from the coherent nature of the inflaton condensate. They include parametric resonances and tachyonic instabilities, all of which lead to exponential growth in the occupation numbers of the fields the inflaton decays to (i.e., the decay products). These kinds of rapid decay are called preheating, with the decay products in a highly non-thermal state. Finally, we discuss the implications from coupling these decay products to additional matter fields for the energy transfer from the inflaton condensate. %; we show that it leads to an effect known as instant preheating.

\subsection{Perturbative treatment of reheating}
\label{sec:PerReh}

Originally, reheating was studied as a perturbative process \cite{ABBOTT198229} in which individual inflaton particles were assumed to decay independently of each other. Interaction rates and decay rates were calculated in the usual manner, using perturbative coupling expansions. For illustrative purposes we consider inflaton interactions of the form $S_{\rm{matter}}\supset\int d^4x\sqrt{-g}(-\sigma \phi\chi^2-h\phi\bar{\psi}\psi)$, where $\chi$ and $\psi$ are some scalar and fermion decay products. These sort of couplings arise in gauge theories with spontaneously broken symmetries. We avoid tachyonic instabilities in $\chi$ by assuming that its mass, $m_{\chi}$, is greater than $\sqrt{\sigma|\phi|}$. The inflaton potential is assumed to be $V=m^2\phi^2/2$. To tree-level order, for decay products much lighter than the inflaton quanta, the decay rates are \cite{Peskin:257493}
\Beq
\label{eq:Gammas}
\Gamma_{\phi\rightarrow\chi\chi}=\frac{\sigma^2}{8\pi m}\,,\qquad \Gamma_{\phi\rightarrow\bar{\psi}\psi}=\frac{h^2m}{8\pi}\,.
\Eeq
The total width, $\Gamma_{\rm{tot}}\equiv\Gamma_{\phi\rightarrow\chi\chi}+\Gamma_{\phi\rightarrow\bar{\psi}\psi}$, is supposed to determine the decay rate of the number of inflaton quanta in a fixed co-moving volume
\Beq
\label{eq:dndtGamma}
\frac{d\left(a^3n_{\bar{\phi}}\right)}{dt}=-\Gamma_{\rm{tot}}a^3n_{\bar{\phi}}\,,
\Eeq
where $n_{\bar{\phi}}=\rho_{\bar{\phi}}/m$ is the number density of inflaton particles in the condensate. Hence, $a^3(t)n_{\bar{\phi}}(t)\sim\exp(-\Gamma_{\rm{tot}}t)$. Since after inflation $m\gg H\sim t^{-1}$, the homogeneous inflaton undergoes many oscillations during one Hubble time. If $\Gamma_{\rm{tot}}^{-1}\gg m^{-1}$, then we can approximate $\bar{\phi}\approx\bar{\Phi}(t)\cos(mt)$, where $\bar{\Phi}(t)$ varies much more slowly than the phase. Using eq. \eqref{eq:RhoPInfl}, we then find that $n_{\bar{\phi}}\approx m\bar{\Phi}^2/2$. Thus, $\bar{\Phi}(t)\sim a^{-3/2}(t)\exp(-\Gamma_{\rm{tot}}t/2)$, which agrees with eq. \eqref{eq:PrehBackAppr} to leading order up to an extra exponential factor. We can check that this additional exponential decrease due to particle production can be roughly taken into account by including a friction term into the background equation of motion
\Beq
\label{eq:EoMGamma}
\ddot{\bar{\phi}}+(3H+\Gamma_{\rm{tot}})\dot{\bar{\phi}}+m^2\bar{\phi}=0\,.
\Eeq
Having $m\gg H\sim\Gamma_{\rm{tot}}$, one can write the WKB ansatz $\bar{\phi}\approx\bar{\Phi}(t)\cos(mt)$ assuming the phase varies much faster than the amplitude. Neglecting $\ddot{\bar{\Phi}}$ and $H\dot{\bar{\Phi}}$ terms, we then find that $2\dot{\bar{\Phi}}+(3H+\Gamma_{\rm{tot}})\bar{\Phi}=0$ as required. Even if $m\gg H\gg\Gamma_{\rm{tot}}$ one can still show that the second order WKB solution is $a^{-3/2}(t)\exp(-\Gamma_{\rm{tot}}t/2)\cos(mt)$.

For small coupling constants, as required for radiative corrections to not spoil the flatness of the potential during inflation, typically $\Gamma_{\rm{tot}}\ll H$. At the beginning of the oscillatory phase, the inflaton condensate mainly loses energy due to the expansion of space. Once the Hubble rate has decreased to $H\lesssim\Gamma_{\rm{tot}}$, the particle production becomes effective. Thus, the energy density transferred into decay products is $\sim3\mpl^2\Gamma_{\rm{tot}}^2$. Note that $H\lesssim\Gamma_{\rm{tot}}$ is one of the conditions for establishing thermal equilibrium between the inflaton particles and (at least one of) the decay products. Setting the decay rates into individual species to be comparable to each other, i.e., $\Gamma_{\phi\rightarrow\chi\chi}\sim\Gamma_{\phi\rightarrow\bar{\psi}\psi}$, all decay products can be in thermal equilibrium provided they have sufficiently high concentrations. Thereby, we can find an upper bound on the reheating temperature. It is safe to assume that most of the energy has been transferred into the light (with respect to $m$) decay products. Assuming they are relativistic as well, the energy density of the universe is $\pi^2g_*T^4/30$, where $g_*$ is the number of relativistic degrees of freedom, of order $10^2$ for the Standard Model. The maximal reheating temperature is
\Beq
T_{\rm{reh}}\sim\left(\frac{90}{g_*\pi^2}\right)^{1/4}\sqrt{\mpl\Gamma_{\rm{tot}}}\,.
\Eeq
Recalling eq. \eqref{eq:Rslowroll} and the CMB bound on the tensor-to-scalar ratio we find that $T_{\rm{reh}}<10^{15}\,\rm{GeV}$, implying that the GUT symmetries cannot be restored after inflation and the solution to the monopole problem is not in danger. However, this does not rule out the production of other dangerous massive relics such as gravitinos. They could ruin the predictions of the successful big-bang nucleosynthesis by leading to an unwanted matter dominated state of expansion at the beginning of the epoch or by releasing excessive amounts of entropy close to it. One needs to make sure that in this sort of models, the reheating temperature is low enough to avoid the thermal production of such relics. 

We should point out that since each $\Gamma$ is proportional to the square of the small coupling constants, the perturbative decay is actually quite slow and can take many {\it e}-folds of expansion after inflation before the Hubble rate becomes small enough for perturbative particle production to become efficient.

%We should also remark that equations \eqref{eq:dndtGamma} and \eqref{eq:EoMGamma} are applicable only during the oscillatory phase and cannot be extended back to the slow-roll phase. Working at zeroth order in slow-roll parameters the expression in eq. \eqref{eq:dndtGamma} has to be modified by removing the $a^3$ factors\footnote{Consider eq. \eqref{eq:EnergyConservationFRW} and assume an equation of state of $-1$.}, implying $\bar{\Phi}(t)\sim \exp(-\Gamma_{\rm{tot}}t/2)$. The zeroth order slow-roll solution to eq. \eqref{eq:EoMGamma} is different: $\bar{\Phi}\sim(\bar{\phi}_{\rm{init}}-\dot{\bar{\phi}}t)\exp(m^2\Gamma_{\rm{tot}}t/(9H^2))$.
%Using eq. \eqref{eq:RhoPInfl} and eq.\eqref{eq:PrehBackAppr}, assuming $mt\gg1$, we find that 

\subsubsection{Limitations}

There are many issues with the above perturbative analysis. The heuristic equation of motion in eq. \eqref{eq:EoMGamma}, while capturing the qualitative behaviour, does not provide a consistent description of even the perturbative decay of the condensate. It violates the fluctuation dissipation theorem which states that dissipation inevitably leads to fluctuations within the system at hand. The effects of these fluctuations on the effective mass of the inflaton condensate are not included in eq. \eqref{eq:EoMGamma} \cite{Kofman1997}. 

Another problem with the above perturbative approximation is that it does not account for the Bose condensation effects. Even if the couplings of the inflaton to bosons, e.g., to $\chi$, are small enough to allow for a perturbative coupling expansion, if the phase space of bosonic decay products, e.g., of $\chi$ particles, is densely populated Bose condensation effects can greatly enhance the decay rate. We discuss this situation in Section \ref{sec:BosePertEff}.

Most importantly, for larger couplings (but still small enough for radiative corrections to be negligible) the perturbative methods fail. Particle production has to be treated as a non-perturbative effect. The inflaton condensate is a coherent oscillating homogeneous field, implying that particle production has to be treated as a collective process in which many inflaton particles decay simultaneously, not independently of each other. Due to the large occupation number, we can treat the condensate classically. However, the decay products have to be described quantum mechanically, since they have vanishing occupation numbers at the end of inflation (due to the enormous dilution of space during the accelerated expansion). It is justified to use their vacuum state as an initial condition for the ensuing quantum mechanical particle production in the classical inflaton background. The periodic time-dependence of the effective masses of the decay products in the classical oscillating background can have a powerful effect on their production rates in the form of a parametric resonance, which will be the subject of Section \ref{sec:ParRes}.

Despite all of these problems, the perturbative analysis in this section can be applied to the late stages of reheating, e.g., to the decay of remnant inflaton particles after most of the energy has been transferred into relativistic species. Note that such decay channels are crucial to include, so that the energy transfer can be completed. Otherwise, we can face another relic problem.

\subsubsection{Bose condensation of decay products in the perturbative limit}
\label{sec:BosePertEff}

We finish this section with a short discussion of the Bose condensation effects in the perturbative limit, $\sigma\ll m^2/\bar{\Phi}$. By a perturbative limit, we mean that the tree-level order Feynman diagram gives the dominant contribution to the decay of the condensate into $\chi$ particles. %A single decay event produces a pair of $\chi$ particles. 
Higher-order Feynman diagrams are subdominant. They can describe the simultaneous decay of more than one inflaton particles from the condensate %production of more than two $\chi$ particles per decay event 
and are negligible in the perturbative limit to leading order. To avoid significant radiative corrections to the Lagrangian we also put $\sigma\ll m$. Taking into account that the condensate is comprised of particles at rest with large occupation number $n^{\phi}_{\bf{0}}$, the decay rate to a pair of $\chi$ particles at tree-level order is proportional to
\Beq
|\langle n^{\phi}_{\bf{0}}-1, n^{\chi}_{\bf{k}}+1, n^{\chi}_{-\bf{k}}+1|\left(\hat{a}^{\chi}_{\bf{k}}\right)^{\dagger}\left(\hat{a}^{\chi}_{-\bf{k}}\right)^{\dagger}\hat{a}^{\phi}_{\bf{0}}| n^{\phi}_{\bf{0}}, n^{\chi}_{\bf{k}}, n^{\chi}_{-\bf{k}}\rangle|^2=(n^{\chi}_{\bf{k}}+1)(n^{\chi}_{-\bf{k}}+1)n^{\phi}_{\bf{0}}\,,
\Eeq
whereas the rate of the inverse process is proportional to
\Beq
|\langle n^{\phi}_{\bf{0}}+1, n^{\chi}_{\bf{k}}-1, n^{\chi}_{-\bf{k}}-1|\left(\hat{a}^{\phi}_{\bf{0}}\right)^{\dagger}\hat{a}^{\chi}_{\bf{k}}\hat{a}^{\chi}_{-\bf{k}}| n^{\phi}_{\bf{0}}, n^{\chi}_{\bf{k}}, n^{\chi}_{-\bf{k}}\rangle|^2=n^{\chi}_{\bf{k}}n^{\chi}_{-\bf{k}}(n^{\phi}_{\bf{0}}+1)\,.
\Eeq
Note that the occupation number is the number of occupied states per $(2\pi)^3$ phase space volume. The only exception is the inflaton condensate for which $n^{\phi}_{\bf{0}}/V_{\rm{com}}=n_{\bar{\phi}}=m\bar{\Phi}^2/2$, whereas for the $\chi$ particles the number density, $n_{\chi}$, is related to the occupation number in the standard way $n_{\chi}=\int d^3k\,n_{\bf{k}}^{\chi}/(2\pi)^3$. Note that $n^{\chi}_{\bf{k}}=n^{\chi}_{-\bf{k}}$ and are independent of the direction of $\bf{k}$. From now on we put them to be equal to $n^{\chi}_{k}$. Roughly speaking, due to energy and momentum conservation, a stationary inflaton particle decays into a pair of $\chi$ particles, each of which has energy $m/2$ and momentum $[m^2/4-m_{\chi}^2-2\sigma\bar{\phi}(t)]^{1/2}$. Since $m\gg m_{\chi}>\sigma\bar{\Phi}$, all particles are produced within a thin spherical momentum shell in phase space, centred near $m/2$ and of width $4\sigma\bar{\Phi}/m$. Hence, $n_{k=m/2}^{\chi}=n_{\chi}/[4\pi(m/2)^2(4\sigma\bar{\Phi}/m)/(2\pi)^3]=(\pi^2\bar{\Phi}/\sigma)(n_{\chi}/n_{\bar{\phi}})$. Then the rate of change of $\chi$ particles within a given co-moving volume is
\Beq
\label{eq:chiProduction}
\frac{d\left(a^3n_{\chi}\right)}{dt}&=\frac{2a^3}{V_{\rm{com}}}\Gamma_{\phi\rightarrow\chi\chi}\left[(n^{\chi}_{\bf{k}}+1)(n^{\chi}_{-\bf{k}}+1)n^{\phi}_{\bf{0}}-n^{\chi}_{\bf{k}}n^{\chi}_{-\bf{k}}(n^{\phi}_{\bf{0}}+1)\right]\\
&\approx 2a^3\Gamma_{\phi\rightarrow\chi\chi}n_{\bar{\phi}}\left[1+2n_k^{\chi}\right]\approx2a^3\Gamma_{\phi\rightarrow\chi\chi}n_{\bar{\phi}}\left[1+\frac{2\pi^2\bar{\Phi}}{\sigma}\frac{n_{\chi}}{n_{\bar{\phi}}}\right]\,.
\Eeq
where $n^{\phi}_{\bf{0}}\gg\{n_k^{\chi},1\}$ and $|{\bf{k}}|=m/2$. For $n_{k}^{\chi}>1$, i.e, $n_{\chi}>\sigma n_{\bar{\phi}}/(\pi^2\bar{\Phi})$, the second term inside the brackets in the last line in eq. \eqref{eq:chiProduction} becomes important, which is a manifestation of Bose condensation effects becoming relevant. Since $\rho_{\chi}/n_{\chi}\sim m=\rho_{\bar{\phi}}/n_{\bar{\phi}}$, Bose effects should be considered for fractions of energy stored in the decay product satisfying $\rho_{\chi}/\rho_{\bar{\phi}}>\sigma/\bar{\Phi}$. For small coupling constants and large amplitudes, the right-hand side can be much less than unity and the equality can be satisfied shortly after inflation. Bose effects become important and the perturbative treatment presented in the beginning of this section breaks down. For high occupancies, $n_{k}^{\chi}\gg1$, after ignoring the expansion of space, and using eq. \eqref{eq:Gammas} we can integrate eq. \eqref{eq:chiProduction} to get
\Beq
\label{eq:PertExpGrowth}
n_{\chi}\sim \exp\left(\frac{\pi\sigma\bar{\Phi}t}{2m}\right)\,.
\Eeq

Bose effects lead to an exponential increase of the decay efficiency. We have shown it for small enough couplings which allow for a perturbative treatment. When couplings are increased, non-perturbative effects become important, but the exponential increase in the decay efficiency remains. This is shown in the next two sections. A discussion of the effects on the particle production rate due to the expansion of the universe is also included.

\subsection{Parametric resonance}
\label{sec:ParRes}

As shown at the end of the previous section, Bose condensation effects can exponentially enhance the rate at which energy is transferred from the oscillating inflaton condensate to the bosonic fields it is coupled to. We worked in the perturbative limit in which the coupling is small, e.g., $\sigma\bar{\Phi}\ll m^2$ in the trilinear interaction model $V(\phi,\chi)=m^2\phi^2/2+m_{\chi}^2\chi^2/2+\sigma\phi\chi^2$, with $m\gg m_{\chi}>\sqrt{\sigma\bar{\Phi}}$. In this limit a perturbative coupling expansion makes sense. If the amplitude of inflaton oscillations and/or the coupling constant become large, e.g., $\sigma\bar{\Phi}>m^2$ in the trilinear case, high-order Feynman diagrams give comparable predictions to the lowest-order ones and the problem has to be approached non-perturbatively. Note that simultaneous decays of more than one inflaton particles from the condensate are described by high-order diagrams. Such decays are a consequence of the coherent nature of the oscillating condensate and the non-perturbative calculation presented below captures them, unlike the perturbative one in the previous section. It turns out that Bose effects still exponentially enhance the rate of energy transfer. It is more efficient than in the perturbative limit, due to contributions from the simultaneous decays of more than one inflaton particle. The phenomenon can be understood most easily in the language of parametric resonance. Of course, the method can be applied to the perturbative case as well.

At the end of inflation, matter fields can be treated as fluctuations on top of the oscillating homogeneous inflaton background. Typically, they start in the vacuum state, since inflation has diluted the corresponding particle densities to vanishing values. Ignoring the expansion of space for now, the linearised equations of motion take the form
\Beq
\label{eq:Hill}
\ddot{\hat{\chi}}_{\bf{k}}+\omega^2(k,t)\hat{\chi}_{\bf{k}}(t)=0\,,
\Eeq
where the angular frequency is periodic, i.e., $\omega^2(k,t)=\omega(k,t+T)^2$; $T$ is the period of oscillations of the condensate. In the trilinear model, $\omega^2(k,t)=k^2+m_{\chi}^2+2\sigma\bar{\Phi}\cos(mt)$ and $T=2\pi/m$. Unlike the previous section, here we do not assume anything about the relative values of $m$, $m_{\chi}$ and $\sqrt{\sigma\bar{\Phi}}$. The equation of the form given in eq. \eqref{eq:Hill} with $\omega$ a periodic function of time is known as the Hill's equation \cite{magnus2004hill,teschl2012ordinary}. In the triliniear case $\omega$ evolves harmonically and the equation can be reduced to the Mathieu equation form
\Beq
\label{eq:Mathieu}
\frac{d^2}{dz^2}\hat{\chi}_{\bf{k}}+\left[A_k+2q\cos(2z)\right]\hat{\chi}_{\bf{k}}(z)=0\,,
\Eeq
with $A_k$, $q$ and $z$ dimensionless and determined by the form of $\omega$. In the trilinear model, $A_k=4(k^2+m_{\chi}^2)/m^2$, $q=4\sigma\bar{\Phi}/m^2$ and $z=mt/2$.\footnote{Another popular model that can be described with the Mathieu equation is $V(\phi,\chi)=m^2\phi^2/2+g^2\chi^2\phi^2/2+m_{\chi}^2\chi^2/2$ \cite{Kofman1997}, for which $z=mt$, $A_k=(k^2+m_{\chi}^2)/m^2+2q$, $2q=g^2\bar{\Phi}^2/(2m^2)$.}

\subsubsection{Floquet theory}
\label{sec:FloqIntro}

The action leading to the Hill's equation, eq. \eqref{eq:Hill}, is that of a harmonic oscillator with a periodic angular frequency
\Beq
\label{eq:ActionChi}
S_{\chi}^{(2)}=\int dtd^3x\left[\frac{|\dot{\chi}_{\bf{k}}|^2}{2}-\omega^2(k,t)\frac{|\chi_{\bf{k}}|^2}{2}\right]\,.
\Eeq
We can follow the quantization procedure outlined after eq. \eqref{eq:2ndOrderAction}. Now the mode functions of $\hat{\chi}_{\bf{k}}$ obey the Hill's equation
\Beq
\label{eq:HilluIntro}
\ddot{u}_{k}+\omega^2(k,t)u_{k}(t)=0\,.
\Eeq
The Floquet theorem \cite{magnus2004hill} states that the most general solution of the Hill's equation is given by
\Beq
\label{eq:FloqSolIntro}
u_{k}(t)=e^{\mu_{k}t}\mathcal{P}_{k+}(t)+e^{-\mu_{k}t}\mathcal{P}_{k-}(t)\,,
\Eeq
where $\mu_{k}$ is called the Floquet exponent and $\mathcal{P}_{k\pm}(t)=\mathcal{P}_{k\pm}(t+T)$. If $\Re(\mu_{k})\neq0$ one of the two terms increases exponentially with time. This is called parametric resonance. Let's prove eq. \eqref{eq:FloqSolIntro} \cite{landau1976mechanics,Amin2014} and show how to find numerically $\mu_{k}$ \cite{Frolov:2010sz,Amin2014}, knowing the form of $\omega(k,t)$. 

If $u_{k}(t)$ is a solution to eq. \eqref{eq:HilluIntro}, then so must be $u_{k}(t+T)$. Hence, if $u_{k1}(t)$ and $u_{k2}(t)$ are two linearly independent solutions, their time-shifted counterparts must be linear combinations of them, i.e., $u_{ki}(t+T)=\sum_{j=1}^2B_{ij}u_{kj}(t)$ with $B_{ij}$  a constant $2\times2$ invertible matrix. We can diagonalize the expression to get $v_{ki}(t+T)=\sum_{j=1}^2\lambda^B_i\delta_{ij}v_{kj}(t)$ where $\lambda^B_i$ are the two eigenvalues of $B_{ij}$ and $v_{ki}(t)$ are independent linear combinations of $u_{ki}(t)$. From this follows that $v_{ki}(t+T)=\lambda^B_iv_{ki}(t)$, i.e., a time sift $t\rightarrow t+T$ leads to a rescaling by an eigenvalue. The most general solutions having this property are $v_{ki}(t)=(\lambda^B_i)^{t/T}P_{ki}(t)$, where $P_{ki}(t+T)=P_{ki}(t)$. Since the Wronskian, $W[u_{k1},u_{k2}]\equiv u_{k1}\dot{u}_{k2}-\dot{u}_{k1}u_{k2}$, of the Hill's equation, eq. \eqref{eq:HilluIntro}, is constant, $\dot{W}[u_{k1},u_{k2}]=0$, so must be $\dot{W}[v_{k1},v_{k2}]=0$. On the other hand, $W[v_{k1},v_{k2}](t+T)=\lambda_1^B\lambda_2^BW[v_{k1},v_{k2}](t)$, implying $\lambda_1^B=1/\lambda_2^B\equiv\lambda^B$. This completes the proof of eq. \eqref{eq:FloqSolIntro}. The Floquet exponent is simply $\mu_{k}=\ln(\lambda^B)/T$, whereas each of the periodic functions $\mathcal{P}_{k\pm}(t)$ is some linear combination of $P_{k1,2}(t)$.

To find the Floquet exponent, we just need to calculate the eigenvalues of $B_{ij}$, which, as we just showed, has a unit determinant. To do that we choose two orthogonal initial conditions $\{u_{k1}(t_0),\dot{u}_{k1}(t_0)\}=\{1,0\}$ and $\{u_{k2}(t_0),\dot{u}_{k2}(t_0)\}=\{0,1\}$ at some initial time $t_0$. This implies that $\{B_{i1},B_{i2}\}=\{u_{ki}(t_0+T),\dot{u}_{ki}(t_0+T)\}$. Hence, after evolving the Hill's equation forward for one period $T$ for the two sets of initial conditions we can find the eigenvalues\footnote{Using the fact that for our choice of initial conditions $W[u_{k1},u_{k2}](t_0)=1$ and that $W[u_{k1},u_{k2}](t_0+T)=\lambda^B_1\lambda^B_2W[u_{k1},u_{k2}](t_0)$, one can easily show that this expression is consistent with $B_{ij}$ having a unit determinant.}
\Beq
\lambda^B_{1,2}=\frac{1}{2}\{&u_{k1}(t_0+T)+\dot{u}_{k2}(t_0+T)\\
                                                          &\pm\sqrt{[u_{k1}(t_0+T)-\dot{u}_{k2}(t_0+T)]^2+4\dot{u}_{k1}(t_0+T)u_{k2}(t_0+T)}\}\,.
\Eeq

The initial conditions are relevant for the efficiency of the parametric resonance. Essentially, if both the initial field and field velocities are zero, parametric resonance does not lead to any growth. We can see this most easily by re-writing eq. \eqref{eq:FloqSolIntro} as a linear combination of the linearly independent $v_{k1,2}(t)=\exp(\pm\mu_kt)P_{k1,2}(t)$, i.e., $u_{k}(t)=c_1v_{k1}(t)+c_2v_{k2}(t)$. If both $u_{k}(t_0)$ and $\dot{u}_{k}(t_0)$ are zero, then the only possibility for the constant pre-factors is $c_1=c_2=0$. Hence, unlike ordinary resonance where the forcing term leads to a rapid growth even if initially the field displacement and velocity are zero, parametric resonance does not allow for any resonant excitations if no energy is stored in the fluctuations initially. That is why vacuum fluctuations, albeit small, play a crucial role for particle production after inflation as seeds for parametric resonance.

\subsubsection{Narrow resonance}

%~~~~~~~~~~~~~~~~~
\begin{figure*}[t] %  figure placement: here, top, bottom, or page
   \centering
   \includegraphics[width=2.47in]{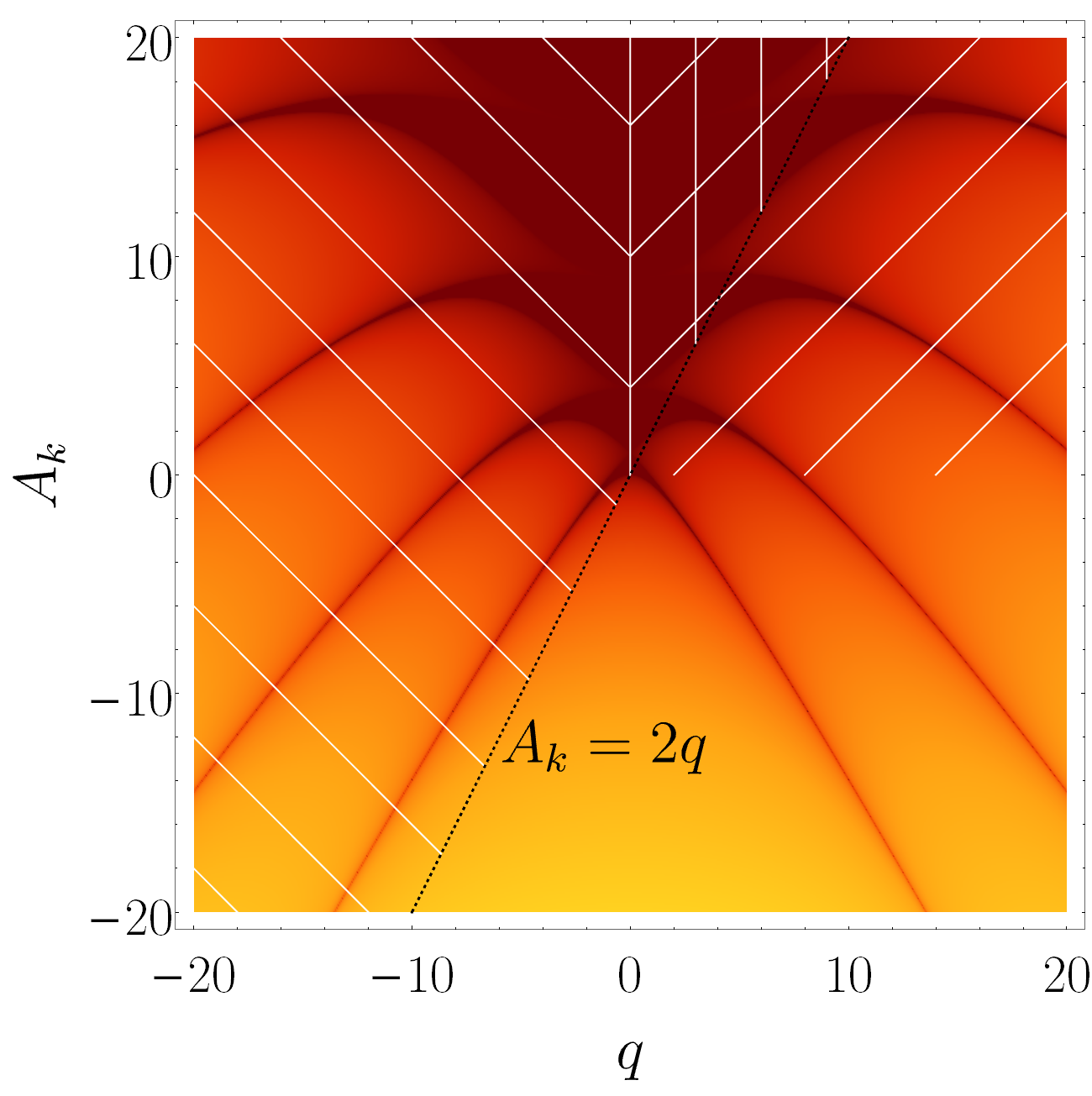} 
     \raisebox{0.097\height}{\includegraphics[width=0.48in]{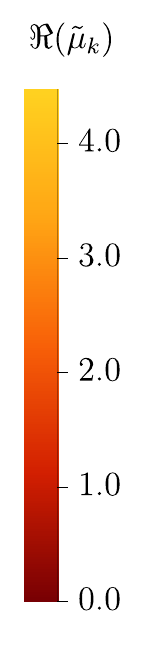}}
   \caption{The instability chart of the Mathieu equation, eq. \eqref{eq:Mathieu}. %, with solutions of the form given in eq. \eqref{eq:FloqSolIntro}. 
The dark areas correspond to vanishing real part of the Floquet exponent and are regions of stability. Narrow parametric resonance occurs for $A_k=n^2$ and $|q|\rightarrow0$ ($n$ is an integer) and broad resonance takes place when $A_k<2|q|$ as well as $A_k-2|q|\ll |q|^{1/2}$. Resonant decay of the inflaton condensate into $\chi$ particles in the trilinear model, $V(\phi,\chi)=m^2\phi^2/2+m_{\chi}^2\chi^2/2+\sigma\phi\chi^2$, can be understood in terms of the Mathieu equation -- the equation of motion for $\chi_k$ (after ignoring the expansion of space) can be mapped onto the Mathieu equation with $A_k\geq0$ and $q\geq0$, i.e., the square region with northeast white lines (see also Fig. \ref{fig:FloqPhiChi2}). Resonant production of $\chi$ particles in another common toy model, $V(\phi,\chi)=m^2\phi^2/2+m_{\chi}^2\chi^2/2+g^2\phi^2\chi^2/2$, can be also mapped onto this chart -- for $g^2>0$, the wedge-shaped region with vertical white lines%($A_k>2q\geq0$)
, whereas for $g^2<0$ the region with northwest white lines, cover the relevant ranges of $A_k$ and $q$ (see also Figs. \ref{fig:FloqPhi2Chi2Positive} and \ref{fig:FloqPhi2Chi2Negative}).}
   \label{fig:FloqMathieu}   
\end{figure*}
%~~~~~~~~~~~~~~~~~

\begin{figure*}[t] %  figure placement: here, top, bottom, or page
   \centering
   \includegraphics[width=2.22in]{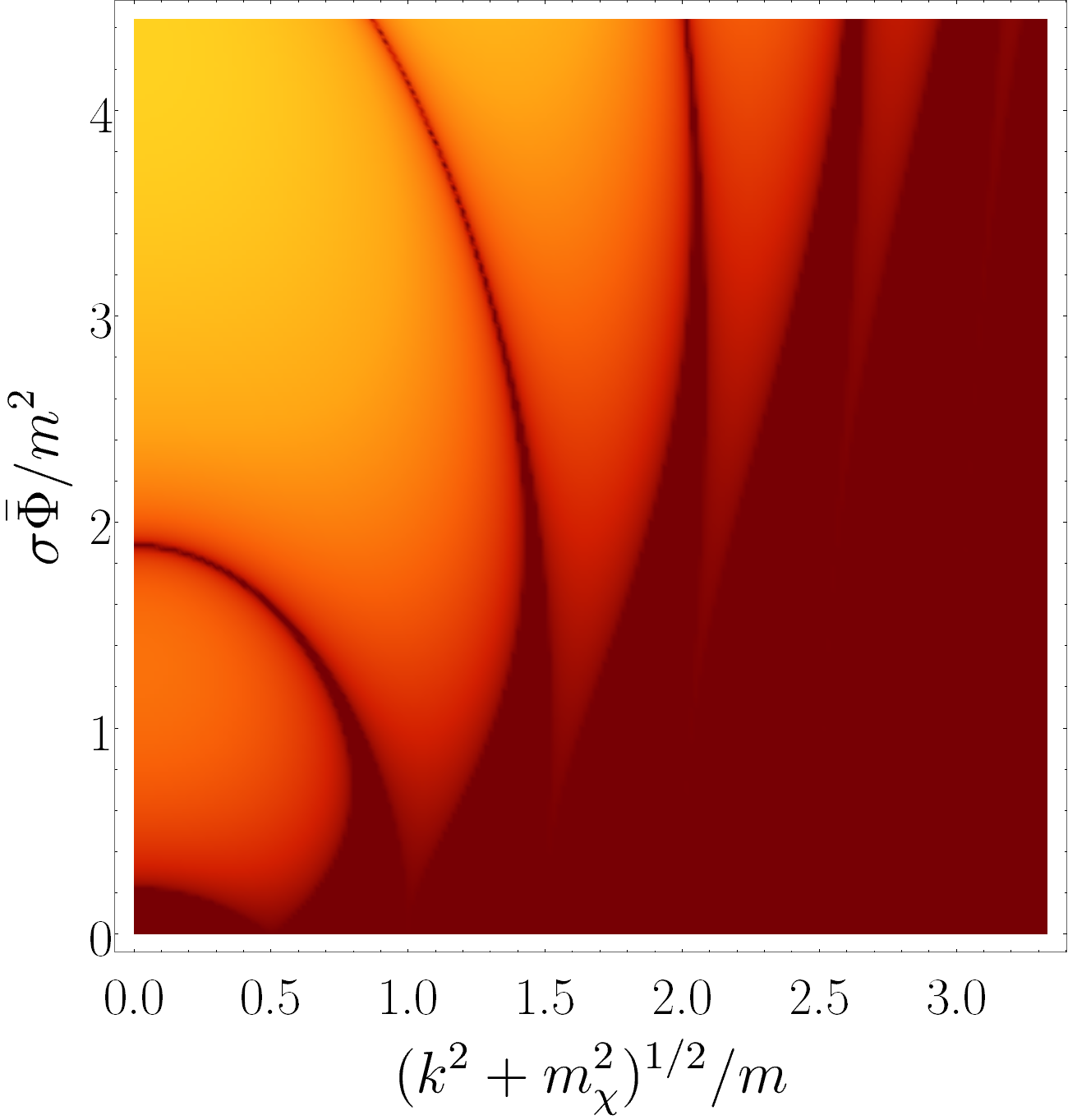} 
     \raisebox{0.13\height}{\includegraphics[height=1.92in]{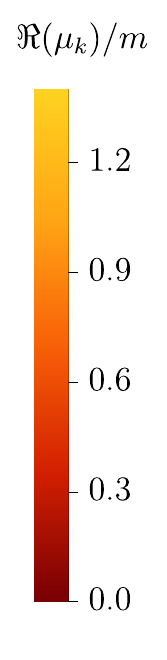}}
   \includegraphics[width=2.22in]{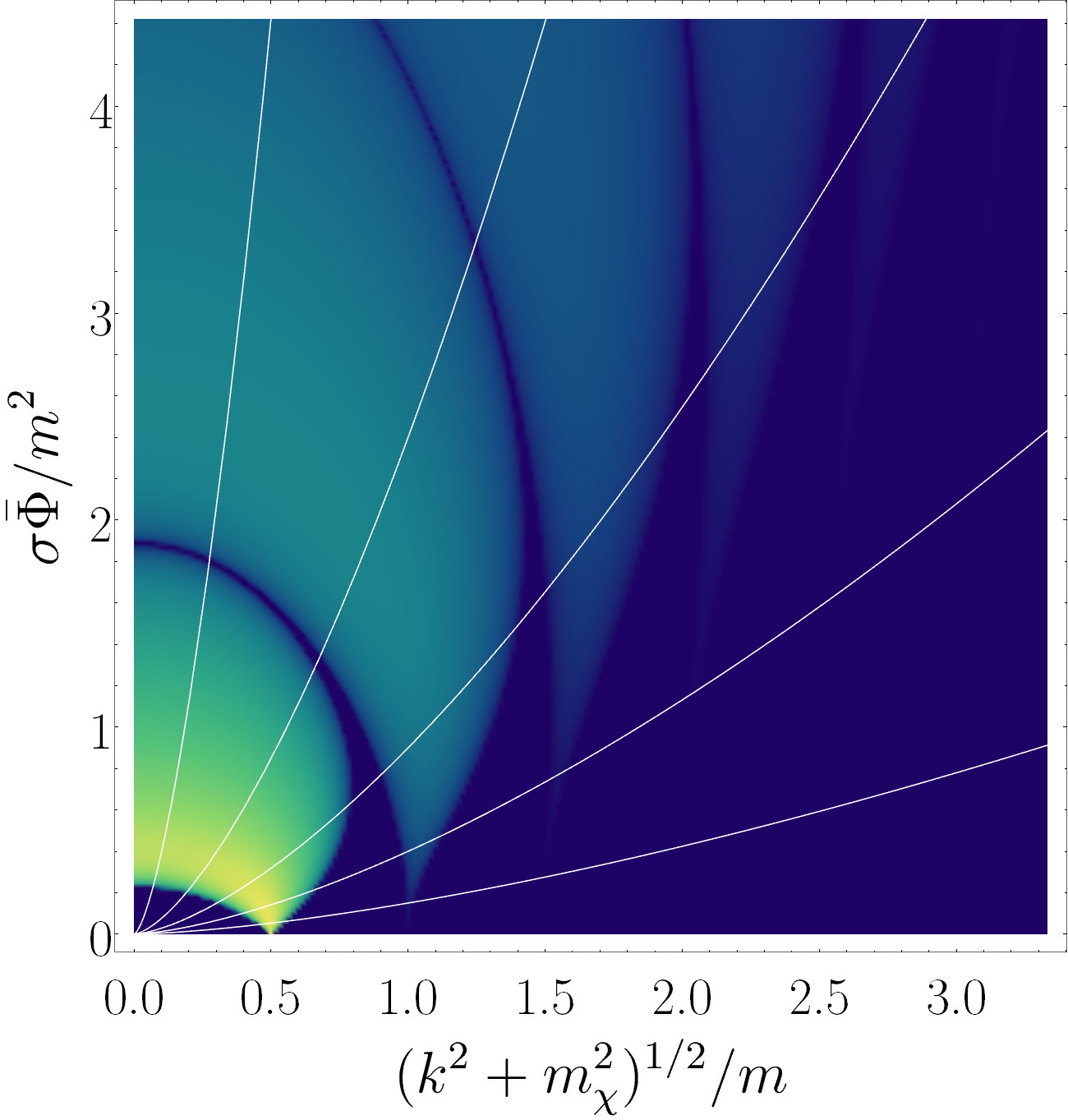} 
     \raisebox{0.13\height}{\includegraphics[height=1.92in]{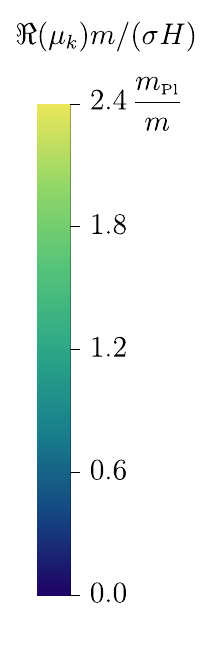}}
   \caption{The instability chart featuring the real part of the Floquet exponent normalized by the inflaton mass (left) and the Hubble rate (right), characterizing the $\chi$ particle production rate in the trilinear model, $V(\phi,\chi)=m^2\phi^2/2+m_{\chi}^2\chi^2/2+\sigma\phi\chi^2$. The equation of motion for $\chi_k$ can be reduced to the Mathieu equation, eq. \eqref{eq:Mathieu}, with $A_k=4(k^2+m_{\chi}^2)^{1/2}/m$, $q=4\sigma\bar{\Phi}/m^2$, where $\bar{\Phi}$ is the amplitude of inflaton oscillations (see also Fig. \ref{fig:FloqMathieu}). In FRW space-time $\bar{\Phi}\propto a^{-3/2}$ and $k\propto a^{-1}$, implying that a given co-moving mode flows towards the bottom left corner of the chart as the universe expands as indicated with the white lines in the second chart (drawn for $m_{\chi}=0$ for simplicity). Note that resonance is efficient if $\Re(\mu_k)/H\sim \sigma\mpl/m^2\gg1$.}
   \label{fig:FloqPhiChi2}   
\end{figure*}

As an exercise, we can now calculate the dimensionless Floquet exponent, $\tilde{\mu}_k$, of the Mathieu equation eq. \eqref{eq:Mathieu}. The magnitude of the real part of $\tilde{\mu}_k$ is plotted in Fig. \ref{fig:FloqMathieu} as a function of the parameters $A_k$ and $q$. We call plots of this type instability charts. There is a series of regions of stability in which $\Re(\tilde{\mu}_k)=0$. They are surrounded by `unstable' regions in which $\Re(\tilde{\mu}_k)>0$. For $|q|\ll1$ and $A_k>0$, the regions of instability become narrow and approach $A_k^{(n)}=n^2$ as $q\rightarrow0$ ($n$ is an integer). In the first narrow band the peak value of the Floquet exponent is $\Re\left(\tilde{\mu}_{k}\right)_{\rm{max}}^{(1)}\approx |q|/2$, while $A_k^{(1)}\approx1\pm |q|$ \cite{magnus2004hill}. For the triliniear model, this corresponds to resonant production of $\chi$ particles with momentum in the range $m/2\pm\sigma\bar{\Phi}/m$  (assuming $m_{\chi}=0$) and mode functions growing as $\exp(\sigma\bar{\Phi}t/m)$ (see also Fig. \ref{fig:FloqPhiChi2}). Since the $\chi$ particles are described with the action of a time-dependent simple harmonic oscillator, eq. \eqref{eq:ActionChi}, the energy stored in a given $\bf{k}$ mode is simply 
\Beq
\label{eq:occupationNumberSHO}
E_{\bf{k}}^{\chi}=\frac{1}{(2\pi)^3}\left(n_{\bf{k}}^{\chi}+\frac{1}{2}\right)\omega(k,t)=\frac{|\dot{u}_k|^2}{2}+\omega^2(k,t)\frac{|u_k|^2}{2}\,,
\Eeq
where $n_{\bf{k}}^{\chi}$ can be interpreted as the mean occupation number (mean, because it is evaluated by taking the expectation value of the Hamiltonian with respect to the Bunch-Davies vacuum). Hence, for $q=4\sigma\bar{\Phi}/m^2\ll1$, modes lying near the peak in the first narrow instability band have occupation numbers growing as $n_{|{\bf{k}}|\approx m/2}^{\chi}\propto\exp(2\sigma\bar{\Phi}t/m)$. This is in good agreement with the perturbative treatment of Bose condensation from the previous section, see, e.g., eq. \eqref{eq:PertExpGrowth}. Thus, in the perturbative limit, $\sigma\bar{\Phi}/m^2\ll1$, the Bose effects, due to the population of $\chi$ modes, in the leading order $\phi\rightarrow\chi\chi$ Feynman diagram can be described as a parametric resonance due to the first, $n=1$, narrow, $q\ll1$, instability band. Higher order, $n>1$, narrow bands lead to production of particles with momentum in the range $k^{(n)}=nm/2\geq m$. They correspond to higher order Feynman diagrams describing the simultaneous decay of $n$ $\phi$ particles from the condensate into a pair of $\chi$s, taking into account the Bose effects due to the dense populations of the $\chi$ modes, $n^{\chi}_{|{\bf{k}}|=nm/2}>1$. Since this happens in the perturbative limit, one should be able to describe it using the methods from the previous section, leading to eq. \eqref{eq:PertExpGrowth}. In summary, resonance from the narrow bands, $|q|\ll1$, describes perturbative decays of particles from the inflaton condensate in the trilinear model, taking into account the occupation of $\chi$ modes. This type of parametric resonance is known as narrow resonance. It corresponds to $|q|\ll1$ for the Mathieu equation, but for the general Hill's equation it corresponds to the parametric resonance in some region of parameter space which features a narrow instability band. 

%~~~~~~~~~~~~~~~~~
%~~~~~~~~~~~~~~~~~

\subsubsection{Broad resonance}

Similarly, the term broad resonance is used to describe parametric resonance in broad instability bands in parameter space. For instance, it occurs if $|q|\gtrsim1$ for the Mathieu equation, see Fig. \ref{fig:FloqMathieu}. This corresponds to the non-perturbative limit in the trilinear model. In this limit, the only means for calculating the particle production is by solving the mode equation, eq. \eqref{eq:HilluIntro}, and a very intuitive way of describing its solutions is the Floquet analysis we have developed. Broad resonance is much more efficient than narrow resonance since a broad, continuous range of $k$ modes is excited. The typical rate of excitation is comparable to the background oscillation rate, $|\Re(\mu_k)|\sim T^{-1}$, and is much greater than in narrow resonance. The reason why the period of inflaton oscillations is the characteristic time-scale for particle production can be understood from the fact that in broad resonance, particles are produced in bursts, rather than smoothly as in the narrow resonance. Those bursts are separated in time by $\sim T$. They occur every time the adiabadicity condition
\Beq
\label{eq:AdiabCond}
\frac{\dot{\omega}(k,t)}{\omega^2(k,t)}\ll1\,,
\Eeq
%~~~~~~~~~~~~~~~~~
\begin{figure*}[t] %  figure placement: here, top, bottom, or page
   \centering
   \includegraphics[width=2.22in]{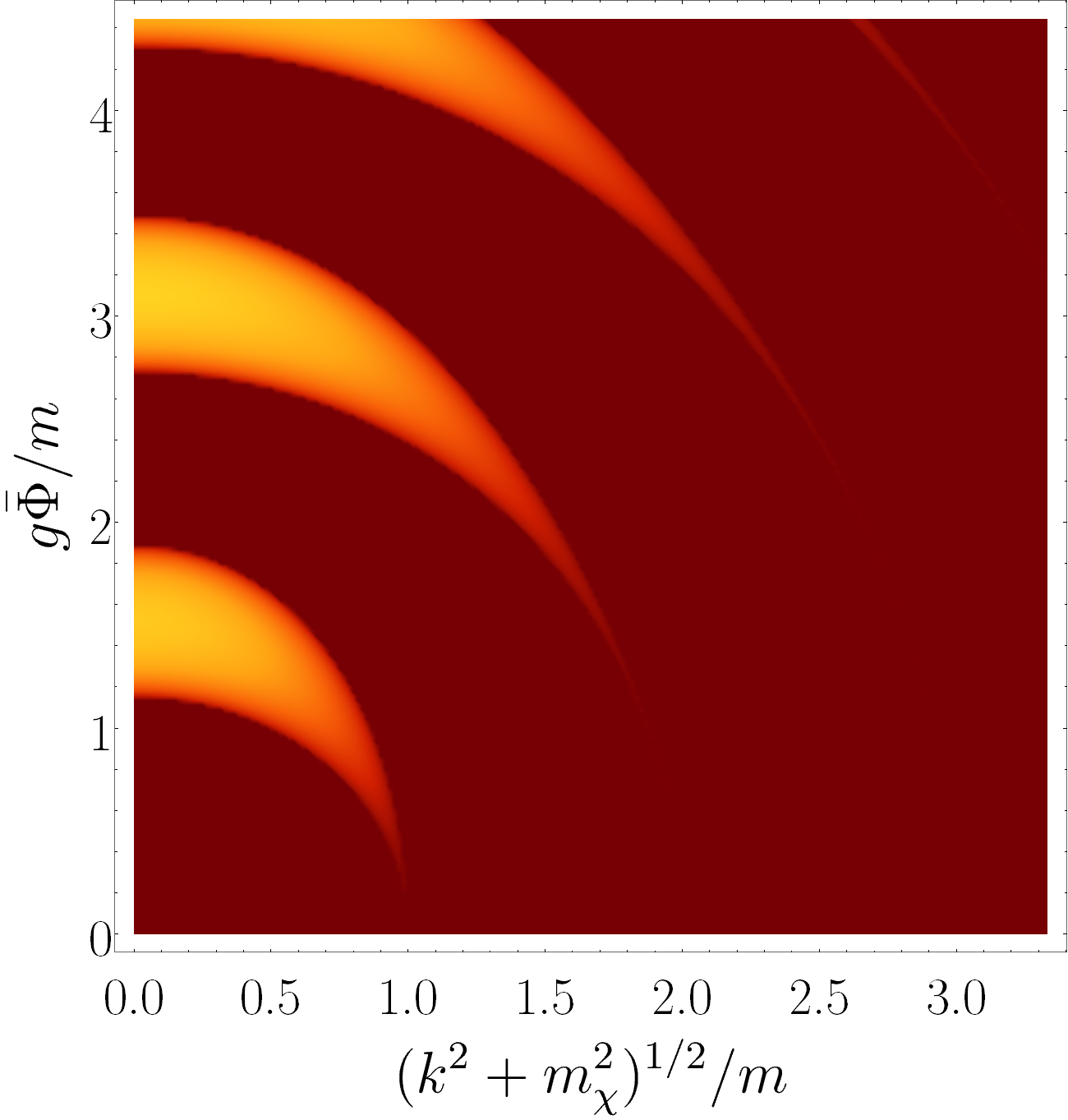} 
     \raisebox{0.13\height}{\includegraphics[height=1.92in]{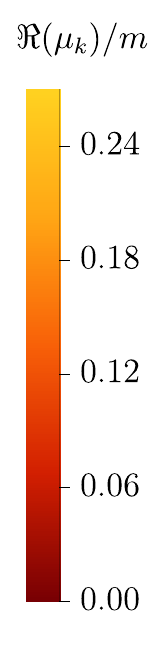}}
   \includegraphics[width=2.22in]{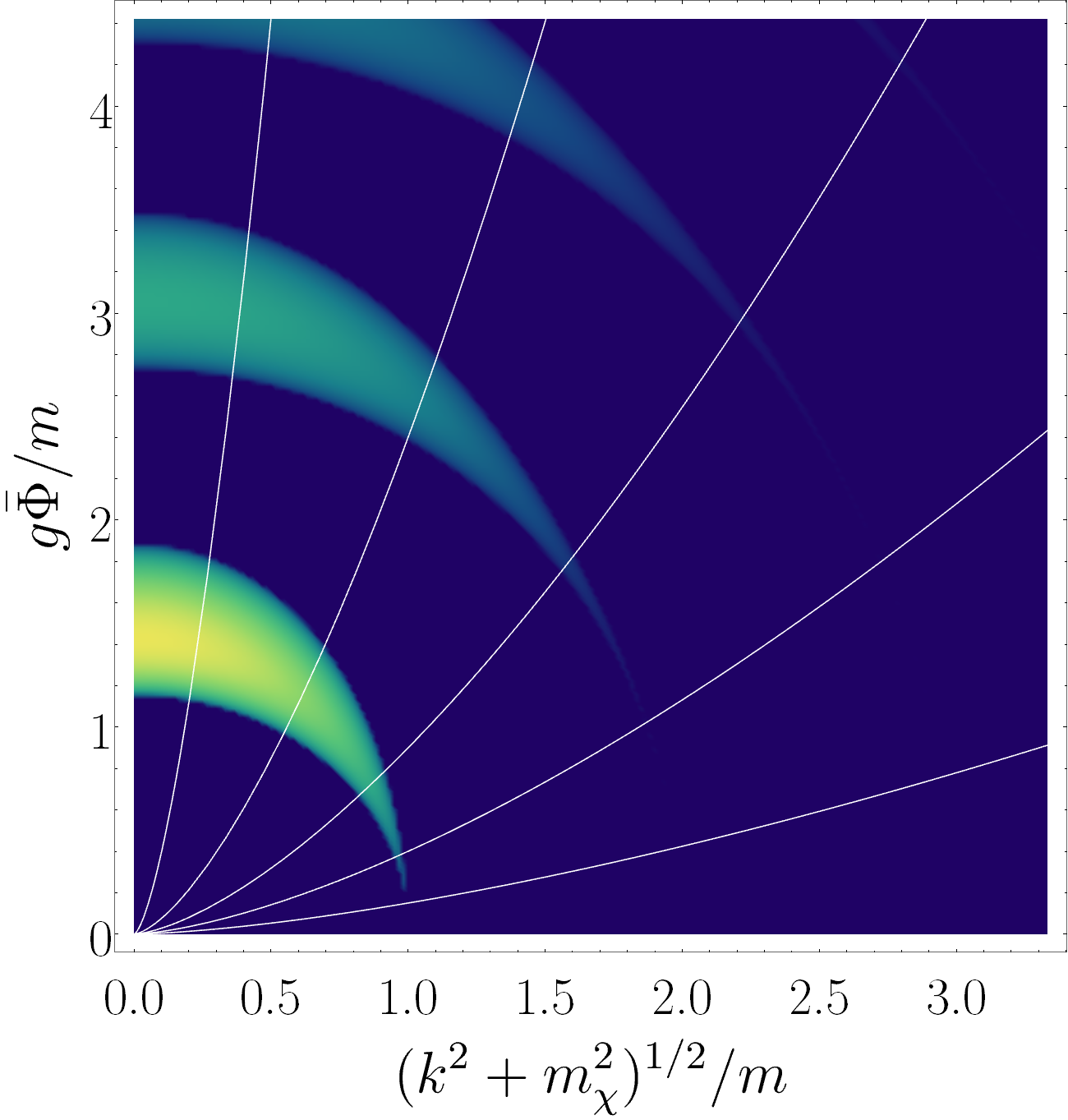} 
     \raisebox{0.13\height}{\includegraphics[height=1.92in]{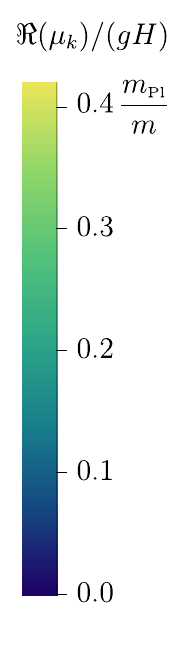}}
   \caption{Same as Fig. \ref{fig:FloqPhiChi2}, but for $V(\phi,\chi)=m^2\phi^2/2+m_{\chi}^2\chi^2/2+g^2\phi^2\chi^2/2$, for which $A_k=(k^2+m_{\chi}^2)/m^2+2q$, $2q=g^2\bar{\Phi}^2/(2m^2)$. The charts are for $g^2>0$, see Fig. \ref{fig:FloqMathieu}. Broad resonance occurs only for $A_k-2q\ll q^{1/2}$, i.e., for low-momentum modes $(k^2+m_{\chi}^2)/m^2\ll g\bar{\Phi}/m$, for specific ranges of $g\bar{\Phi}/m$.}
   \label{fig:FloqPhi2Chi2Positive}   
\end{figure*}
%~~~~~~~~~~~~~~~~~
is violated. Since broad resonance occurs in the non-perturbative regime, where interactions with the inflaton background determine $\omega(k,t)$, and since their magnitude varies with period $T$, the adiabadicity condition is violated each time the background value of the inflaton is such that the interaction terms vanish -- then $\dot{\omega}(k,t)\gg\omega^2(k,t)$. For an oscillating field, this happens twice a period, implying a rate of particle production comparable to $T$. Note that in the narrow resonance the adiabadicity condition is always satisfied, since interactions are weak (they can be treated perturbatively) and $\omega^2(k,t)\approx k^2=\rm{const}$ always. The only reason for resonance is the dense occupation of $\chi$ modes, which leads to a smooth exponential increase in the occupation numbers of particular modes. The reason why the case of broad resonance is different can be understood qualitatively by considering the mode functions in the adiabatic and non-adiabatic regimes. In the adiabatic limit, the WKB solutions to eq. \eqref{eq:HilluIntro} are
\Beq
\label{eq:WKBSolintro}
u_k(t)=\frac{1}{(2\pi)^{3/2}}\left[\frac{\alpha_k}{\sqrt{2\omega(k,t)}}e^{-i\int\omega(k,t)dt}+\frac{\beta_k}{\sqrt{2\omega(k,t)}}e^{i\int\omega(k,t)dt}\right]\,.
\Eeq
The vacuum state mode functions which minimize the Hamiltonian corresponding to the action in eq. \eqref{eq:ActionChi} and which are such that the commutators in eqs. \eqref{eq:Introcommutators} and \eqref{eq:CommaadagIntro} are satisfied for $\hat{\chi}_{\bf{k}}$ and $\hat{\pi}_{\bf{k}}^{\chi}$, and $a^{\chi}_{\bf{k}}$ and $a^{\chi\dagger}_{\bf{k}}$, respectively, have $|\alpha_k|=1$ and $\beta_k=0$ \cite{Mukhanov:2007zz}. More generally, mode functions satisfying the field-momentum commutator in eq. \eqref{eq:Introcommutators}, given eq. \eqref{eq:CommaadagIntro}, correspond to $|\alpha_k|^2-|\beta_k|^2=1$. One can show that this expression is consistent with the constancy of the Wronskian, since $W[u_k,u_k^*]=i(|\alpha_k|^2-|\beta_k|^2)/(2\pi)^3$ and is equal to $i/(2\pi)^3$ if we start in the vacuum. The mean occupation number, see eq. \eqref{eq:occupationNumberSHO}, is simply $n_{\bf{k}}^{\chi}=|\beta_k|^2$, i.e., an adiabatic invariant and equal to zero in the vacuum state. We should note that for $|\beta_k|>0$, the Bunch-Davies vacuum is no longer an eigenstate of the Hamiltonian. 

\begin{figure*}[t] %  figure placement: here, top, bottom, or page
   \centering
   \includegraphics[width=2.22in]{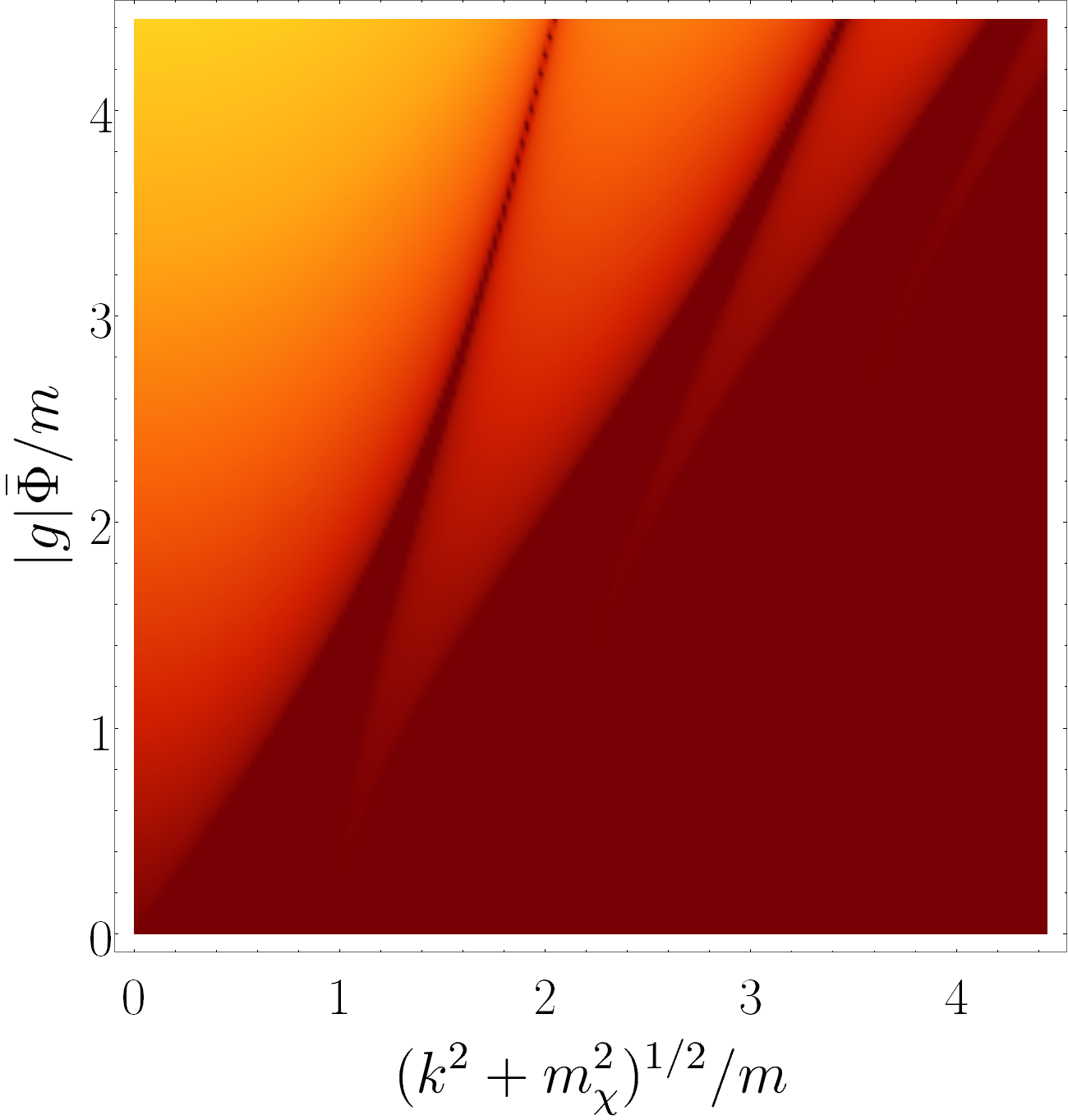} 
     \raisebox{0.13\height}{\includegraphics[height=1.92in]{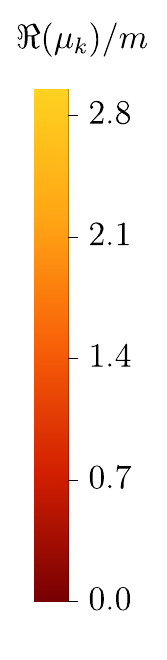}}
   \includegraphics[width=2.22in]{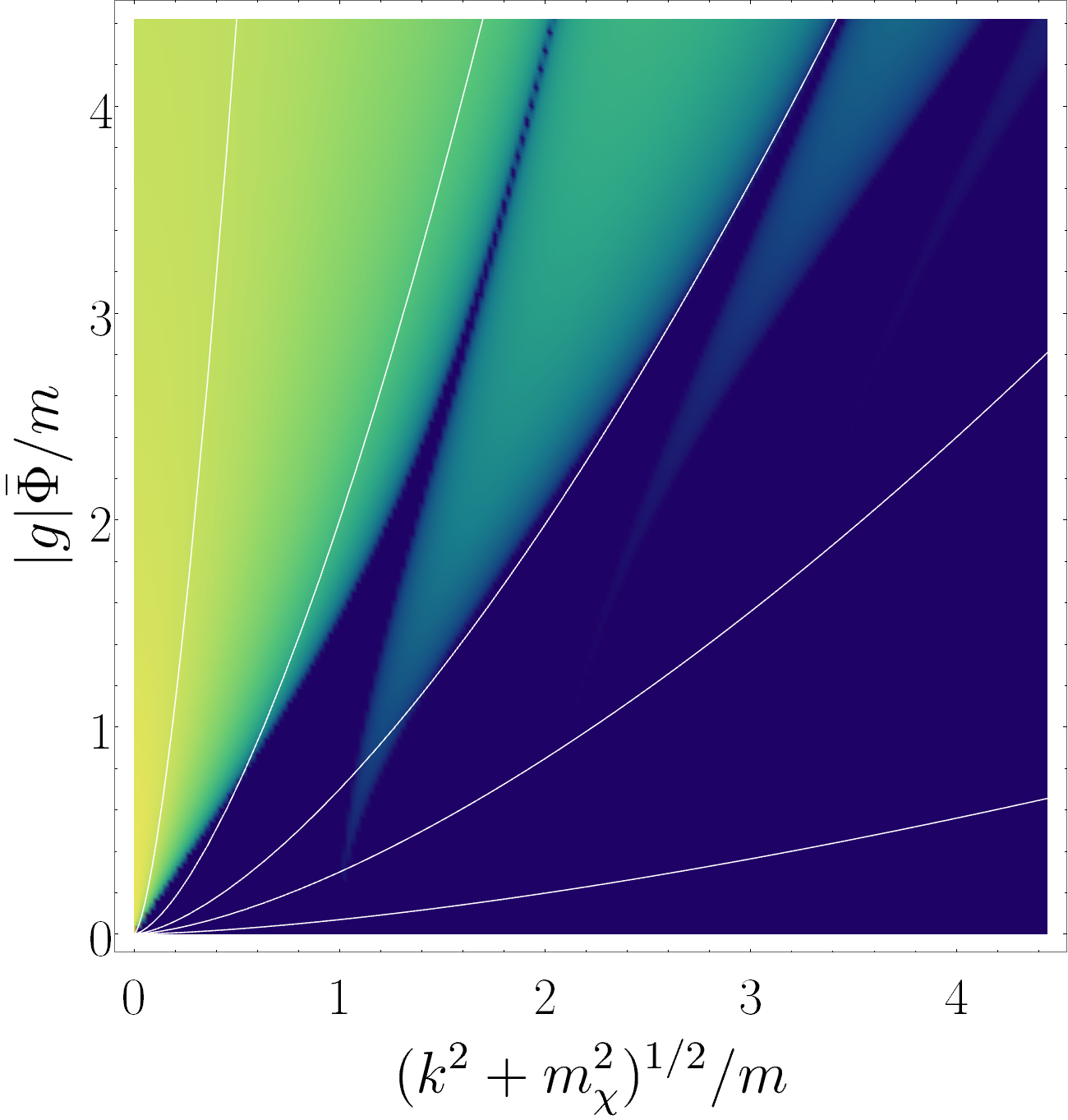} 
     \raisebox{0.13\height}{\includegraphics[height=1.92in]{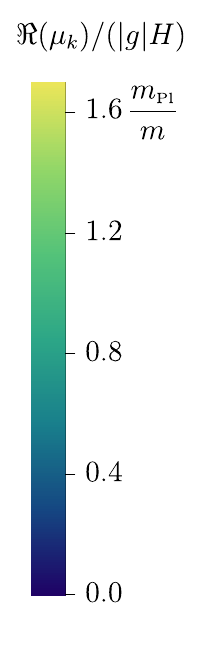}}
   \caption{Same as Fig. \ref{fig:FloqPhi2Chi2Positive}, but for $g^2<0$, (see also Fig. \ref{fig:FloqMathieu}). Note the prominent broad resonance bands, corresponding to $A_k\lesssim2q<0$, not present in the instability chart given in Fig. \ref{fig:FloqPhi2Chi2Positive} for $g^2>0$.}
   \label{fig:FloqPhi2Chi2Negative}   
\end{figure*}

For instance, the adiabaticity condition, eq. \eqref{eq:AdiabCond}, for the Mathieu equation reduces to
\Beq
\frac{2q\sin(2z)}{(A_k+2q\cos(2z))^{3/2}}\ll1\,,
\Eeq
implying that if $A_k\lesssim2|q|$, the inequality is not satisfied near $z_j=\pi/4,3\pi/4,...$ and the WKB solution, see eq. \eqref{eq:WKBSolintro}, does not hold. Away from these $z_j$, the WKB solution is a good approximation. If $A_k\gtrsim2q>0$, adiabadicity can be also violated for $A_k-2q\ll q^{1/2}$ near $z_j=\pi/2,3\pi/2,...$ (similar expressions hold for $A_k\gtrsim-2q>0$; this provides a qualitative explanation of the broad bands in Figs. \ref{fig:FloqMathieu}, \ref{fig:FloqPhiChi2}, \ref{fig:FloqPhi2Chi2Positive}, \ref{fig:FloqPhi2Chi2Negative}). In general, since $|\alpha_k^j|^2-|\beta_k^j|^2=|\alpha_k^{j+1}|^2-|\beta_k^{j+1}|^2=1$, where the superscript $j$ labels the coefficients between the $j$th and $(j+1)$th violation of adiabadicity, etc., the connection between these Bogolyubov type coefficients is
%~~~~~~~~~~~~~~~~~
%~~~~~~~~~~~~~~~~~
\Beq
\label{eq:ReflTransBogol}
\begin{pmatrix} \alpha_k^{j+1}e^{-i\theta_k^j} \\ \beta_k^{j+1}e^{i\theta_k^j} \end{pmatrix}=\begin{pmatrix} 1/D_k^j \qquad R_k^{j*}/D_k^{j*} \\ R_k^j/D_k^j \qquad 1/D_k^{j*} \end{pmatrix}\begin{pmatrix} \alpha_k^je^{-i\theta_k^j} \\ \beta_k^je^{i\theta_k^j} \end{pmatrix}\,,
\Eeq
where $\theta_k^j=\int_{t_0}^{t_j}\omega(k,t) dt$ is the accumulated phase until the $j$th violation of adiabadicity, and the reflection and transmission coefficients must obey $|R_k^j|^2+|D_k^j|^2=1$, to preserve the Bogolyubov nature of the $\alpha$s and the $\beta$s. The calculation of the reflection and transmission coefficients is tedious -- one has to derive connection formulae for the WKB solution on both sides of the non-adiabatic region \cite{landau1977quantum} for a given form of $\omega(k,t)$. Nevertheless, one can use the general expression in eq. \eqref{eq:ReflTransBogol} to study particle production. If we assume that we start in the vacuum state, i.e., $\beta_k^{j=0}=0$, there will be particle production after the first violation of adiabadicity -- $n_{\bf{k}}^{\chi,j=1}=|\beta_k^{j=1}|^2=|R_k^{j=1}/D_k^{j=1}|^2$. In general,
\Beq
n_{\bf{k}}^{\chi,j+1}=\left|\frac{R_k^{j}}{D_k^j}\right|^2(n_{\bf{k}}^{\chi,j}+1)+\left|\frac{1}{D_k^j}\right|^2n_{\bf{k}}^{\chi,j}+2\left|\frac{R_k^{j}}{D_k^jD_k^{j*}}\right|\sqrt{n_{\bf{k}}^{\chi,j}(n_{\bf{k}}^{\chi,j}+1)}\cos(\theta_k^j+\Delta\theta_k^j)\,,
\Eeq
where $\Delta\theta_k^j=\arg(R_k^j\alpha_k^j\beta_k^{j*})$. In the limit $n_{\bf{k}}^{\chi,j}\gg1$, we can write $n_{\bf{k}}^{\chi,j+1}=e^{2\mu_k^j}n_{\bf{k}}^{\chi,j}$, where
\Beq
\label{eq:FloqNonAdiab}
\mu_k^j=\ln\left|\frac{1+|R_k^j|e^{i(\theta_k^j+\Delta\theta_k^j)}}{\sqrt{1-\left|R_k^j\right|^2}}\right|\,.
\Eeq
The argument of the logarithm can take values greater or smaller than unity. $\mu_k^j>0$ corresponds to non-adiabatic particle production at event $j$. We note that violation of adiabaticity is a necessary, but not sufficient condition for particle production. The actual form of $\omega(k,t)$ must be such that $R_k^j$ and $\theta_k^j+\Delta\theta_k^j$ allow for $\mu_k^j>0$, at least on average. That is why regions in parameter space in which the adiabatic condition is not satisfied can still contain stability bands, e.g., see the narrow regions of stability for $A_k\lesssim2q$ in Fig. \ref{fig:FloqMathieu} for the Mathieu equation.

\subsubsection{Classical limit}

The last point we wish to make for parametric resonance after inflation and preheating in general, is that the exponentially amplified modes can be treated classically. Intuitively, this can be understood from the large occupation numbers of these modes. Quantitatively, we can see it by considering the field-momentum commutator
\Beq
\label{eq:CommClassicalParRes}
\hat{\chi}_{\bf{k}}(t)\hat{\pi}_{\bf{q}}(t)=\hat{\pi}_{\bf{q}}(t)\hat{\chi}_{\bf{k}}(t)+i(2\pi)^{-3}\delta(\bf{k}-\bf{q})\,.
\Eeq
The expectation values of the operator products on the left and right hand sides of the expression grow as $e^{2|\Re(\mu_k)| t}$ and can become much greater than unity. Their difference, however, remains small and constant. It is equal to the delta-function term. We can check this by evaluating the commutator explicitly
\Beq
\left[\hat{\chi}_{\bf{k}}(t),\hat{\pi}_{\bf{q}}(t)\right]=W[u_k(t),u_k^*(t)]\times[a_{\bf{k}},a_{\bf{q}}^{\dagger}]=W[u_k(t),u_k^*(t)]\delta(\bf{k}-\bf{q})\,.
\Eeq
Since the Wronskian $W[u_k(t),u_k^*(t)]=\rm{const}$ for any equation of the form given in eq. \eqref{eq:HilluIntro}, and since we start with vacuum fluctuations ($\beta_k=0$) $W[u_k(t),u_k^*(t)]=i/(2\pi)^3$ always, even if each of the terms has grown exponentially. This proves that even though the commutation relation is respected, the quantum correction delta-function term affects the expectation value of $\hat{\chi}_{\bf{k}}(t)\hat{\pi}_{\bf{q}}(t)$ negligibly for the resonantly amplified $\bf{k}$. Hence, the quantum expectation value with respect to the Bunch-Davies vacuum of any function of the densely populated $\hat{\chi}_{\bf{k}}$ mode can be treated as a classical ensemble average over field realizations drawn from a Gaussian probability distribution. The variance of a (zero-mean) field in the WKB regime at some time $t$ is 

\Beq
\langle\chi^2(t,{\bf{x}})\rangle_{\rm{ens}}&=\int d^3kd^3q\langle\chi_{\bf{k}}(t)\chi_{\bf{q}}(t)\rangle_{\rm{ens}}e^{i({\bf{k}}+{\bf{q}})\boldsymbol{\cdot}\bf{x}}\\
                                           &\approx\langle0|\hat{\chi}^2(t,{\bf{x}})|0\rangle=\int d^3kd^3q\delta({\bf{k}}+{\bf{q}})u_k^*(t)u_q(t)e^{i({\bf{k}}+{\bf{q}})\boldsymbol{\cdot}\bf{x}}\,,
\Eeq
where $(2\pi)^3|u_k(t)|^2=[1+2n_{\bf{k}}^{\chi}+2\sqrt{n_{\bf{k}}^{\chi}(n_{\bf{k}}^{\chi}+1)}\cos(\gamma_k)]/(2\omega(k,t))\approx2n_{\bf{k}}^{\chi}\cos^2(\gamma_k/2)/\omega(k,t)$ for $n_{\bf{k}}^{\chi}\gg1$, where $\gamma_k=\arg(\alpha_k\beta_k^*)$. Similar considerations apply to more complicated functions which depend on time derivatives of $\hat{\chi}_{\bf{k}}(t)$ as well.

\bigskip
\bigskip

In the following section we discuss how gravity and additional oscillating background fields can affect the resonant particle production described here.
%Note that the $\mathcal{P}_{k\pm}$ are determined by the initial conditions for $u_{k}$ and $\dot{u}_{k}$.

\subsection{Stochastic resonance}
\label{sec:StochRes}

In the previous section we showed that parametric resonance can play an important role in the preheating phase. We considered the growth of matter fields, represented as fluctuations in an oscillating background, by applying Floquet analysis to the linear equations of motion with periodic coefficients. In doing so, we made several simplifying assumptions. In this section we re-introduce some of the ignored effects and show that they lead to a phenomenon known as stochastic resonance.

Neglecting gravity and assuming that the inflaton is the only field that has a background value allows for the possibility of having strictly periodic linear equations of motion, with exponentially growing solutions. One expects that any extension beyond this set-up can spoil the exact periodicity and, in general, counteract the growth of perturbations. 

\subsubsection{Metric fluctuations}

Actually, incorporating gravity is not difficult. The metric perturbations remain negligible while particle production takes place. One can see that from the generalized Poisson equation (which follows from a combination of the Einstein equations)
\Beq
\label{eq:GeneralizedPoisson}
\frac{\Delta\Psi}{a^2}=\frac{\delta\rho_{\rm{m}}}{2\mpl^2}\,,
\Eeq
where, $\Psi$, is the Bardeen potential, see eq. \eqref{eq:BardeenIntro}, and $\delta\rho_{\rm{m}}\equiv\delta\rho+[\bar{\rho}'(\tau)/a(\tau)]\delta u^{\parallel}$ is the co-moving, gauge-invariant, density perturbation.\footnote{Under a diffeomorphism, eq. \eqref{eq:DiffDeff}, $\delta u^{\parallel}$ transforms according to eq. \eqref{eq:DiffTransfPsiVel}, while from eq. \eqref{eq:DiffEnergyMomentumTensor} follows $\Delta\delta\rho=\bar{\rho}'(\tau)\xi_0$} After defining the co-moving overdensity field $\delta_{\rm{m}}=\delta\rho_{\rm{m}}/\bar{\rho}$, we can say that the linearized equations of motion governing the perturbations hold for small $\delta_{\rm{m}}\ll1$. The Fourier transform of eq. \eqref{eq:GeneralizedPoisson} is $\Psi_{\bf{k}}=(3/2)(aH/k)^2\delta_{{\rm{m}}\bf{k}}$, implying $\Psi_{\bf{k}}\rightarrow0$ as $aH/k\ll1$ for small $\delta_{{\rm{m}}\bf{k}}$. Hence, metric perturbations remain vanishingly small on sub-Hubble scales during the preheating phase. During this phase, the super-horizon metric perturbations also do not grow in single-field models of inflation according to Weinberg's adiabatic theorem \cite{Weinberg:2008zzc}. 

\subsubsection{Expansion of space}

Unlike the metric perturbations, the background space-time curvature cannot be easily neglected during preheating. The FRW expansion of space causes the amplitude of inflaton oscillations to decay, while co-moving wave-numbers are red-shifted to smaller physical values. Going back to our parametric resonance approach, we can see that the equation of motion for the scalar matter fields, eq. \eqref{eq:Hill}, can still be reduced to the form of a simple harmonic oscillator with a time varying frequency. Using the canonically-normalized field $\hat{\chi}_c(t)=a(t)^{3/2}\hat{\chi}(t)$, where $t$ is cosmic time, we obtain
\Beq
\label{eq:EoMChiConf}
\ddot{\hat{\chi}}_{c\bf{k}}+\omega^2(k,t)\hat{\chi}_{c\bf{k}}(t)=0\,.
\Eeq
In the trilinear model, see Sections \ref{sec:PerReh} and \ref{sec:ParRes}, $\omega^2(k,t)=(k/a)^2+m_{\chi}^2+2\sigma^2\bar{\Phi}(t)\cos(mt)-(3H/2)^2-(3/2)\dot{H}$, implying that this is not the Hill's equation any more. Nevertheless, one can depict qualitatively the effects from FRW expansion on particle production by adding flow lines to the Floquet chart, tracing the evolution of particular co-moving modes. Since, in $m^2\phi^2/2$, $\bar{\Phi}(t)\sim a(t)^{-3/2}$ and $3H^2\approx-2\dot{H}$ a given co-moving mode $k$ flows exactly along $\bar{\Phi}\sim k_{\rm{phys}}^{3/2}\equiv(k/a)^{3/2}$ curve in the $k_{\rm{phys}}-\bar{\Phi}$ plane, see right panel in Fig. \ref{fig:FloqPhiChi2} (see also Figs. \ref{fig:FloqPhi2Chi2Positive}, \ref{fig:FloqPhi2Chi2Negative} for other models). Empirically, a condition for parametric resonance (both narrow and broad) to result in significant particle production is
\Beq
\label{eq:mubyH}
\frac{|\Re(\mu_k)|}{H}\gg1\,,
\Eeq
for sufficiently long times. This is another way of saying that particle production occurs only in those bands in which the resonant growth is rapid on the Hubble time-scale. Using the heuristic picture of Floquet theory, we can conclude that depending on the model, broad resonance can be enhanced or shut off by the expansion of space. When more and more co-moving modes are redshifted towards a broad instability band, we observe a temporary increase in the net particle production, see bottom left corner of right panel in Fig. \ref{fig:FloqPhiChi2}, but as they eventually leave the instability regions the resonance gets completely shut-off. 

As we showed in the previous section, broad resonance can be described as a series of particle creation events in which the adiabaticity condition, eq. \eqref{eq:AdiabCond}, is violated. Taking into account the effects of the expansion of space, implies that the quantities appearing in eq. \eqref{eq:FloqNonAdiab} will be time-dependent. The reflection coefficient, $R_k^j$, should have some model dependent and usually monotonic time-dependence, whereas the phase, $\theta_k^j+\Delta\theta_k^j$, can be assumed to vary randomly in the interval $[0,2\pi)$ . The fact that the Floquet index $\mu_k^j$ in eq. \eqref{eq:FloqNonAdiab} can change stochastically between successive particle creation events is the reason why broad resonance in an expanding space is called {\it stochastic resonance}. On average $\mu_k^j\approx(1/2)\ln[(1+|R_k^j|^2)/(1-|R_k^j|^2)]>0$, implying an increasing number of particles, in agreement with entropic arguments. Note that due to the randomness of the phase, $\mu_k^j$ on average can be smaller than in the Minkowski space-time limit. This is a curious feature of stochastic resonance, where particle production occurs on time-scales much shorter than the Hubble time, but still the expansion of space affects the final result.

On the other hand, the efficiency of narrow resonance is severely degraded by the FRW expansion. As one can see in the Floquet charts in Figs. \ref{fig:FloqPhiChi2}, \ref{fig:FloqPhi2Chi2Positive}, \ref{fig:FloqPhi2Chi2Negative}, co-moving modes cross the narrow instability bands much faster than in the broad resonance regime. Thus, expansion takes particles out of the thin resonance layers and the occupation numbers boosting the Bose condensation effect become smaller than in the Minkowski limit. If the rate of escape of particles is greater than the rate of their production, i.e., eq. \eqref{eq:mubyH} does not hold, then Bose effects play no role. The efficiency of narrow resonance is sensitive to other suppressing effects such as the re-scattering of the newly created particles out of the resonance layer, as well as the shift of the resonance region from its original location due to the change of the inflaton effective mass as a consequence of particle production.

We also note that after including the expansion of space we are still allowed to treat the heavily populated modes classically. In particular, the analysis after eq. \eqref{eq:CommClassicalParRes} still holds for the canonically-normalized field $\hat{\chi}_{c}(t)$.

\subsubsection{Multi-field preheating}
\label{sec:MFpreh}

The periodicity of the time-dependent background can be violated also if there are several oscillating homogeneous fields. Even without expansion of space, unless the motion at the background level in the multi-field space occurs along special trajectories such as Lissajous curves or effectively one-dimensional oscillatory trajectories, the time-dependent coefficients in the linear equations of motion governing the fluctuations are not exactly periodic. This can again lead to stochastic resonance if the adiabaticity condition, eq. \eqref{eq:AdiabCond}, is violated \cite{Amin:2015ftc}. Note that this time both the reflection coefficient, $R_k^j$, and the phase, $\theta_k^j+\Delta\theta_k^j$, can be assumed to vary randomly between successive non-adiabatic events. Even the length of the time intervals separating such events can vary randomly. Nevertheless, just like in the case of an expanding space, we could approximate the motion in field space at the background level as being periodic to check if substantial instability (both broad and narrow) bands exist.

We should point out that there is an alternative description of resonant particle production when the number of oscillating homogeneous fields is much greater than one. In this case the effective masses of the daughter fields evolve with a random component to a very good approximation. This reduces the efficiency of the particle production, but resonance still takes place. It occurs at all wavenumbers, not only within particular resonance bands. The alternative way to see why this happens is to note that there is a duality between the equation of motion of daughter fields, see eq. \eqref{eq:EoMChiConf}, and the time-independent one-dimensional Schrodinger equation. The duality interchanges time and space, the mode-function with the wavefunction, the time-dependent effective mass squared with the space-dependent one-dimensional potential energy and $k^2$ with the eigenenergy. Then recalling the celebrated condensed matter phenomenon of Anderson localization, in which small random impurities make eigenfunctions exponentially localized in space, we expect that in the case of preheating, time-dependent masses with random components give rise to exponentially growing modes at all wavelengths; for more details on the condensed matter analogue and the random resonance see \cite{Amin:2015ftc}. % We note that it is quite natural for a large number of scalar fields to acquire VEVs during inflation in supersymmetric models.

\bigskip
\bigskip

We have shown that realising a strictly periodic motion at the end of inflation is difficult. The FRW expansion and the possibility of having more than one oscillating homogeneous fields can lead to a quasi-periodic motion at the background level. This can lead to stochastic resonance if the adiabaticity condition, eq. \eqref{eq:AdiabCond}, is not respected. Even if it is, there could be still some particle production due to perturbative decays. However, as opposed to the strictly periodic case, the Bose enhancement of decays into scalar fields is normally not significant. Despite all that, Floquet analysis remains an important first step towards understanding the instabilities in the evolution of matter fields during preheating.

\subsection{Tachyonic decay}
\label{sec:Tach}

So far we have assumed that the effective frequency, $\omega^2(k,t)$, of the matter fields, $\chi_c$, changes (quasi) periodically with time due to the inflaton oscillations. This need not be the case always. For instance, towards the end of Hybrid inflation \cite{Linde:1993cn}, $V(\phi,\chi)=\lambda_{\chi}(\chi^2-v^2)^2+g^2\phi^2\chi^2+V_{\rm{infl}}(\phi)$, as the inflaton becomes smaller than a critical value, $\phi^2<\lambda_{\chi}v^2/g^2$, but is not oscillating, the sign of $\omega^2(k,t)$ changes from positive to negative for long-wavelength modes and can remain such for an extended period of time. Since one of the two imaginary frequency solutions to eq. \eqref{eq:EoMChiConf} is exponentially growing with time, $\chi_c\propto e^{|\omega|t}$, we again have exponential particle production. A negative squared frequency, $\omega^2(k,t)=(k/a)^2+m_{\chi{\rm{eff}}}^2<0$, implies an imaginary effective mass, $m_{\chi{\rm{eff}}}^2<-(k/a)^2<0$. That is why this mechanism for particle production is dubbed tachyonic preheating. Importantly, all modes whose momenta are less than the magnitude of the imaginary effective mass are unstable, and in the limit $k\rightarrow0$ the exponential index approaches the maximal value of $|m_{\chi,{\rm{eff}}}|$. Tachyonic instabilities in fluctuations always occur in symmetry breaking models for small background field values, e.g., in Hybrid inflation for small enough $\phi$. Tachyonic instabilities can be also observed in the fluctuations of the inflaton field itself, e.g., when it has a symmetry breaking self-interaction potential or in field ranges where the self-interaction potential is shallower than quadratic.

Just like in the case of resonant particle production, to have efficient tachyonic decay of the inflaton condensate,
\Beq
\frac{|m_{\chi,{\rm{eff}}}|}{H}\gg1\,
\Eeq 
must hold for a sufficiently long time. Otherwise, the expansion of space drives $m_{\chi,{\rm{eff}}}^2$ to its equilibrium, positive value (implying positive $\omega^2(k,t)$) before substantial energy can be transferred from the condensate to fluctuations.

In general, tachyonic instabilities can be achieved in models with negative couplings. For instance, in the trilinear model in Sections \ref{sec:PerReh} and \ref{sec:ParRes}, the interaction term $\sigma\phi\chi^2$ implies that even if the inflaton is oscillating, half of the period small $k$ modes will be tachyonic. This corresponds to the $0<A_k<2q$ region in the Mathieu instability chart in Fig. \ref{fig:FloqMathieu}, and explains why there the stability bands are so narrow (see also Fig. \ref{fig:FloqPhiChi2}). They correspond to the small parameter region in which effectively only the exponentially decaying imaginary frequency solution is excited. Note that the expansion of space blurs the boundaries between different regions in the Floquet chart and the narrow stability bands in the tachyonic region go away. Another example of negative coupling resonance is the models with a $g^2\phi^2\chi^2/2$ interaction, where $g^2<0$, see Fig. \ref{fig:FloqPhi2Chi2Negative}. This implies $q<0$ in the notation of the Mathieu equation, see Fig. \ref{fig:FloqMathieu}. Note that in models like this, where interaction terms are always negative to ensure stability we should add higher order positive potential terms. In this case, we can add quartic potential terms, that dominate at large field values, but are unimportant during preheating. In terms of the Mathieu instability chart the tachyonic region corresponds to $2|q|\geq A_k\geq2q$, where the latter bound comes from the $q$-dependence of $A_k$ in this model, see Figs. \ref{fig:FloqMathieu}, \ref{fig:FloqPhi2Chi2Negative}. Compared with the standard resonant preheating scenario ($g^2>0$) where $\mu_k^{\rm{max}}\lesssim m$, see Fig. \ref{fig:FloqPhi2Chi2Positive}, tachyonic preheating can be much more efficient, with maximal exponential index $\sim|g|\bar{\Phi}$. Even if the couplings are small, $|g|\ll1$, to ensure negligible radiative corrections, we can still have $|g|\bar{\Phi}\gg m$ at the end of inflation. Typically, $\bar{\Phi}\sim\mpl$, and even with small couplings it can take less than one oscillation of the condensate for the tachyonic growth of the long-wavelength modes to lead to interesting non-linear dynamics.

%but have not discussed the importance of its sign.

\subsection{Instant preheating}
\label{sec:InstPreh}

The time-dependent nature of the effective mass of fluctuations can give rise to another preheating mechanism. Normally, for a coupling of $\chi$ to some fermion $\psi$ of the Yukawa form, $h_{\chi}\chi\bar{\psi}\psi$, the decay $\chi\rightarrow\bar{\psi}\psi$ is kinematically forbidden if the corresponding bare masses are such that $m_{\chi}<2m_{\psi}$. However, if the scalar is coupled to the inflaton via $g^2\phi^2\chi^2/2$ (assume $g^2>0$) then the effective mass, $m_{\chi{\rm{eff}}}^2=m_{\chi}^2+g^2\phi^2$, can become significantly bigger. And even for a scalar of vanishing bare mass, the decay can be kinematically allowed. For an oscillating inflaton with large enough amplitude, $\bar{\Phi}>2m_{\psi}/g$, the decay rate, see also eq. \eqref{eq:Gammas},
\Beq
\Gamma_{\chi\rightarrow\bar{\psi}\psi}=\frac{h_{\chi}^2g|\bar{\phi}|}{8\pi}\,,
\Eeq
vanishes when $\bar{\phi}\approx0$, and is maximal as the oscillating inflaton reaches its maximal value $|\bar{\phi}|=\bar{\Phi}$. In the large coupling limit, $\sqrt{q}=g\bar{\Phi}/m\gg1$, we have broad resonance, or in other words non-adiabatic particle production every time the non-adiabaticity condition given in eq. \eqref{eq:AdiabCond} is violated. This happens when $\bar{\phi}\approx0$, implying that $\Gamma_{\chi\rightarrow\bar{\psi}\psi}$ is maximal half-way between two consecutive particle production events. Hence, even if a significant amount of $\chi$ particles are produced at each creation event, they can all decay into fermions before the next one. This mechanism is called instant preheating. In it, the back-reaction of $\chi$ particles on the $\phi$ condensate is slowed down and the efficiency of the resonance maintained for very long times. Furthermore, for $g\sim10^{-2}$ and $\bar{\Phi}\sim\mpl$ the light inflaton, $m\sim10^{-6}\mpl$, can decay to heavier scalars and fermions, as heavy as the GUT scale $\sim10^{16}\,\rm{GeV}$. The return of the GUT scale into play obviously presents a threat to inflationary models. Far-from-equilibrium production of topological defects can take place, thus allowing cosmological observations to place bounds on different preheating scenarios.

\newpage

%-----------------------------------------------------------------------------------------------------------------------------
%				Non-linear reheating
%-----------------------------------------------------------------------------------------------------------------------------
\section{Non-linear reheating}
\label{sec:NonLinRehThesis}

%\hfill\begin{minipage}{\dimexpr\textwidth}
%\hfill
\hfill\begin{minipage}{\dimexpr\textwidth-3.7cm}
`{\it Using a term like non-linear science is like referring to the bulk of zoology as the study of non-elephant animals.}'
%\xdef\tpd{\the\prevdepth}
\end{minipage}

\hfill {\it Stanislaw Ulam}

\bigskip

As inflation ends, non-perturbative phenomena such as stochastic resonances and tachyonic preheating can amplify quantum fluctuations of the matter fields, creating particles in a far-from-equilibrium state. The instabilities grow exponentially fast on cosmological time-scales. Such exponential growth cannot proceed forever. Eventually, the produced particles back-react on the preheating process. Mode-mode couplings and non-linear interactions become important. Soon the inflaton condensate fragments and non-linear dynamics takes over. The subsequent evolution of the bosonic fields can be rather non-trivial and a lot of interesting things can happen. Towards the end of this out-of-equilibrium evolution, the fields must thermalize, marking the end of reheating and setting the scene for big-bang nucleosynthesis.

This section begins with a discussion of the end of preheating. We talk about the various places back-reaction can arise in and terminate preheating. We then focus on the non-linear dynamics following the initial burst of particle production and the fragmentation of the inflaton condensate. We survey the different numerical techniques available for tackling the non-linear evolution, and also review the various non-trivial structures that have been studied. We finish with a discussion of the approach to thermalization which can include the turbulent evolution of scalar fields. 

\subsection{Back-reaction: the end of preheating}

Resonant particle production and tachyonic instabilities can be terminated in various ways. If the expansion of space does not intercept the non-perturbative particle production, then the back-reaction of the produced particles eventually shuts it off. Back-reaction effects are associated with higher order in field fluctuations correction terms to the equations of motion in the approximate picture of preheating in which the inflaton condensate is treated as a time-dependent classical background with quantum field fluctuations on top of it.

\subsubsection{Back-reaction at the background level}

The equation of motion describing the evolution of the classical inflaton background, eq. \eqref{eq:FriedKG}, can have corrections due to non-vanishing spatial averages of interaction terms. For instance, in the $V(\phi,\chi)=m^2\phi^2/2+g^2\phi^2\chi^2/2$ model, the presence of $\chi$ particles alters the effective squared mass of the inflaton condensate oscillations by $\Delta m_{\bar{\phi}}^2=g^2\langle\chi^2\rangle$. The angle brackets represent a volume average of the classical $\chi$ (classical in the sense described at the end of Section \ref{sec:ParRes}). From now on, when discussing back-reaction and non-linear dynamics, we shall treat all bosonic fields classically and drop their hats. If there are exponentially unstable modes, then 
\Beq
\langle\chi^2\rangle=\int\frac{d\ln k}{2\pi^2}k^3|\chi_{k}|^2\propto e^{2\mu t}\,,
\Eeq
where $\mu$ is some effective growth index, close to the maximal one $\mu_k^{\rm{max}}$. \footnote{Note that according to the Ergodic theorem \cite{Weinberg:2008zzc}, the spatial average of $\chi^2$ is also equal to the ensemble average over realizations of the stochastic field. This is what the vacuum expectation value of the quantum field tends to, since $\langle0|\hat{\chi}(t)^2|0\rangle=\int k^3|u_{k}(t)|^2d\ln k/(2\pi^2)\propto e^{2\mu t}$, where we integrate the mode function.} The coefficient of proportionality varies slowly with time (apart from an oscillating modulation, it decays monotonically due to the expansion of space). Hence, back-reaction effects become important, $\Delta m_{\bar{\phi}}^2\sim m^2$, within, up to logarithmic factors,
\Beq
\Delta t_{\rm{br}}\sim\mu^{-1}\,,
\Eeq
from the beginning of particle production. For broad resonance, $\mu\sim m\gg H$, and so $\Delta t_{\rm{br}}$ is very short in comparison to the Hubble expansion time-scale. The effect on the condensate from the increase in its effective mass is that its amplitude of oscillations, $\bar{\Phi}$, decreases whereas its frequency increases. In the Mathieu equation notation, the resonance parameter $q=g^2\bar{\Phi}^2/m_{\bar{\phi}{\rm{}}}^2$ rapidly decreases and soon it is difficult for the resonant production of $\chi$ particles to continue further.

\subsubsection{Re-scattering and non-linearity}

The equations of motion describing the field fluctuations are also affected by the particle production. Working in the mean-field/Hartree approximation in which different modes and fields evolve independently (are uncorrelated in time), i.e., %$\int{\bf{d}}^3{\bf{k}}\chi_{\bf{k-q}}^*\chi_{\bf{k}}=0$ if $\bf{q}\neq\bf{0}$, $\int{\bf{d}}^3{\bf{k}}\delta \phi_{\bf{k-q}}^*\chi_{\bf{k}}=0$ for all $\bf{q}$, etc., 
$\langle\chi_{\bf{k-q}}^*\chi_{\bf{k}}\rangle_{\rm{time}}\approx0$ if $\bf{q}\neq\bf{0}$, $\langle\delta \phi_{\bf{k-q}}^*\chi_{\bf{k}}\rangle_{\rm{time}}\approx0$ for all $\bf{q}$, etc.,\footnote{The time average is taken over several oscillations of the more slowly oscillating Fourier transform.} there are only correction mass terms, $\Delta m^2_{\chi}=g^2\langle\delta\phi^2\rangle$ and $\Delta m^2_{\delta\phi}=g^2\langle\chi^2\rangle$, for $\chi$ and the inflaton fluctuations, respectively. They may change the evolution of the field fluctuations slightly, e.g., shift $\chi$ particles out of resonance bands. However, as the number of particles increases, the mean-field/Hartree approximation stops being a good description. The coupling between different Fourier modes becomes important, heralding the true beginning of the non-linear stage. The mode-mode coupling between different momentum modes is called {\it re-scattering} and is what actually leads to the fragmentation of the inflaton condensate. For instance, there is an additional non-vanishing source term in the equation of motion for the inflaton fluctuations $\sim g^2\Phi\int{\bf{d}}^3{\bf{k}}\langle\chi_{\bf{k-q}}^*\chi_{\bf{k}}\rangle_{\rm{time}}\propto e^{2\mu t}$. Having an inhomogeneous equation with exponentially growing source term, implies that its particular solution also grows exponentially, i.e., $\delta\phi_q\propto e^{2\mu t}$. Thus, due to the interactions of pairs of $\chi$ particles with particles in the condensate, inflaton fluctuations grow twice as fast. The growth is a manifestation of inflaton particles being scattered out of the inflaton condensate. They are low-momentum excitations, predominantly. When $\langle\delta\phi^2\rangle\gtrsim\Phi^2$, we say that the condensate is substantially fragmented and if $\langle\delta\phi^2\rangle\gg\Phi^2$ we say that it is completely fragmented or destroyed. Re-scattering also re-distributes the energy stored in the $\chi$ particles. Parametric resonance leads to the excitation of $\chi$ momentum modes lying in instability bands. The re-scattering transfers energy from the amplified modes to modes with momenta lying in the stability regions. This may slow down the resonant particle production. Even if it completely shuts off the resonance, re-scattering becoming important is a sign of the ensuing non-linear evolution and fragmentation of the inflaton condensate.

We should point out that having a fragmented inflaton condensate, $\langle\delta\phi^2\rangle\gtrsim\Phi^2$, does not necessarily imply that the energy stored in it is negligible. However, we can say with certainty that re-scattering and fragmentation kick in when the energy stored in interaction terms and/or fluctuations is comparable to the energy of the classical background. Hence, non-perturbative particle production ends and non-linear evolution begins with either most or at least a non-negligible fraction of the total energy being stored in field fluctuations.

\bigskip
\bigskip

Before moving forward to different approaches for studying the non-linear stage, we consider the possibility of having a second field that has a small, but non-vanishing background value, e.g., $\Phi\gg|\bar{\chi}|>0$. While the back-reaction mechanisms remain largely unchanged, the preceding linear evolution during preheating can exhibit novel behaviour. Essentially, there are additional mixing terms, $\sim g^2\Phi\bar{\chi}\delta\phi_k$ and $\sim g^2\Phi\bar{\chi}\chi_k$, in the equations for $\chi_k$ and the inflaton fluctuations, respectively. % If the inflaton condensate oscillates, 
They lead to chaotic evolution of the field fluctuations. The strong dependence on the initial value of $\bar{\chi}$ can give rise to observational signatures of preheating, as we will discuss in Section \ref{sec:MetrFluc}.

We now proceed with the non-linear stage of reheating, following the back-reaction of the produced particles on the inflaton condensate and the breakdown of the linear analysis.

\subsection{Non-linear evolution}
\label{sec:NonLinIntro}

Preheating ends when the occupation numbers of excited bosonic field modes become large and back-reaction effects render the linearized approximation not applicable. The inflaton and the fields it is coupled to start evolving as a combined system. Non-linear interactions lead to the transfer of power between different wavenumbers. This non-linear phase is dynamically rich and can be studied numerically.

\subsubsection{Numerical approach}

The standard approach in numerical analysis is to solve the classical evolution equations, e.g.,
\Beq
\Box\phi+\partial_{\phi}V(\phi,\chi)=0\,,\qquad\Box\chi+\partial_{\chi}V(\phi,\chi)=0\,,\qquad R_{\mu\nu}-\frac{1}{2}g_{\mu\nu}R=\frac{T_{\mu\nu}}{\mpl^2}\,.
\Eeq
There are several publicly available codes created for this purpose. Most of them use a finite-difference method for solving the equations. The fields are discretized on a cubic co-moving spatial grid, with periodic boundary conditions. The time evolution is then a matter of evolving forward a system of coupled ordinary differential equations. For numerical integration in time LATTICEEASY \cite{Felder:2000hq} uses the simplest symplectic integrator -- the leapfrog scheme, which is fast (no need for storage of field and field time derivatives simultaneously during a time step) and second order accurate in time. DEFROST \cite{Frolov:2008hy} and HLATTICE \cite{Huang:2011gf} use higher order symplectic integrators. GABE \cite{Child:2013ria} uses a second order Runge-Kutta method which stores field and field time derivatives on the same time slices, unlike symplectic integrators. Although, this requires more time to run the simulations and more physical memory, it allows for non-canonical kinetic terms. CUDAEASY \cite{Sainio:2009hm} and PYCOOL \cite{Sainio:2012mw} are GPU-accelerated codes based on DEFROST. A pseudo-spectral code, PSpectre \cite{Easther:2010qz}, is also available, which evolves the Fourier transforms of the fields. In it, unlike in finite-difference codes, Laplace terms are dealt with straightforwardly. Each contributes a single term, e.g., $k^2\phi_{\bf{k}}$, to the Fourier transformed equations of motion with no computational cost, whereas $\Delta_{\bf{x}}\phi(\bf{x})$ in finite-difference codes is more costly, since one has to compute the differences with neighbouring points for each lattice site. However, non-linear interaction terms in the equations of motion, e.g., $g^2\phi^2(\bf{x})\chi(\bf{x})$, are easy to deal with in finite-difference codes, whereas for pseudo-spectral codes they present a problem, since there one has to calculate multidimensional integrals. 

We should also point out that most publicly available codes do not include metric perturbations, i.e., they evolve the fields in pure FRW space-time. In addition to the Klein-Gordan equations, they solve one equation for the evolution of the scale factor, $a(t)$. Note that the Einstein equations yield two equations for the evolution of $a(t)$, namely the Friedmann and Raychaudhuri equations given in eq. \eqref{eq:FRWeqs}, with $\rho(t)=\langle\rho\rangle$ and $p(t)=\langle p\rangle$ averaged over the simulation box. Programs typically evolve the Raychaudhuri equation and treat the Friedmann equation as a constraint that has to be satisfied after each time step. Empirically, violations of the Friedmann equation $\geq0.1$ \% indicate poor energy conservation and render the simulations unreliable. Some studies simplify matters further, by assuming a fixed time-dependence of $a(t)$. This means that the expansion of space is not calculated self-consistently, e.g., by solving the Raychaudhuri. Common choices are $a\propto t^{n}$ with $n=2/3,\,1/2$ for matter and radiation-dominated backgrounds, respectively. But still the Friedmann equation is treated as a constraint that has to be checked after each time step. Approximating the space-time to be FRW is justified, since the lattice size of typical simulations is sub-horizon and just like during preheating, metric perturbations are suppressed on these scales and do not affect the non-linear evolution of the fields. The reason why sub-horizon scales are of main interest are the causal mechanisms which drive the non-linear evolution of the fields. Causally disconnected Hubble patches evolve independently and almost identically, implying that it is sufficient to capture one Hubble volume in numerical simulations. Otherwise, the only publicly available code that can include metric perturbations is HLATTICE. 

Of course, the FRW approximation is non-viable if large sub-horizon inhomogeneities, that can lead to the formation of primordial black holes, are present during preheating. However, such inhomogeneities rarely form due to matter field instabilities \cite{Frolov:2010sz,Bassett:2005xm}. They either occur in models in which significant super-horizon inhomogeneities generated during inflation re-enter the horizon during preheating, or are induced by gravitational instabilities which become important long after the end of inflation.

We should also point out that all publicly available codes are written for scalar fields. The GABE code can be adapted for gauge field dynamics, but its ability to respect the gauge constraints has not been fully tested yet, especially with charged scalar fields. We should also note that DEFROST differs from the other finite-difference codes. In it, instead of directly discretizing the equations of motion, the Lagrangian is discretized and then the corresponding equations of motion are evolved numerically. % We shall adopt this approach to study non-linear dynamics in the Abelian-Higgs system in section \ref{ch:GThaw}. 

\subsubsection{Non-linear dynamics}

\begin{figure*}[t] %  figure placement: here, top, bottom, or page
   \centering
   \includegraphics[width=2.22in]{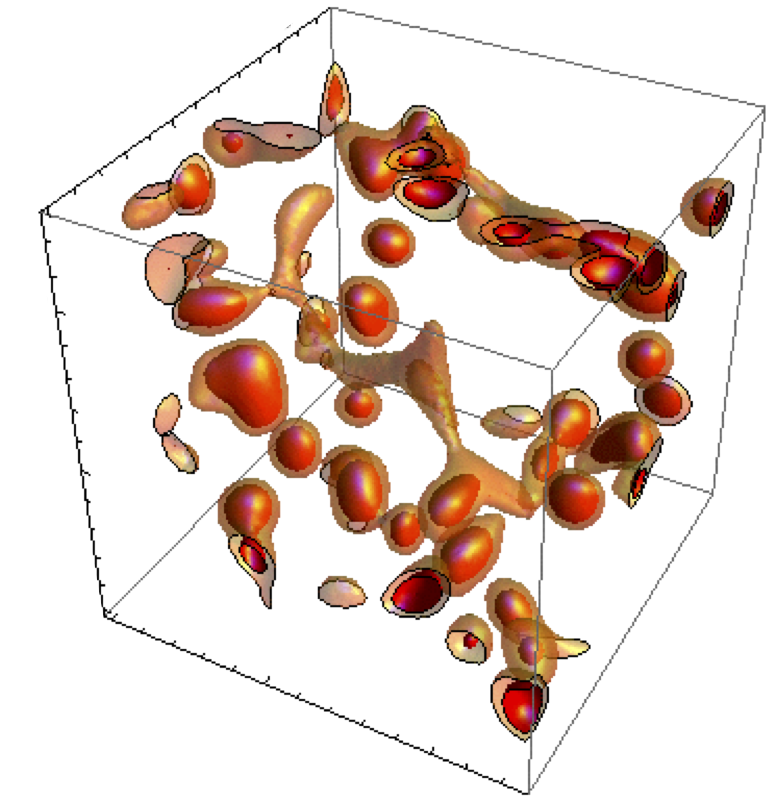} 
   \includegraphics[width=2.22in]{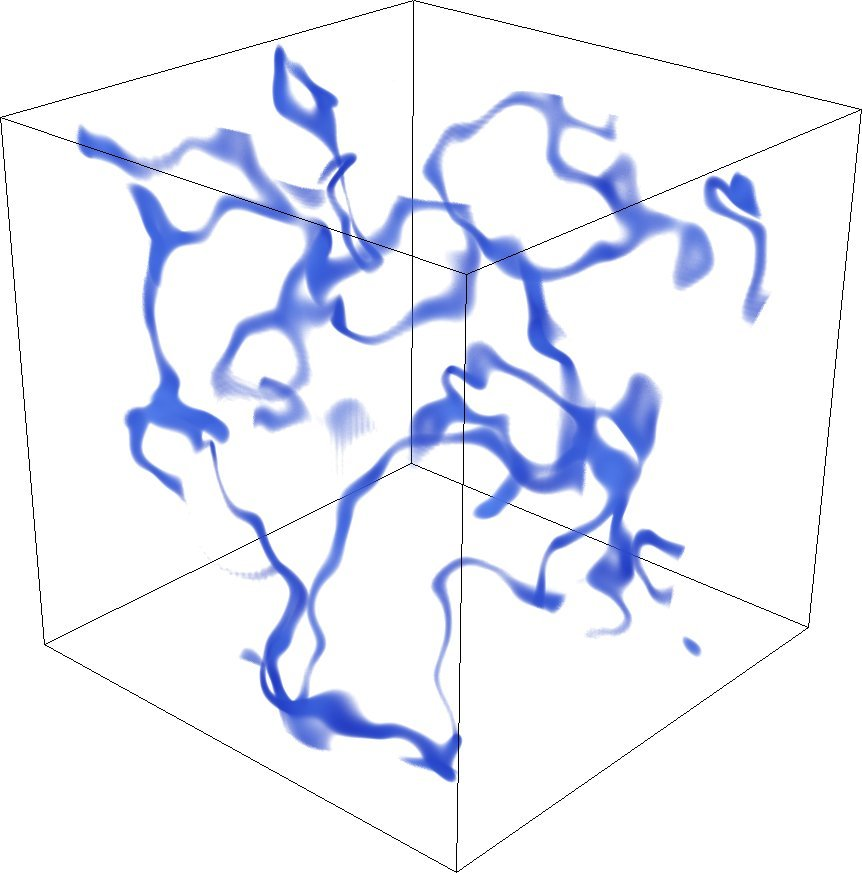}\\
   \includegraphics[width=2.22in]{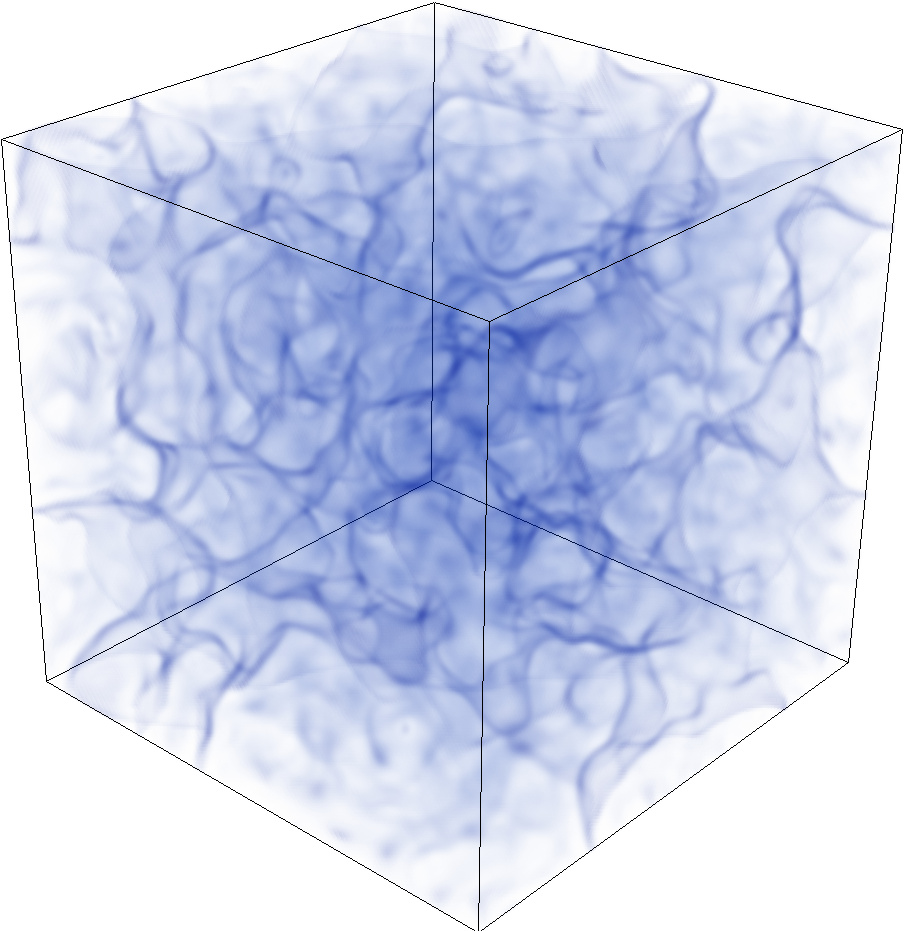} 
   \includegraphics[width=2.22in]{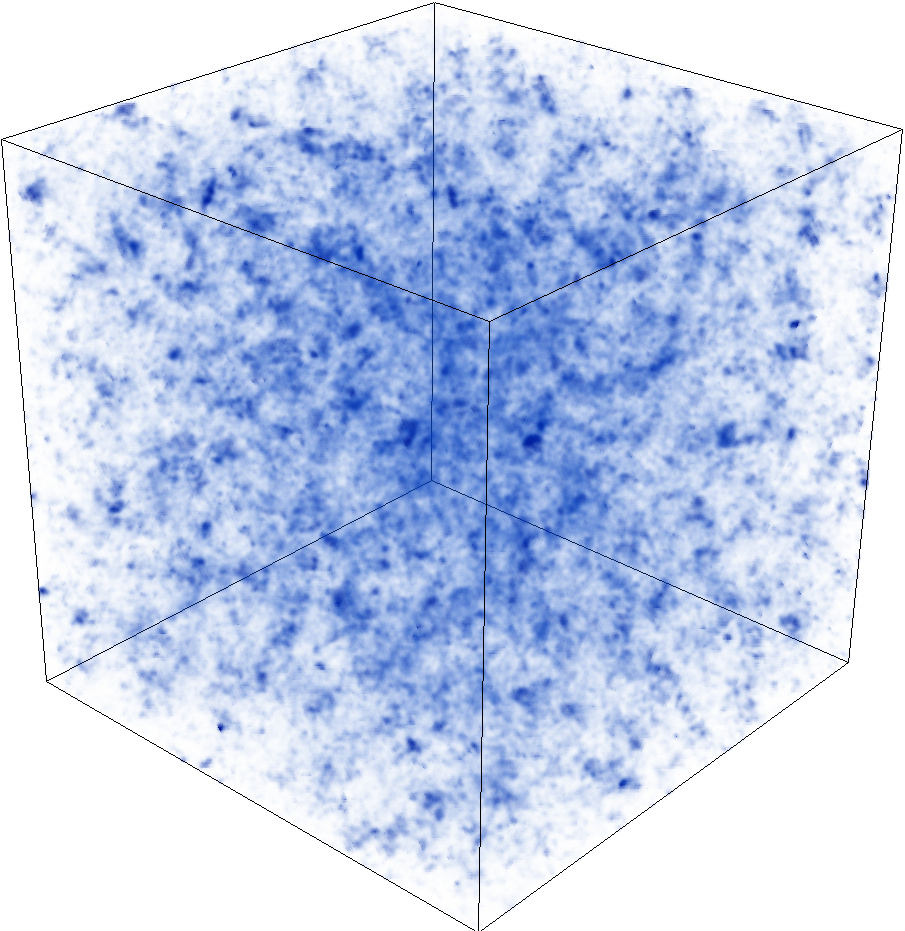}\\
   \includegraphics[clip, trim=5.5cm 4.7cm 23.5cm 2.7cm, height=1.7in]{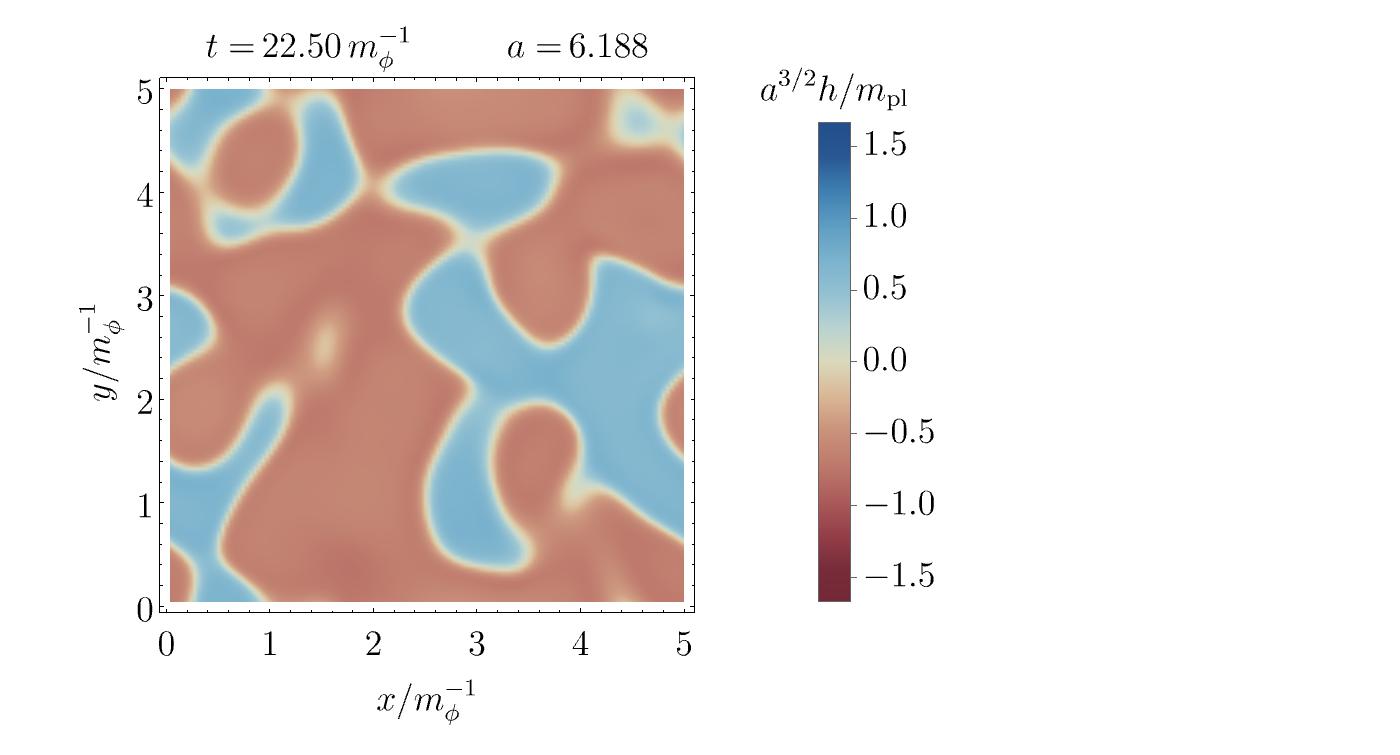} 
   \raisebox{-0.1\height}{\includegraphics[width=2.22in]{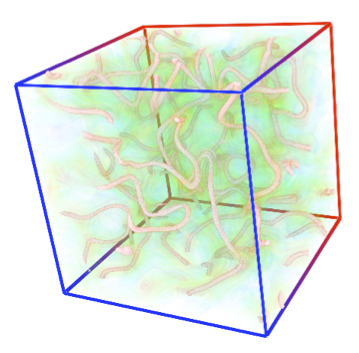}}
   \caption{The formation of non-linear structures during reheating. Top row: oscillons (left) \cite{Amin:2010dc} and global strings (right) \cite{Frolov:2010sz}. Middle row: formation of bubbles (left) \cite{Frolov:2008hy} and their decay products (right). Bottom row: domain walls (left) \cite{Amin:2018kkg} and cosmic strings (right) \cite{Dufaux2010}.}
   \label{fig:NonLinearObjects}   
\end{figure*}

Non-linear effects can become important even in the simplest models of reheating in which the interactions of the inflaton with other fields are negligible, see Fig. \ref{fig:NonLinearObjects}. If self-interaction terms, e.g., $\propto\phi^{n}$, $n\neq2$, become important, the inflaton condensate can fragment after self-resonance. It can form non-trivial field configurations such as oscillons which can lead to long periods of matter-dominated state of expansion \cite{Amin:2011hj}, or form Q-balls if the inflaton is a complex scalar \cite{Kusenko:2008zm}. Oscillons (as well as Q-balls) can also affect predictions in baryogenesis models with a complex inflaton \cite{LozAmin}. %, as we will show in Chapter \ref{ch:baryogenesis}. 
If the inflaton is very light, but self-interacting, it inevitably fragments and attains a radiation-like equation of state \cite{Lozanov:2016hid}. % This can have important observational consequences as will be shown in Chapter \ref{ch:eos}.
Gravitational waves can also be generated due to fragmentation induced by self-interactions \cite{Zhou:2013tsa,Antusch:2016con}. Even if the inflaton is not self-interacting, the condensate inevitably fragments due to gravitational instabilities \cite{Easther:2010mr}.

%~~~~~~~~~~~~~~~~~
%~~~~~~~~~~~~~~~~~

Coupling the inflaton to other fields can lead to very rich phenomenology, see Fig. \ref{fig:NonLinearObjects}. Interaction with scalar fields can lead to the formation of metastable bubble-wall-like configurations \cite{Felder:2006cc}, whose size is not much smaller than $H^{-1}$. After they collide the density is transferred to much smaller scales. This could be interpreted as upscattering of modes to higher momenta due to non-linear interactions. Scalar field theories can also feature the formation of solitons and defects, such as domain walls and metastable global cosmic strings \cite{Tkachev:1998dc}. Models that include gauge fields lead to new phenomena as well \cite{Dufaux2010}. Non-equilibrium phase transitions can lead to the formation of stable topological defects, whereas models with conformal couplings feature large-scale magnetic fields and a faster approach to a radiation dominated state of expansion \cite{Deskins2013,Adshead:2015pva}.

The reason why topological defects can be produced during preheating, in theories which allow them, is non-equilibrium restoration of broken symmetries \cite{Rajantie:2000fd,Rajantie:2000nj,Copeland:2002ku}. Just like in phase transitions, a negative bare mass squared can receive significant positive contributions from large field variances leading to a positive effective mass squared and temporary restoration of a symmetry. This can happen either during back-reaction and re-scattering, or after the fields enter the full non-linear regime. Once the expansion of space dilutes the energy enough, the symmetry is re-broken and topological defects can be produced. For instance, in models allowing the formation of cosmic strings, one has to wait for the mean energy density to become less than the potential energy in the central unstable maximum, filling the ring at the bottom of the potential. This mechanism leads to the production of strings on both sub and super horizon scales \cite{Tkachev:1998dc}. The sub-horizon strings are transient -- they evaporate due to the emission of classical radiation. In general, the density of topological defects arising in such non-equilibrium phase transitions is determined by the correlation length of the fields shortly before the symmetry is restored. For inhomogeneous configurations, the correlation length scale can be considerably smaller than the Hubble scale unlike in the Kibble mechanism. Thus, preheating provides a mechanism for copious production of dangerous topological defects, even at GUT scales, which can have implications for inflationary models.

\subsection{Turbulent scaling}
\label{sec:TurbScal}

The early stage of the non-linear evolution, following back-reaction and re-scattering, in models with scalar fields is dynamically rich and chaotic. Various transient non-trivial field configurations can form, wiping out details on initial conditions from inflation and preheating. Unless long-lived objects form, e.g., oscillons, stable defects, black holes, etc., the state of the fields eventually enters a highly inhomogeneous phase which can persist for very long times, much longer than the preheating and transient phases. It is characterised by a slow, but steady transfer of energy to higher momenta. Essentially, straight after the transient stage the field occupation numbers in the infrared quickly saturate to a power-law, $n(k)\propto k^{-3/2}$, with a UV cut-off not much greater than the typical wavenumber of excited particles during preheating \cite{Micha:2002ey,Micha:2004bv}. The power-law is non-thermal -- for a thermal distribution of relativistic weakly interacting bosons we expect $n(k)\propto k^{-1}$ in the infrared. It then slowly propagates towards higher momenta. Typically, the cascading of the distribution towards the UV can be characterized as turbulent scaling in which the occupation numbers evolve self-similarly, $n(k,\tau)=(\tau/\tau_0)^{-q_1}n_0(k\tau^{-q_2})$, where $q_1$ and $q_2$ are some positive powers, determined by the form of the interactions and $\tau_0$ is the conformal time when the scaling regime begins. This slow fragmentation proceeds until the occupation numbers of the highest $k$-modes belonging to the power-law distribution become of order unity. Then the classical description breaks down and quantum effects become important. Note that the energy density in a given mode is $\rho(k)\propto k^4n(k)$, implying that the high-$k$ modes belonging to the power-law dominate the energy budget. That is why, if present, their quantum behaviour cannot be neglected.

\subsection{Thermalization}
\label{sec:ThermIntro}

\subsubsection{Two stages}

None of the preheating mechanisms described in Section \ref{sec:PreheatIntro}, nor the subsequent non-linear evolution yield a thermal spectrum of decay products. However, measurements of the anisotropies in the CMB and the relative abundances of light elements tell us that the Standard Model degrees of freedom were in thermal equilibrium at the beginning of the big-bang nucleosynthesis and that the universe at that time was in a radiation-dominated state \cite{Ade:2015xua}. The moment when the universe achieves thermal equilibrium for the first time after the end of inflation, at some reheating temperature, $T_{\rm{reh}}\geq T_{\rm{BBN}}\sim1\,\rm{MeV}$, in a radiation-dominated state of expansion, $w\approx1/3$, marks the end of thermalization and the reheating epoch. The value of $T_{\rm{reh}}$ can have an impact on the production of dangerous relics, such as gravitinos, or on the formation of topological defects from thermal phase transitions and the gravitational waves they generate. The expansion history of the universe during reheating, and in particular the moment when the equation of state approaches $1/3$ can have important implications for the uncertainties in predictions of inflationary models \cite{Lozanov:2016hid,Lozanov:2017hjm}. % as will be shown in Chapter \ref{ch:eos}. 

Thermalization can be a long process, much longer than the preceding preheating, transient and turbulent phases. In principle, the universe can attain a radiation-like equation of state during or shortly after the turbulent stage, i.e., it can satisfy one of the two criteria for thermalization quite early. However, reaching a state of Local Thermal Equilibrium (LTE) can take much longer and involve particle fusion and off-shell processes. We say that the universe is in a prethermalized state if $w\approx1/3$, but LTE is not established yet. Prethermalization can be delayed by the formation of long-lived objects like oscillons and Q-balls. They behave as pressureless dust and therefore must decay into relativistic matter to achieve a radiation-like equation of state before BBN. Similar considerations apply to massive scalar field condensates. For instance, if there is some remnant inflaton condensate, even if subdominant in energy during thermalization, it can make the universe re-enter a matter-dominated state of expansion before BBN. To avoid this, one must ensure the complete decay of the condensate. Introducing perturbative decays through three-leg interactions like Yukawa couplings, $h\phi\bar{\psi}\psi$, proves to be a reliable way for the absolute removal of $\bar{\phi}$. That is why, albeit unimportant during the early non-perturbative stages of reheating, perturbative decays of $\bar{\phi}$ are vital for the late stage of thermalization.

In a state of LTE the local value of the entropy, i.e., the entropy per unit volume, $s$, is maximized. LTE is achieved by particle species which are both in kinetic and chemical equilibrium. This requires both the re-distribution of momentum and energy between different particles, as well as an increase in their total number. Hence, both number-conserving and number-violating (off-shell process, particle fusion) reactions are involved. Negligible interactions between different species lead to Bose-Einstein and Fermi-Dirac distributions for bosons and fermions, respectively, in kinetic equilibrium. Kinetic equilibrium entails efficient exchange of energy and momentum between particles, i.e., it is sufficient to have number-conserving interactions only. On the other hand, chemical equilibrium can be achieved only by changing the number of particles. If number-violating interactions are suppressed (this could occur if the particles mediating the number-violating interactions acquire a large mass at early times) the state of kinetic equilibrium is also known as a quasi-thermal state. However, as number-violating processes become efficient and particles flow to lower chemical potentials until the sum of chemical potentials of reacting particles becomes equal to the sum of the chemical potentials of the products in every reaction, $s$ can be truly maximized and full LTE reached.

\subsubsection{Perturbative limit}

If after the end of inflation Bose effects and non-adiabatic particle production are unimportant, i.e., the inflaton condensate undergoes perturbative decays as described at the beginning of Section \ref{sec:PerReh}, then the decay is completed when $\rho\sim\Gamma^2\mpl^2$. In the trilinear model, $V(\phi,\chi)=m^2\phi^2/2+\sigma\phi\chi^2$, the perturbative decay rate is $\Gamma=\Gamma_{\phi\rightarrow\chi\chi}$, see eq. \eqref{eq:Gammas}, and to ensure no parametric resonance $q=\sigma\Phi/m^2\lesssim\sigma\mpl/m^2\ll1$. Hence, $\rho\sim(\sigma^4/m^2)\mpl^2\ll m^4(m/\mpl)^2\ll m^4$. The momentum of the massless $\chi$ particles will be $m/2$, implying a particle energy $\langle E\rangle= m/2\gg \rho^{1/4}$. On the other hand, the $\chi$ particles number density is $n=\rho/\langle E\rangle\ll\rho^{3/4}$. Note that in LTE $\langle E\rangle_{\rm{LTE}}\sim T$ and $\rho_{\rm{LTE}}\sim T^4$, implying $n_{\rm{LTE}}=\rho_{\rm{LTE}}/\langle E\rangle_{\rm{LTE}}\sim T^3\sim \rho_{\rm{LTE}}^{3/4}$. Thus, perturbative preheating leads to a non-equilibrium dilute universe containing very energetic particles. To ensure the completion of thermalization, we should now introduce number-conserving and number-violating interactions, which are efficient even in a dilute plasma. When the universe reaches a radiation-dominated state of expansion, and the rate of these interactions is $>\!\!H$, reheating is completed.

\subsubsection{Non-perturbative effects}

Resonant and/or tachyonic decays of the inflaton condensate are also highly non-thermal processes. They yield non-equilibrium spectra, with peaks lying in instability bands $n(k)\propto e^{2\mu_k t}$ at the end of preheating. After the phase of exponential growth is terminated by back-reaction and re-scattering, and after a brief period of chaotic evolution of inhomogeneous field configurations, the spectrum of a scalar field typically relaxes into a continuous band, $n(k)\propto k^{-1}$, going all the way to $k\rightarrow0$ and having an increasing UV cut-off, $k_{\rm{c}}(\tau)=k_{\rm{c}0}(\tau_0)(\tau/\tau_0)^{q_2}$, as described in Section \ref{sec:TurbScal}. After the front of the distribution, which dominates the energy budget of the universe, starts behaving quantum mechanically, $n(k_{\rm{c}}(\tau))=\mathcal{O}(1)$, the matter spectra should relax into Bose-Einstein and Fermi-Dirac (for fermions weakly coupled to the scalars) distributions. We can then use the results obtained from the classical field theory analysis to put a lower bound on the duration of reheating. Assuming the radiation-dominated state of expansion begins soon after the end of inflation, $a(\tau)/a(\tau_0)=\tau/\tau_0=\rho(\tau_0)^{1/4}/\rho(\tau)^{1/4}$ and putting $(k_{\rm{c}}(\tau_{\rm{reh}})/a(\tau_{\rm{reh}}))^4\sim T_{\rm{reh}}^4$ (recall $n(k_{\rm{c}}(\tau_{\rm{reh}}))=\mathcal{O}(1)$ and $\rho(k)\sim (k/a)^4n(k)$) and $k_{\rm{c}0}/a(\tau_0)\sim m$
\Beq
\left(\frac{\tau_{\rm{reh}}}{\tau_0}\right)^{q_2}&=\frac{k_{\rm{c}}(\tau_{\rm{reh}})}{k_{\rm{c}}(\tau_0)}\sim\frac{T_{\rm{reh}}}{m}\frac{a(\tau_{\rm{reh}})}{a(\tau_0)}\sim\frac{\rho(\tau_0)^{1/4}}{m}\\
                                             &=\left(\frac{a(\tau_{\rm{reh}})}{a(\tau_0)}\right)^{q_2}\sim\left(\frac{\rho(\tau_0)^{1/4}}{T_{\rm{reh}}}\right)^{q_2}\,,
\Eeq
we find that 
\Beq
T_{\rm{reh}}\sim \left(\frac{m}{\rho(\tau_0)^{1/4}}\right)^{1/q_2}\rho(\tau_0)^{1/4}\,.
\Eeq
Putting $m=10^{-6}\mpl$ and $\rho(\tau_0)^{1/4}\sim10^{15}\,\rm{GeV}$ and $q_2=1/7$ \cite{Micha:2002ey,Micha:2004bv} yields $T_{\rm{reh}}\sim 10^3\,\rm{eV}$. This estimate gives an unacceptably low reheating temperature, implying that additional interactions, e.g., decays into fermions, become important before the non-linear evolution of the scalar fields drives them into thermal equilibrium. The calculation of the reheating temperature in such models with highly inhomogeneous scalar field configurations remains an open challenge.

\newpage

%-----------------------------------------------------------------------------------------------------------------------------
%				Reheating and High Energy Physics models
%-----------------------------------------------------------------------------------------------------------------------------
\section{Reheating and High-Energy Physics models}
\label{sec:RehHEPmodels}

\hfill\begin{minipage}{\dimexpr\textwidth-3.7cm}%-8.4cm}
`{\it Is the universe `elegant', as Brian Greene tells us? Not as far as I can tell, not the usual laws of particle physics, anyway. I think I might find the universal principles of String Theory most elegant -- if I only knew what they were.}'%I do not keep up with the details of particle physics.}'
%\xdef\tpd{\the\prevdepth}
\end{minipage}

\hfill {\it Leonard Susskind}%Murray Gell-Mann}

\bigskip

Accelerator experiments such as the LHC have given us information about the governing particle theory up to $\mathcal{O}(10\,\rm{TeV})$. This is many orders of magnitude below the highest reheating scale allowed by observations. Measurements of the CMB anisotropies \cite{Ade:2015lrj} constrain $r<0.11$, implying that the energy scale at the end of inflation, see eq. \eqref{eq:SlowRollInflatonPot}, must be $V_{\rm{end}}^{1/4}<10^{15}\,\rm{GeV}$. This is also the upper bound on $T_{\rm{reh}}$. At such high energy scales we could ignore GUT-mass particles (approximately) and stringy states, but have to include all other degrees of freedom. Hence, there is a huge theoretical uncertainty regarding the actual model of reheating.

Since there is a great ambiguity regarding the degrees of freedom and their interactions at the high energy scales relevant to reheating we can just focus on the particle content of the Standard Model for simplicity. Ignoring the effects from extensions that account for baryogenesis and the generation of dark matter, one can study the evolution of the Standard Model degrees of freedom during preheating, assuming they were all spectator fields during inflation. Furthermore, coupling non-minimally the Standard Model Higgs field to gravity allows it to play the role of the inflaton field. This model is known as Higgs-inflation \cite{Bezrukov:2007ep}. It is quite interesting since in it all couplings of the inflaton to the Standard Model degrees of freedom are known, allowing for a complete calculation of the thermal history of the visible universe. Studies of the non-perturbative preheating dynamics (in the linear approximation) \cite{GarciaBellido:2008ab,Bezrukov2008} have shown that non-linear effects become important soon after the end of Higgs-inflation and a further detailed numerical investigation is required for the calculation of $T_{\rm{reh}}$.

The exploration of extensions of the Standard Model, motivated by e.g., supersymmetry and/or supergravity, can introduce many new degrees of freedom. Unfortunately, they come with new interactions and parameters, many of which are poorly constrained, if at all. The vast landscape of string theory is a good example of the level of theoretical uncertainty one has to deal with when building models of reheating. To make further progress in constraining the particle physics of reheating we should turn to observations. A determination of the exact model of inflation through observations of the CMB could give us some insight into the physical laws governing the dynamics of reheating. Another possibility is the detection of a reheating signal that cannot be mimicked by any inflationary model. The observational consequence of reheating are the subject of the next section.

Despite the fact that we do not know the exact particle physics model describing reheating, it is safe to say that in any realistic scenario there will be a large number of scalar fields, fermions, vector fields, and perhaps non-minimal couplings to gravity and operators that are suppressed below some energy cut-off scale%, e.g., higher-derivative and non-canonical kinetic terms. 
. In the rest of this section we briefly discuss their effects on various aspects of the non-perturbative linear dynamics of preheating by considering simple toy models. We also talk about some generic models of (p)reheating.

%~~~~~~~~~~~~~~~~~
\begin{figure*}[t] %  figure placement: here, top, bottom, or page
   \centering
   \includegraphics[width=2.22in]{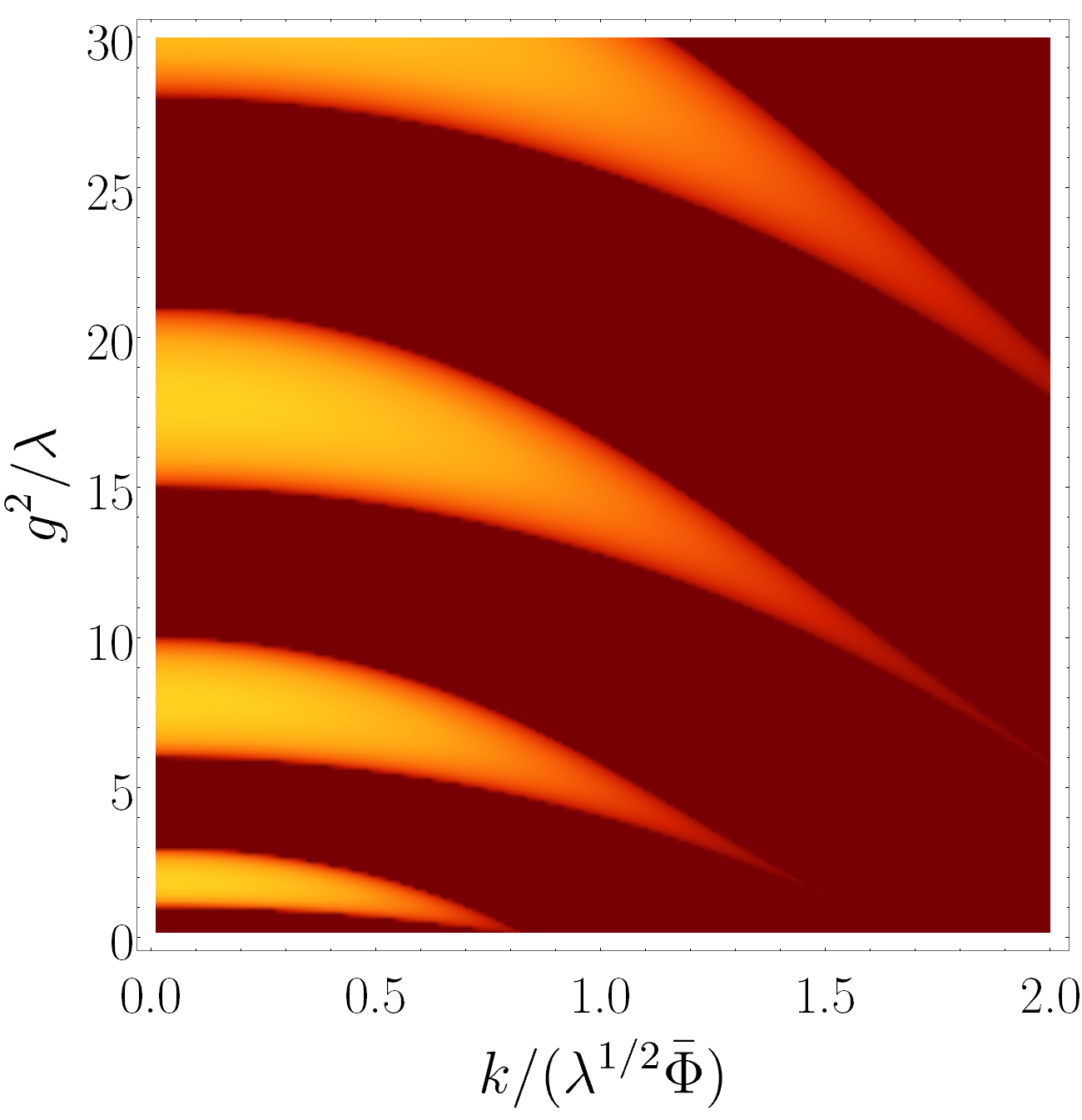} 
     \raisebox{0.13\height}{\includegraphics[height=1.92in]{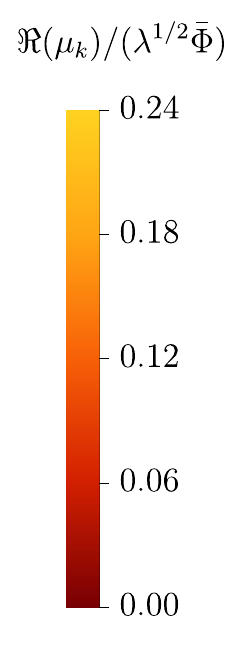}}
   \includegraphics[width=2.22in]{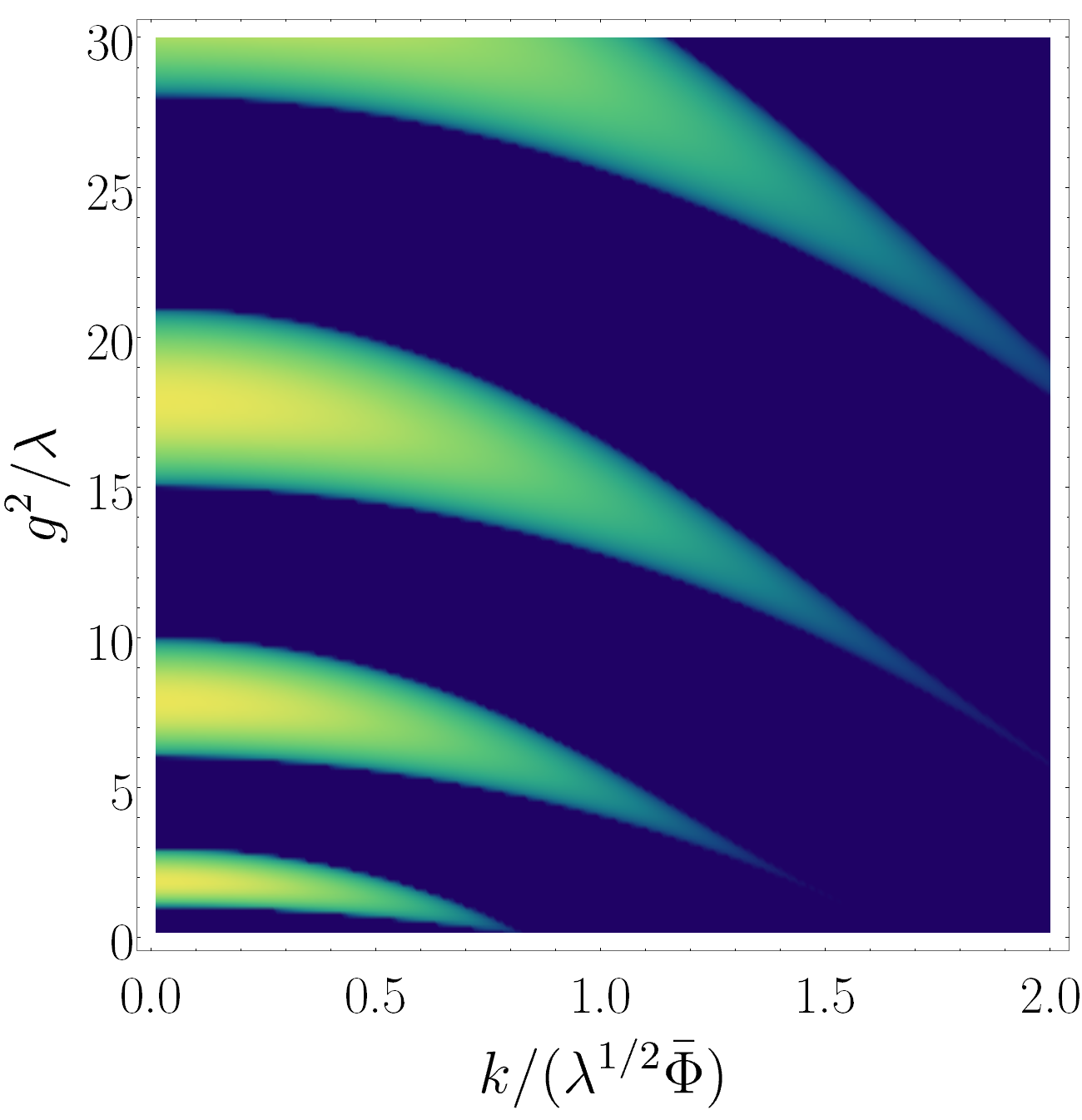} 
     \raisebox{0.13\height}{\includegraphics[height=1.92in]{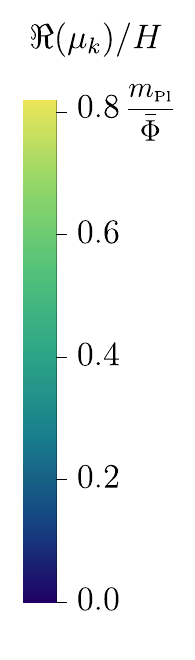}}
   \caption{The instability chart featuring the real part of the Floquet exponent normalized by the effective inflaton mass (left) and the Hubble rate (right), characterizing the $\chi$ particle production rate in the Vanilla model of preheating, $V(\phi,\chi)=\lambda\phi^4/4+g^2\phi^2\chi^2$, $g^2>0$ \cite{Greene:1997fu}. In FRW space-time $\bar{\Phi}\propto a^{-1}$ and $k\propto a^{-1}$, implying that co-moving modes do not flow across the chart as the universe expands unlike in Figs. \ref{fig:FloqPhiChi2}, \ref{fig:FloqPhi2Chi2Positive}, \ref{fig:FloqPhi2Chi2Negative}. The resonance is virtually unaffected by the expansion of space.}
   \label{fig:FloqLamePhi2Chi2}   
\end{figure*}
%~~~~~~~~~~~~~~~~~

\subsection{Scalar fields}

Historically, non-perturbative effects during preheating were first studied in the context of scalar field dynamics \cite{Dolgov:1989us,Traschen:1990sw,Kofman:1994rk,Kofman1997}. In Sections \ref{sec:ParRes} and \ref{sec:StochRes} we also used scalar fields to introduce the concept of resonant particle production. We showed that oscillations of the scalar condensate induce a time-dependence in the effective mass of the daughter scalar fields. Depending on the strength of the interactions, the occupation numbers of the decay products can grow either gradually in narrow momentum ranges for small couplings or in bursts in broad momentum ranges for large couplings, but in both regimes exponentially fast. We showed that this kind of particle production can be understood qualitatively using Floquet theory and, in particular, if the inflaton is massive -- in terms of the instability chart of Mathieu equation. In this section we just wish to point out that although it is often enough to approximate the inflaton potential during the oscillatory phase of preheating by Taylor expanding around its minimum to quadratic order, there are models with massless inflatons that do not fall into this category. A famous example is the Vanilla model, $V(\phi,\chi)=\lambda\phi^4/4+g^2\phi^2\chi^2$. The interesting thing about this model is that the Floquet analysis provides a virtually exact description of the resonant particle production \cite{Greene:1997fu}. Essentially, a certain choice of field and time redefinitions can lead to the absorption of all terms containing the scale factor, $a(t)$, and higher order derivatives of it (provided we ignore the small oscillations in $a(t)$). The expansion of space can be transformed away, converting the equations of motion into Minkowski form. In this case the background equation is strictly periodic, whereas the equation for the field fluctuations is known as the Lame equation. The instability chart of the Lame equation, see Fig. \ref{fig:FloqLamePhi2Chi2}, describes quite accurately the instabilities in the daughter fields and there is no need for the introduction of any flow lines, unlike the case of the Mathieu equation.

As we discussed in Section \ref{sec:StochRes}, stochastic resonance arises when we include the background expansion of space (apart from the Vanilla model). The effective masses of the daughter fields vary quasi-periodically with time and are modulated by powers of $a(t)$. The momentum range of unstable modes is increased, at the expense of decreasing the rate of particle production. The same phenomenon is observed if several scalar fields are oscillating at the background level. If any of the ratios of their frequencies is different from one the effective masses of the daughter fields rarely go to zero (non-adiabatic events are rare) and if the ratio is an irrational number, the motion is not periodic at all. When the number of the oscillating background fields is $\gg1$, we still get resonant particle production, see Section \ref{sec:MFpreh}. We note that it is quite natural for a large number of scalar fields to acquire VEVs during inflation in supersymmetric models.

\subsection{Fermions}

Interactions of the inflaton with fermion fields is a natural thing to consider. As already discussed at the end of Section \ref{sec:ThermIntro}, they can have important implications for the last stage of reheating. Interactions of, e.g., Yukawa form, are needed to ensure that any massive remnants of the inflaton decay into pairs of fermions and anti-fermions at late times, making it possible for the universe to become radiation-dominated. 

We should point out that while the inflaton condensate oscillates, the fermions acquire a periodically varying mass and this can lead to fermionic preheating \cite{Greene19996}. The resonance is not as efficient as in the case with daughter scalar fields, since the Pauli exclusion principle enforces the occupation number of a given mode to be $\leq1$. Nevertheless, the resonance can excite a broad range of modes, enhancing the decay rate in comparison with the standard perturbative estimate. In supergravity models, the gravitino can be non-perturbatively produced during reheating. Thermal production can take place after that as well. The danger of overproducing this massive relic can put constraints on its interactions and $T_{\rm{reh}}$ \cite{Copeland:2005qe}.

\subsection{Gauge fields}
\label{sec:IntroGaugeFields}

Similar to scalars, gauge bosons can be resonantly amplified quite efficiently during preheating. If the inflaton is a gauge singlet, it can be coupled to gauge fields through conformal factors
\Beq
\label{eq:SingletGaugeInt}
S_{\rm{matter}}\supset\int d^4x\sqrt{-g}\left[-W_1(\phi)F_{\mu\nu}F^{\mu\nu}-W_2(\phi)\epsilon^{\mu\nu\eta\sigma}F_{\mu\nu}F_{\eta\sigma}\right]\,,
\Eeq
without violating the gauge invariance of the action. The first term can lead to very efficient resonant transfer of energy to the massless gauge fields \cite{Deskins2013} during preheating. The second term violates parity and can generate chiral gravitational waves \cite{Cook:2011hg}. The second term naturally arises in models where the inflaton is an axion, e.g., in Natural inflation, with $W_2(\phi)\propto \phi$. Axions are the Goldstone bosons, appearing whenever an axial symmetry is spontaneously broken. An axion possesses an almost exact shift symmetry, so it naturally couples to total derivative terms such as $\epsilon^{\mu\nu\eta\sigma}F_{\mu\nu}F_{\eta\sigma}$ with $\epsilon^{\mu\nu\eta\sigma}$ the totally anti-symmetric tensor. Note that the mass dimension of both terms in eq. \eqref{eq:SingletGaugeInt} is $>4$ and they must be suppressed by some energy cut-off. 

If the inflaton is charged under a gauge symmetry, the covariant derivative can give rise to novel types of interaction. For instance, if the inflaton is a complex scalar, charged under an Abelian $U(1)$ symmetry, the kinetic term in the action
\Beq
S_{\rm{matter}}\supset\int\!d^4x\sqrt{-g}\,D_{\mu}\phi (D^{\mu}\phi)^*=\int\!d^4x\sqrt{-g}\left[\partial_{\mu}\phi\partial^{\mu}\phi^*+2g_{\!_A}\Im (\phi\partial_{\mu}\phi^*)A^{\mu}+g_{\!_A}^2|\phi|^2A_{\mu}A^{\mu}\right]\,,
\Eeq
yields a term that couples the complex phase of the inflaton with the gauge fields, in addition to a $g^2\phi^2\chi^2$ type of term. It turns out that the two transverse components of the spatial part of the gauge field appear only in the final term and their evolution during preheating is identical to that of $\chi$ and has been studied extensively \cite{Kari,Finelli:2000sh,Dimopoulos:2001wx,PhysRevD.63.103515}. However, the longitudinal spatial component of the gauge field, the complex phase of the inflaton and $A_0$ are all coupled through the $2g_{\!_A}\Im (\phi\partial_{\mu}\phi^*)A^{\mu}$ term. Due to the complexity of the interaction, their evolution during preheating used to be approximated or ignored \cite{Kari,Finelli:2000sh,Dimopoulos:2001wx,PhysRevD.63.103515} until very recently. Our paper \cite{Lozanov:2016pac} provided the first accurate treatment of the resonant particle production of these degrees of freedom, taking into account the redundancy introduced by the gauge freedom. It showed that all approximate treatments were insufficient for capturing the dynamics. % The rigorous analysis will be presented in Chapter \ref{ch:GaugeFields} and examples involving Abelain and non-Abelian gauge fields will be considered. 

Preheating of a $U(1)$ gauge field can be applied to the generation of the observed large scale magnetic fields \cite{Kandus:2010nw}. Conformal couplings like the ones in eq. \eqref{eq:SingletGaugeInt} can generate strong magnetic fields soon after the end of inflation \cite{Fujita:2016qab,Kobayashi:2014sga}. They can act as primordial seeds for the galactic dynamo mechanism which can amplify them to the observed values today \cite{Kandus:2010nw}. On the other hand, parametric resonance in models with an electrically charged inflaton fail to produce strong enough seed fields \cite{Finelli:2000sh,Dimopoulos:2001wx,PhysRevD.63.103515,Lozanov:2016pac}. It is worth pointing out that at the high energy scales relevant to preheating, the $U(1)$ symmetry of electromagnetism is unified with the weak force, making it necessary to consider the full electroweak gauge theory $SU(2)\times U(1)$ \cite{Dimopoulos:2001wx,Lozanov:2016pac}. % This is included in Chapter \ref{ch:GaugeFields}.

\subsection{Non-minimal couplings to gravity}
\label{sec:NonMinGrav}

The inflaton and the rest of the matter fields can have non-minimal couplings to gravity, which can become important at the high energies relevant to inflation and preheating. The simplest interaction one can consider is of the form $\xi\chi^2R$, where $\chi$ could be the inflaton or a daughter scalar field, $R$ is the Ricci scalar and $\xi$ is a dimensionless coupling constant. If it was the inflaton, than for field values $\gtrsim\mpl/\sqrt{\xi}$ the expansion of space will be affected by the interaction term, e.g., Higgs-inflation \cite{Bezrukov:2007ep}. Otherwise, in general, since $R$ oscillates during preheating, see eq. \eqref{eq:RicciOsc}, this type of interaction provides a new way for amplifying scalar field fluctuations. If $|\xi R|\sim|\xi| H^2\gtrsim$ the oscillating effective mass squared of field fluctuations (the mass induced by non-gravitational interactions), the parametric resonance could be affected. And if the gravitational interaction provides the dominant contribution to the effective mass, than it can induce resonant particle production on its own. We note that models with inflaton interaction terms that are linear in $R$ can be studied in the Einstein frame, which is related to the original frame, also known as the Jordan frame, via a conformal transformation. The conformal transformation automatically makes coupling constants time-dependent during preheating. The resonant particle production in the Einstein frame has been studied in \cite{GarciaBellido:2008ab,Bezrukov2008} for the Standard Model degrees of freedom, in \cite{DeCross:2015uza,DeCross:2016fdz,DeCross:2016cbs} for scalar field fluctuations after multi-field inflation and in \cite{Ema:2016dny} for gauge fields. 

\subsection{Non-conventional interactions}

We refer to all operators that have some cut-off scale, $\Lambda_{\rm{UV}}$, below which they are suppressed, as non-conventional. They can be important during preheating. For instance the non-minimal coupling to gravity discussed in Section \ref{sec:NonMinGrav} falls in this category and becomes negligible for field values $\ll\mpl/\sqrt{\xi}\equiv\Lambda_{\rm{UV}}$. All terms of higher than $4$ mass dimension also have a cut-off scale, e.g., the terms in eq. \eqref{eq:SingletGaugeInt}. If the inflaton is an axion, we can write $W_2(\phi)=\phi/\Lambda_{\rm{UV}}$; $\Lambda_{\rm{UV}}$ should be associated with the Peccei-Quinn scale and the gauge fields with the gluons to resolve the Strong CP problem. Another example is the case of non-canonical kinetic terms, $f(\phi/\Lambda_{\rm{UV}})\partial_{\mu}\phi\partial^{\mu}\phi$, with $\lim_{x\rightarrow 0} f(x)=1$. They automatically arise in models with non-minimal coupling to gravity, see Section \ref{sec:NonMinGrav}, after transformation to the Einstein frame. Similar patterns are observed in models with non-local interactions where the Fourier transformed kinetic terms are non-canonical $f(k/k_{\rm{UV}})|\partial_{\tau}\phi_{\bf{k}}|^2$. The longitudinal component(s) of gauge field(s) in models with a charged inflaton have kinetic terms of this form with the role of the cut off played by the Compton wavenumber of the gauge field \cite{Lozanov:2016pac}. %, as will be shown explicitly in Chapter \ref{ch:GaugeFields}.
Another situation where suppression can occur is when high-derivative interactions are present $(\partial_{\mu}\phi\partial^{\mu}\phi)^{1+n}/\Lambda_{\rm{UV}}^n$, e.g., DBI inflation \cite{Alishahiha:2004eh}.

A common feature of all non-conventional interactions is that they modify the effective mass of the field fluctuations (after canonical normalization) above the cut off scale. This typically changes the resonance structure -- it alters the shapes of the instability bands, but never degrades the efficiency of the resonant particle production. In fact non-conventional interactions can provide an alternative channel for resonant preheating.

\subsection{Miscellaneous}
\label{sec:MiscIntro}

Inflation and the subsequent stage of reheating allow for the testing of low energy models of particle physics (e.g., supersymmetric models), constrained by colliders. Even if all fields are negligibly coupled to the inflaton during inflation and reheating, they can still exhibit non-perturbative preheating dynamics or even lead to new phenomenology. Fields that have negligible interactions with the inflaton sector are known as spectator fields.

Light spectator scalar fields during inflation (having masses $\ll H$) develop an effective non-zero vacuum expectation value. Basically, the equation of motion for long-wavelength modes is overdamped during inflation. Once vacuum fluctuations cross outside the Hubble radius, they freeze and their amplitude is determined by the Hubble rate which is approximately constant, yielding a nearly scale-invariant power-spectrum
\Beq
\label{eq:ChiPSInflIntro}
\Delta_{\chi}^2\approx\left(\frac{H}{2\pi}\right)^2\bigg|_{k=aH}\,.
\Eeq
The mechanism is similar to the one which generates the curvature perturbations. In fact, the fields evolve identically to the inflaton fluctuations in the spatially-flat gauge in the slow-roll approximation, see eq. \eqref{eq:2ndOrderAction}. 

As inflation ends the preheating dynamics of the spectator field $\chi$ is reminiscent of that of the inflaton. On small scales, comparable to or shorter than the Hubble radius at that time, the field can be approximately separated into background and inhomogeneous parts, i.e., $\chi(\bf{x})=\bar{\chi}+\delta \chi(\bf{x})$, with $\bar{\chi}$ drawn from a Gaussian distribution with variance\footnote{Ignoring any scale-dependence in the inflationary power-spectrum due to departures from perfect de Sitter.} $\Delta_{\chi}^2$, $\delta \chi(\bf{x})$ being vacuum sub-horizon fluctuations and $|{\bf{x}}|<H^{-1}$. This is known as the separate universe approach. The homogeneous value of the spectator, $\bar{\chi}$, does not evolve until the Hubble rate becomes smaller than its effective mass. Then it starts to oscillate about the bottom of its potential.

If $\chi$ is coupled to other fields, e.g., if it is the Standard Model Higgs, which is coupled to the charged leptons, $W$ and $Z$ bosons, its oscillations can lead to resonant particle production \cite{Kari}, followed by a non-linear period \cite{Figueroa2015,Enqvist:2015sua}, generating a stochastic gravitational wave background \cite{Figueroa:2013vif,Figueroa:2016ojl}. More generally, a complex $\bar{\chi}$, embedded within, e.g., a supersymmetric model, can lead to baryogenesis, according to the Affleck-Dine mechanism \cite{Affleck:1984fy}, and the non-linear dynamics following the resonant stage can involve the formation of Q-balls \cite{Enqvist:1997si,Enqvist:1999mv}.

If $\bar{\chi}$ eventually comes to dominate the energy budget of the universe (e.g., if it oscillates about a quadratic minimum with its energy being redshifted as $\sim a^{-3}$), it has to decay into radiation to be in agreement with the big-bang nucleosynthesis scenario. If we assume that $\bar{\chi}\neq0$ on cosmological length scales, than the radiation will have the inhomogeneities of the field imprinted on it. This is the essence of the curvaton scenario \cite{PhysRevD.42.313,Linde:1996gt,Lyth20025}. In it, the final primordial density fluctuations are generated after inflation and depend on the physics during reheating. Observations of the CMB give a constraint on the combination of the initial fluctuations from inflation and the post-inflationary ones coming from the decay of the curvaton field, $\chi$.

Another possibility for a light spectator, $\chi$, to lead to the generation of primordial curvature perturbations after inflation is if it affects the decay rate, $\Gamma$, of the inflaton into other fields. The fluctuations in $\chi$ will lead to a spatial variation in $\Gamma$. Hence, the final primordial curvature perturbation imprinted on the decay products is a consequence of the spatially varying couplings and any initial fluctuations generated during inflation. This mechanism is known as modulated reheating \cite{Kofman:2003nx} %, and its observational aspects and those of the curvaton scenario will be discussed in more detail in the next section. We just note that modulated reheating was first discussed in the context of Superstring theory models \cite{Kofman:2003nx}.
and was first discussed in the context of Superstring theory models.

\newpage

%-----------------------------------------------------------------------------------------------------------------------------
%				Observational implications and signatures of reheating
%-----------------------------------------------------------------------------------------------------------------------------
\section{Observational implications and signatures of reheating}
\label{sec:ObsImplReh}

\hfill\begin{minipage}{\dimexpr\textwidth-3.7cm}
`{\it The recent developments in cosmology strongly suggest that the universe may be the ultimate free lunch.}'
%\xdef\tpd{\the\prevdepth}
\end{minipage}

\hfill {\it Alan Guth}

\bigskip

Despite being a very important and phenomenologically rich period, reheating and the high-energy physics laws governing it are hard to constrain observationally. Just like inflation and all other epochs preceding recombination, reheating cannot be observed directly, since it is hidden by the opaque thermal baryonic plasma. Similarly to the case of inflation, one should look for observational signatures of reheating that survive thermalization and could be inferred from various cosmological measurements. Inflation predicts the stretch of microscopic quantum fluctuations to super Hubble scales, generating a superhorizon curvature perturbation, which is conserved and eventually imprinted on the CMB. Unfortunately, during the decelerating phase of reheating, co-moving modes re-enter the horizon and only the sub-horizon scales are affected by the non-linear dynamics. The length-scales on which the curvature perturbation is affected are so short, that the change is completely concealed by the later non-linear evolution of cosmic structure, making it impossible (for now) to be inferred from the CMB. Another reason why reheating is difficult to connect with observations is that by the time of BBN at the latest, all Standard Model species must be thermalized, hiding away the details of the earlier stages when they were produced. 

Still, reheating can yield signatures, potentially observable in the future. These include the generation of relics and metric perturbations, which could be observed directly. In effect, the early universe takes the role of an accelerator for poor people, allowing us to probe roughly, yet freely, the fundamental physics at otherwise virtually inaccessible energy scales. 

Indirect signatures are also possible. For instance, the mapping of co-moving modes between horizon exit during inflation and re-entry at later times depends on the entire expansion history between the two events. Thus, the confirmation of a particular model of inflation can give us information about reheating, e.g., constrain its expansion history. Or vice versa, a better understanding of reheating can reduce the uncertainties in predictions of simple inflationary models. 

In the rest of this section we discuss different observational implications of reheating, starting with the indirect expansion history effect.

\subsection{Expansion history of reheating and the CMB}
\label{sec:ExpHistIntro}

The expansion history of reheating is largely uncertain. We only know that between the end of inflation and the time the universe thermalizated completely (i.e., achieved chemical and local thermal equilibrium), the mean equation of state is $\int_{t_{\rm{end}}}^{t_{\rm{th}}} dt\, w(t)/(t_{\rm{th}}-t_{\rm{end}})\equiv\bar{w}_{\rm{int}}>-1/3$. This implies various possibilities for $N_{\star}$ -- the number of {\it e}-folds of expansion before the end of inflation, when the pivot scale crossed outside the Hubble radius, $k_{\star}=a_{\star}H_{\star}$. $N_{\star}$ can take a range of different values, depending on the expansion history of reheating. Hence, the predictions of any model of inflation have an inherent uncertainty, due to the poorly constraint period of reheating. Before discussing inflationary observables in detail we consider the uncertainties in $N_{\star}$. Given a co-moving pivot scale that has re-entered the horizon at late times, having some fixed physical wavenumber today, say $k_{\star,\rm{phys}0}=k_{\star}/a_0$, we discuss how $N_{\star}$ depends on the details of the inflaton potential, $\bar{w}_{\rm{int}}$, $\rho_{\rm{th}}$ and $g_{\rm{th}}$ -- the last two being the energy density and the number of relativistic degrees of freedom at thermalization. 

We start with the free parameters
\Beq
N_{\star}\,,\,\{q_i\}\,,
\Eeq
where %$n>i>1$ and 
$\{q_i\}$ are the parameters entering the inflaton potential, i.e., $V=V(\{q_i\},\phi)$. By definition
\Beq
\label{eq:IntroNstPhist}
N_{\star}\equiv\ln\left(\frac{a_{\rm{end}}}{a_{\star}}\right)=\int_{a_{\star}}^{a_{\rm{end}}}d\ln a=\left|\int_{\phi_{\star}}^{\phi_{\rm{end}}}d\phi\frac{H}{\dot{\phi}}\right|\approx\left|\int_{\phi_{\star}}^{\phi_{\rm{end}}}\frac{d\phi}{\mpl}\frac{1}{\sqrt{2\epsilon_V}}\right|\,,
\Eeq
where the last expression follows from eq. \eqref{eq:SlowRollPotential} and the discussion above it. The value of the inflaton at the end of slow-roll inflation, $\phi_{\rm{end}}$, is to a very good approximation insensitive to the initial conditions and the inflationary dynamics, implying $\phi_{\rm{end}}=\phi_{\rm{end}}(\{q_i\})$.\footnote{In fact, the end of inflation, $\ddot{a}=0$, is near $\epsilon_V=1$, implying that $\phi_{\rm{end}}\approx\phi_{\rm{end}}(\{q_i-1\})$, i.e., the number of parameters is reduced by one due to the cancellation inside the squared brackets in eq. \eqref{eq:SlowRollPotential}.} Hence, $\phi_{\star}=\phi_{\star}(N_{\star},\phi_{\rm{end}},\{q_i-1\})=\phi_{\star}(N_{\star},\{q_i\})$. This and eq. \eqref{eq:SlowRollInflatonPot} imply that the measured magnitude of curvature perturbation $A_{\rm{s}}=A_{\rm{s}}(\{q_i\},\phi_{\star})=A_{\rm{s}}(\{q_i\},N_{\star})$, from where we can determine one of the potential parameters, e.g., $q_1=q_1(\{q_i-1\},N_{\star},A_{\rm{s}})$. Given all that we can write
\Beq
\label{eq:VsNs}
V_{\star}=V_{\star}(\{q_i\},\phi_{\star})=V_{\star}(\{q_i-1\},N_{\star},A_{\rm{s}})\,.
\Eeq
Note that for a single-parameter model, e.g., $V=m^2\phi^2/2$, this implies a one-to-one correspondence between $V_{\star}$ and $N_{\star}$ (and likewise for derivatives of $V_{\star}$), since $A_{\rm{s}}$ is known from observations. For a two-parameter model of inflation, e.g., $V=\Lambda^4\tanh^2(\phi/M)$, there is a one-parameter set of solutions, etc.

%~~~~~~~~~~~~~~~~~
\begin{figure}[t] %  figure placement: here, top, bottom, or page
   \centering
   \hspace{0.10in}   
   \includegraphics[width=4.292in]{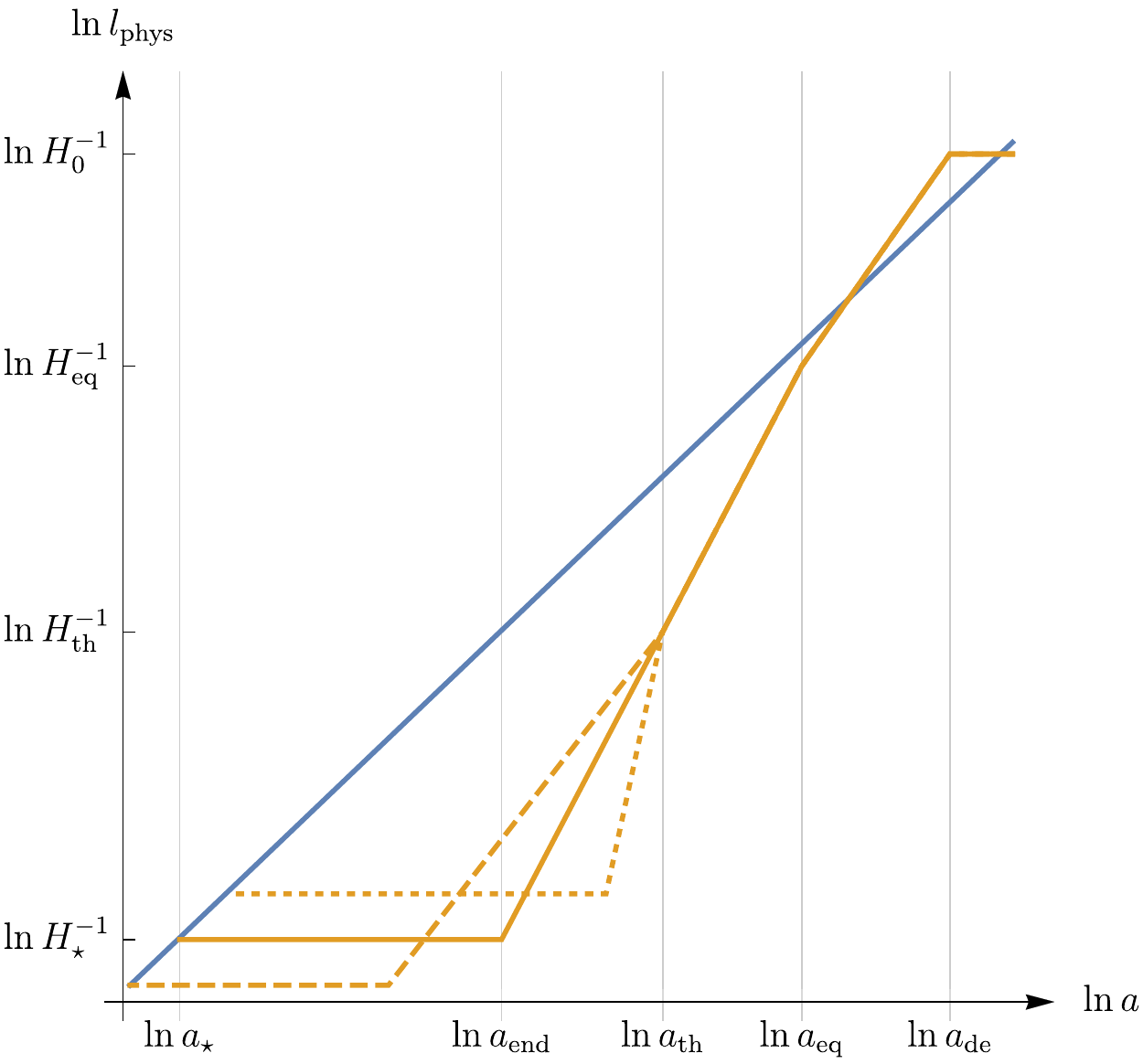} 
   \hspace{-0.00in}   
   \raisebox{2.2\height}{\includegraphics[width=1.295in]{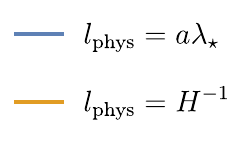}} 
   \caption{The figure illustrates how the uncertainty in the expansion history of reheating is translated on $N_{\star}=\ln (a_{\rm{end}}/a_{\rm{\star}})$ -- the number of {\it e}-folds of expansion before the end of inflation when the pivot scale crossed outside the Hubble radius, $H^{-1}$. For simplicity we fix $\rho_{\rm{th}}$ and $a_{\rm{th}}$, and vary only the mean equation of state of reheating, $w=\bar{w}_{\rm{int}}$ (reheating takes place while $a_{\rm{th}}>a>a_{\rm{end}}$). We consider single-parameter models of inflation, e.g., $V=m^2\phi^2/2$, for which $N_{\star}=N_{\star}(\bar{w}_{\rm{int}})$ and $V_{\star}=V_{\star}(N_{\star}(\bar{w}_{\rm{int}}))=V_{\star}(\bar{w}_{\rm{int}})$. This implies that choosing an energy scale of inflation, $V_{\star}$, uniquely determines $N_{\star}$ and $\bar{w}_{\rm{int}}$, unlike in multi-parameter models of inflation where we have additional degrees of freedom, see Fig. \ref{fig:ExpHistIntro2}. For plotting purposes, we have approximated the inflationary and the dark energy stages as de Sitter expansions, and the two periods preceding and following radiation-matter equality as radiation and matter dominated, respectively. None of the conclusions depend on these simplifications. Note that $\epsilon_H\equiv d(\ln H^{-1})/d(\ln a) =3(1+w)/2$.}
   \label{fig:ExpHistIntro}
\end{figure}
%~~~~~~~~~~~~~~~~~

To find $N_{\star}$ and $V_{\star}$ (given $\{q_i-1\}$) we need to match the pivot scale today to the time it left the horizon, i.e., $k_{\star}/(a_0H_0)=(a_{\star}H_{\star})/(a_0H_0)$. Taking the log of both sides of this equality, one can show that
\Beq
\label{eq:NsVs}
%N_{\star}=66.89-\ln \frac{k_{\star}}{a_0H_0}+\frac{1}{4}\ln\frac{V_{\star}^2}{\mpl^4\rho_{\rm{end}}}-\frac{1}{12}\ln g_{\rm{th}}+\frac{1-3\bar{w}_{\rm{int}}}{12(1+\bar{w}_{\rm{int}})}\ln \frac{\rho_{\rm{th}}}{\rho_{\rm{end}}}\,,
N_{\star}=66.89-\ln \frac{k_{\star}}{a_0H_0}+\frac{1}{4}\ln\frac{V_{\star}^2}{\mpl^4\rho_{\rm{end}}}+\frac{1}{12}\ln \left[\frac{1}{g_{\rm{th}}}\left(\frac{\rho_{\rm{th}}}{\rho_{\rm{end}}}\right)^{\frac{1-3\bar{w}_{\rm{int}}}{1+\bar{w}_{\rm{int}}}}\right]\,,
\Eeq
making the only assumption that entropy, $s\sim g T^3$, is conserved, $sa^3=\rm{const}$, after thermalization, $a>a_{\rm{th}}$. The derivation is given in, e.g., \cite{Lozanov:2017hjm}. Since $\rho_{\rm{end}}=\rho_{\rm{end}}(\{q_i\})$, after substituting $V_{\star}$ from eq. \eqref{eq:VsNs} into eq. \eqref{eq:NsVs}, we find that
\Beq
\label{eq:NstarReh}
N_{\star}=N_{\star}\left(\{q_i-1\},A_{\rm{s}},\rho_{\rm{th}}^{\frac{1-3\bar{w}_{\rm{int}}}{1+\bar{w}_{\rm{int}}}}/g_{\rm{th}}\right)\,.
\Eeq
Reheating affects $N_{\star}$ only through the specific combination of quantities appearing in the last argument. Note that the $g_{\rm{th}}$ dependence contains the information about the time of thermaliztion, $a_{\rm{th}}$. Essentially, the conservation of entropy implies that $g_{\rm{th}}=((\pi^2/30)g_0^{4/3}T_0^4/\rho_{\rm{th}})^{3}(a_{\rm{th}}/a_{0})^{12}$, where $T_0=2.725\,\rm{K}$ and the effective number of relativistic degrees of freedom in entropy is $g_0=43/11$. In other words, $N_{\star}$ depends on a combination of the energy scale, $\rho_{\rm{th}}$, and the time, $a_{\rm{th}}$, of thermalization, as well as the mean equation of state, $\bar{w}_{\rm{int}}$, of reheating (holding for $a_{\rm{th}}>a>a_{\rm{end}}$). We depict this effect in Fig. \ref{fig:ExpHistIntro} for a single-parameter model of inflation, e.g., $V(m,\phi)=m^2\phi^2/2$. For simplicity, we fix $\rho_{\rm{th}}$ and $a_{\rm{th}}$ (fixing these two quantities fixes $g_{\rm{th}}$) and vary only $\bar{w}_{\rm{int}}$. We plot the Hubble radius, $H^{-1}$, in orange and the physical wavenumber corresponding to the pivot scale, $a\lambda_{\star}$, in blue. Both have some fixed values today, $H^{-1}_0$ and $a_0\lambda_{\star}$, respectively. For plotting purposes, we approximate the dark energy dominated universe today and inflation as stages of de Sitter expansion, i.e, for $a>a_{\rm{de}}$ and $a<a_{\rm{end}}$, $H=H_0=\rm{const}$ and $H=H_{\star}=\sqrt{V_{\star}/3}/\mpl=\rm{const}$, respectively. We also assume $w=0$ between radiation-matter equality and dark energy domination, $a_{\rm{de}}>a>a_{\rm{eq}}$, and $w=1/3$ between thermalization and radiation-matter equality $a_{\rm{eq}}>a>a_{\rm{th}}$. Note that for the single-parameter $m^2\phi^2/2$ inflation, $\{q_i-1\}\in\varnothing$. Given that $A_{\rm{s}}=2.2\times10^{-9}$, eq. \eqref{eq:NstarReh} implies $N_{\star}=N_{\star}(\bar{w}_{\rm{int}})$. This is shown in the figure with the three orange solid, dashed and dotted lines corresponding to three different choices of $\bar{w}_{\rm{int}}$. Note that we also have $V_{\star}=V_{\star}(N_{\star}(\bar{w}_{\rm{int}}))=V_{\star}(\bar{w}_{\rm{int}})$, see eq. \eqref{eq:VsNs}. Hence, in single-parameter models of inflation, $\bar{w}_{\rm{int}}$ uniquely defines $V_{\star}$ and $N_{\star}$ (provided $\rho_{\rm{th}}$ and $a_{\rm{th}}$ are fixed). This is a peculiar feature of single-parameter models. In multi-parameter models of inflaton there is a degeneracy, as we discuss below. Note that even in the single-parameter models of inflation, the uncertainty in the equation of state of reheating translates into an uncertainty in the energy scale of inflation. This could be turned the other way round -- a possible confirmation of a single-parameter model of inflation with a given $V_{\star}$ uniquely determines $N_{\star}$ and hence $\bar{w}_{\rm{int}}$ if $\rho_{\rm{th}}$ and  $a_{\rm{th}}$ are known (if they are not, it at least uniquely determines the combination of $\rho_{\rm{th}}$, $a_{\rm{th}}$ and $\bar{w}_{\rm{int}}$ on which $N_{\star}$ depends). So pinning down the model of inflation could give us information about reheating.

%~~~~~~~~~~~~~~~~~
\begin{figure}[t] %  figure placement: here, top, bottom, or page
   \centering
   \hspace{0.10in}   
   \includegraphics[width=4.292in]{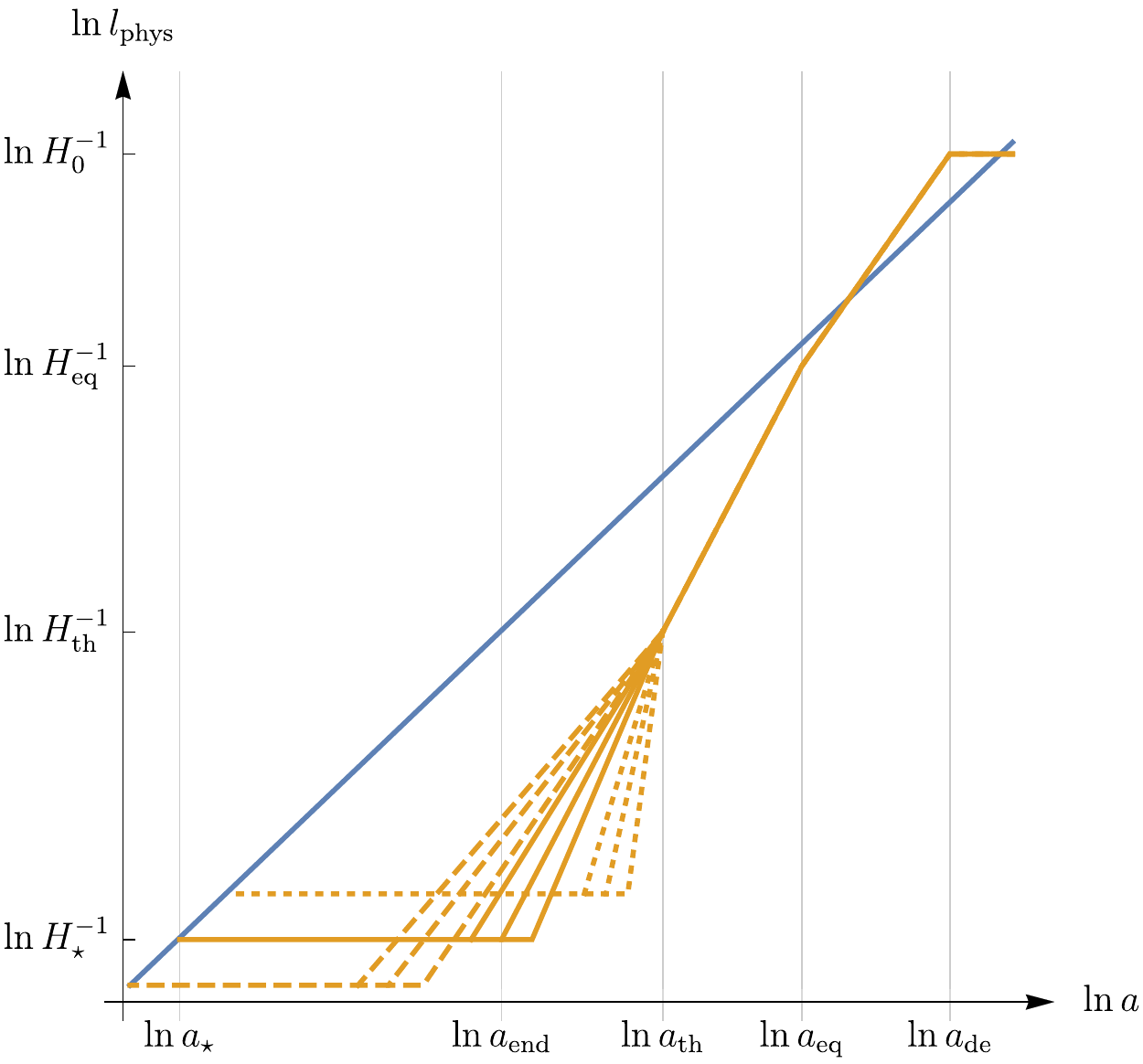} 
   \hspace{-0.00in}   
   \raisebox{2.2\height}{\includegraphics[width=1.295in]{ExpHistLeg.pdf}} 
   \caption{Same as Fig. \ref{fig:ExpHistIntro}, but for two-parameter models of inflation, e.g., $V=\Lambda\tanh^2(\phi/M)$. The extra parameter in the inflaton potential introduces an additional degree of freedom, $N_{\star}=N_{\star}(q_1,\bar{w}_{\rm{int}})$ and $V_{\star}=V_{\star}(q_1,N_{\star}(q_1,\bar{w}_{\rm{int}}))=V_{\star}(q_1,\bar{w}_{\rm{int}})$. This means that unlike in the single-parameter models of inflation, choosing an energy scale of inflation, $V_{\star}$, allows for a range of $N_{\star}$ and $\bar{w}_{\rm{int}}$.}
   \label{fig:ExpHistIntro2}
\end{figure}
%~~~~~~~~~~~~~~~~~

We now repeat the analysis for a two-parameter model of inflation, e.g., $V(M,\Lambda,\phi)=\Lambda^{4}\tanh^2(\phi/M)$. This time, we have one additional degree of freedom, i.e., $\{q_i-1\}=q_1$. For fixed $\rho_{\rm{th}}$ and $a_{\rm{th}}$ this implies $N_{\star}=N_{\star}(q_1,\bar{w}_{\rm{int}})$ and $V_{\star}=V_{\star}(q_1,N_{\star}(q_1,\bar{w}_{\rm{int}}))=V_{\star}(q_1,\bar{w}_{\rm{int}})$. The dependence is depicted in Fig. \ref{fig:ExpHistIntro2}. As mentioned above, unlike the single-parameter models, multi-parameter models possess additional degeneracy due to the extra parameters. It explains the additional lines in Fig. \ref{fig:ExpHistIntro2}. Essentially, there is not a one-to-one correspondence between $V_{\star}$ and $N_{\star}$, i.e., a particular $V_{\star}$ gives a range of $N_{\star}$. Note that in two-parameter models, having fixed $V_{\star}$ and $N_{\star}$, uniquely determines $\bar{w}_{\rm{int}}$. Conversely, one has to measure or calculate $V_{\star}$ and $\bar{w}_{\rm{int}}$ separately, to uniquely determine $N_{\star}$ in two-parameter models of inflation.

%~~~~~~~~~~~~~~~~~
\begin{figure}[t] %  figure placement: here, top, bottom, or page
   \centering
   \hspace{0.10in}   
   \includegraphics[width=4.25in]{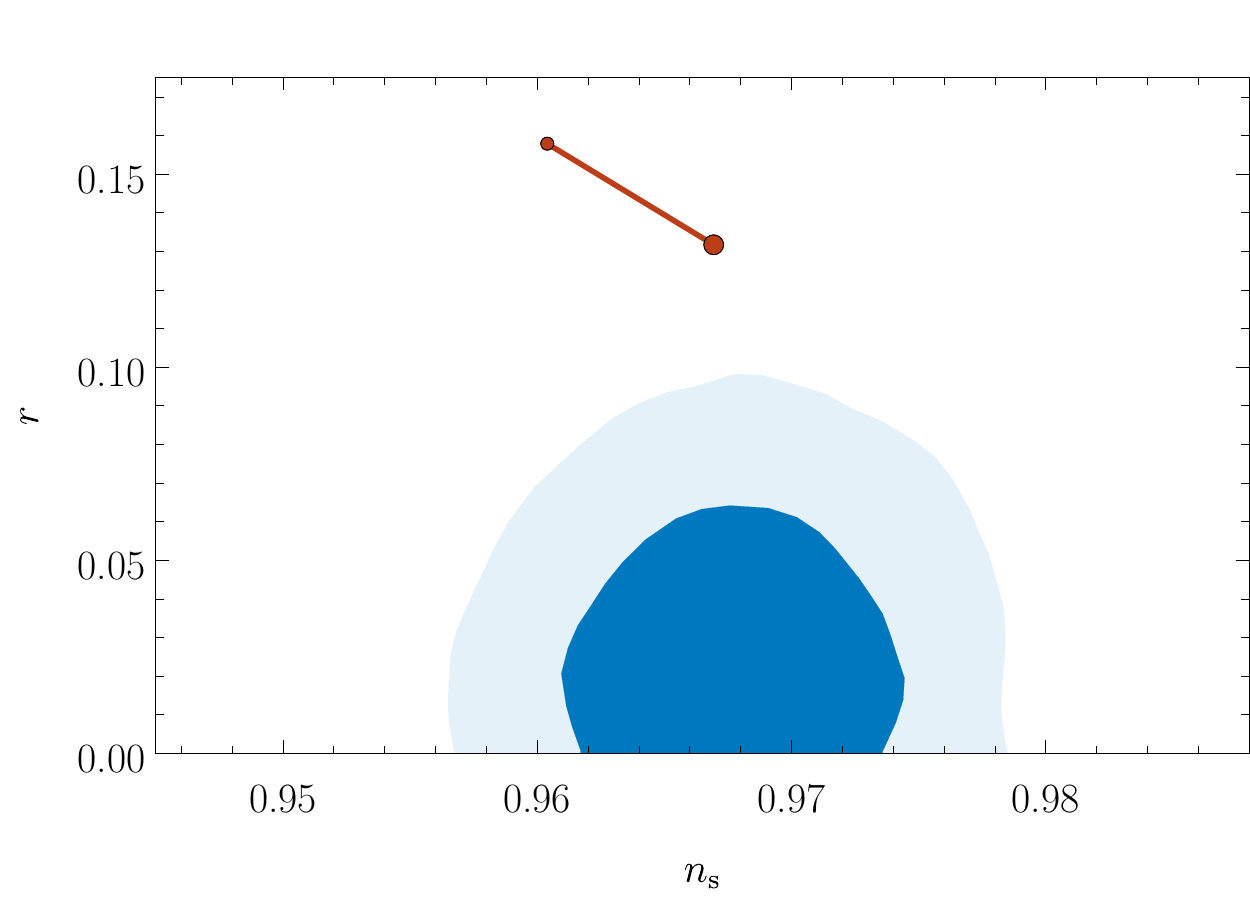} 
   \hspace{-1.00in}   
   \raisebox{3.05\height}{\includegraphics[width=0.9in]{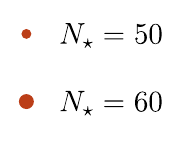}}  
   \caption{Red: the predictions for $n_{\rm{s}}$ and $r$ in $V=m^2\phi^2/2$. Since this is a single-parameter model of inflation, $n_{\rm{s}}=n_{\rm{s}}(N_{\star})$ and $r=r(N_{\star})$, hence the bar-like prediction for $50<N_{\star}<60$. Here the inflaton mass is not a free parameter; instead $m=m(N_{\star},A_{\rm{s}})$. Light and dark blue: the $68$ \% and $95$ \% confidence level regions for $n_{\rm{s}}$ and $r$ from Planck and other data sets \cite{Ade:2015lrj}.}
   \label{fig:nsrsingleparameter}
\end{figure}
%~~~~~~~~~~~~~~~~~

\begin{figure}[t] %  figure placement: here, top, bottom, or page
   \centering
   \hspace{0.10in}   
   \includegraphics[width=4.25in]{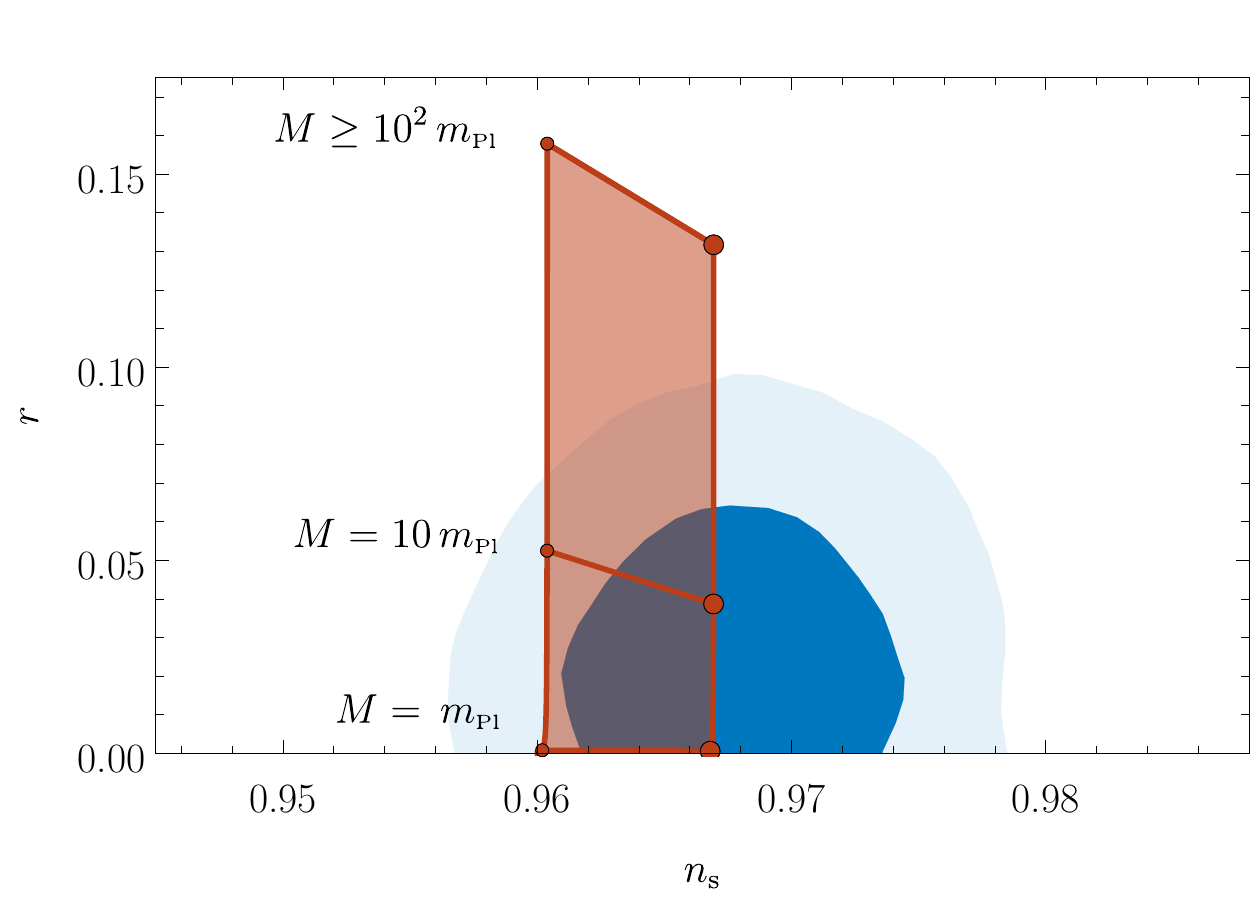} 
   \hspace{-1.00in}   
   \raisebox{3.05\height}{\includegraphics[width=0.9in]{PlanckLeg.pdf}} 
   \caption{Same as Fig. \ref{fig:nsrsingleparameter}, but the predictions for $n_{\rm{s}}$ and $r$ are for $V=\Lambda^4\tanh^2(\phi/M)$. Since this is a double-parameter model of inflation, we have an additional degree of freedom in comparison with single-parameter models, see Fig. \ref{fig:nsrsingleparameter}. This means we can write $n_{\rm{s}}=n_{\rm{s}}(M,N_{\star})$ and $r=r(M,N_{\star})$, with $M$ being a free parameter and $\Lambda=\Lambda(N_{\star},A_{\rm{s}},M)$. Hence, the wide band prediction for $50<N_{\star}<60$. }
   \label{fig:nsrdoubleparameter}
\end{figure}

Hence, the uncertainty in the expansion history of reheating leads to an uncertainty in $N_{\star}$. This has consequences for spectral observables such as $n_{\rm{s}}$ and $r$, see eq. \eqref{eq:SlowRollInflatonPot}. The slow-roll potential parameters at the time of horizon exit are $\epsilon_{V\star}=\epsilon_{V\star}(\phi_{\star},\{q_i-1\})$ and $\eta_{V\star}=\eta_{V\star}(\phi_{\star},\{q_i-1\})$. The reason for having one fewer $q_i$ is because we take ratios of the inflaton potential and corresponding derivatives in eq. \eqref{eq:SlowRollPotential}. Substitution for $\phi_{\star}=\phi_{\star}(N_{\star},\{q_i\})$ implies $\epsilon_{V\star}=\epsilon_{V\star}(N_{\star},\{q_i\})$ and $\eta_{V\star}=\eta_{V\star}(N_{\star},\{q_i\})$. However, the measured magnitude of curvature perturbation reduces the number of free parameters by one, e.g., $q_1=q_1(\{q_i-1\},N_{\star},A_{\rm{s}})$. The first two expressions in eq. \eqref{eq:SlowRollInflatonPot} then imply
\Beq
\label{eq:nsstrstInro}
n_{\rm{s}}=n_{\rm{s}}(N_{\star},\{q_i-1\})\,,\qquad r=r(N_{\star},\{q_i-1\})\,.
\Eeq
In single-parameter models of inflation, $n_{\rm{s}}$ and $r$ are only functions of $N_{\star}$. The reheating related uncertainty in $N_{\star}$ translates into a bar in the $n_{\rm{s}}$-$r$ plane, as shown in Fig. \ref{fig:nsrsingleparameter} for $m^2\phi^2/2$ inflation. In two-parameter models of inflation, $n_{\rm{s}}$ and $r$ are functions of $N_{\star}$ and an additional degree of freedom. This transforms the bars in the $n_{\rm{s}}$-$r$ plane into bands as shown in Fig. \ref{fig:nsrdoubleparameter} for $\Lambda^{4}\tanh^2(\phi/M)$ inflation. 

%~~~~~~~~~~~~~~~~~
%~~~~~~~~~~~~~~~~~

\subsection{Relics}

The out-of-equilibrium dynamics of reheating may lead to the generation of various relics. These include stable ones, e.g., topological defects, that have not been observed yet, tantalizing ones, e.g., dark matter and primordial magnetic fields, for which there is some experimental evidence, and observed ones, such as the baryon asymmetry. In the following we briefly talk about each of these applications of reheating.

\subsubsection{Baryon asymmetry}
\label{sec:BaryogenesisIntro}

Models of baryogenesis try to explain the observed baryon-to-photon ratio
\Beq
\eta\equiv \frac{n_{b}}{n_{\gamma}}\approx6\times10^{-10}\,,
\Eeq
where $n_{b}=n_{B}-n_{\bar{B}}$ and $n_{\gamma}$ are the (net) number densities of baryons and photons, respectively. There are many high-energy physics models that explain the value of $\eta$ with some dynamical mechanism, none of which is singled out by observational tests. It is indeed very interesting to try to connect baryogenesis to reheating. We should point out that since the net number of baryons in the late universe, certainly after BBN, is conserved, then $n_{b}\propto a^{-3}$. Furthermore, since after the epoch of electron-positron annihilation, which also happened around BBN, $n_{\gamma}\propto a^{-3}$, this means that $\eta=\rm{const}$. At earlier times $\eta$ was still conserved, apart from the moments when a relativistic particle species went out of thermal equilibrium and became non-relativistic -- then it changed in a step-like manner by a factor that is determined by the ratio of the old and new number of relativistic degrees of freedom. However, what remained always constant in thermal equilibrium was the ratio $n_{b}/s$ which is $\sim\eta_0$. The Standard Model of particle physics ensures the net conservation of baryons and cannot explain why there are so many photons (or so much entropy) per baryon today, assuming that the universe started in a natural state with no baryons $n_{B}=n_{\bar{B}}=0$, or with no net baron number, $n_{b}=0$. To explain the puzzling observed asymmetry in the amount of matter and antimatter, one has to invoke physics beyond the Standard Model. It must allow for physical processes that meet the following three criteria, known as Sakharov's conditions \cite{Sakharov:1967dj}: (i) non-conservation of the baryon number, $b$; (ii) violation of $C$ and $CP$ invariance; (iii) departure from thermal equilibrium. While condition (i) is obvious -- to generate a net baryon number starting from $n_{b}=0$ we need reactions that violate baryon number conservation -- it is not enough. Condition (ii) is necessary to ensure different decay rates into baryons and anti-baryons. Consider a baryon number violating reaction $X\rightarrow Y+Z$. The violation of $C$ invariance ensures that the rates of the reaction and its charge conjugated counterpart are different
\Beq
\Gamma(X\rightarrow Y+Z)\neq\Gamma(\bar{X}\rightarrow \bar{Y}+\bar{Z})\,,
\Eeq
i.e., the $b$ violating process that creates more baryons than anti-baryons is not counterbalanced by its conjugate that creates more anti-baryons than baryons. Taking into account the helicity of the baryons, e.g., a reaction of the form $X\rightarrow q_L+q_L$, $CP$ violation ensures that $\Gamma(X\rightarrow q_L+q_L)\neq\Gamma(\bar{X}\rightarrow \bar{q}_R+\bar{q}_R)$ and $\Gamma(X\rightarrow q_R+q_R)\neq\Gamma(\bar{X}\rightarrow \bar{q}_L+\bar{q}_L)$. Otherwise, the amount of produced left-handed baryons equals the amount of produced right-handed anti-baryons and vice versa, implying that the net baryon number does not change
\Beq
\Gamma(X\rightarrow q_L+q_L)+\Gamma(X\rightarrow q_R+q_R)=\Gamma(\bar{X}\rightarrow \bar{q}_L+\bar{q}_L)+\Gamma(\bar{X}\rightarrow \bar{q}_R+\bar{q}_R)\,,
\Eeq
despite $C$ invariance being violated and (i). The reason for condition (iii) is slightly less obvious. It comes from the fact that the equilibrium number densities of particles and anti-particles depend on their chemical potentials. In thermal equilibrium $\mu_{q_L}=-\mu_{\bar{q}_L}$, etc (recall that baryons and anti-baryons can annihilate into photons). However, since the baryon number is not conserved by the interactions, $\mu_{q_L}=\mu_{\bar{q}_L}=0$. Hence, particles and anti-particles in thermal equilibrium have equal number densities despite conditions (i) and (ii).

Many high-energy physics models that satisfy the three criteria have been put forward to explain the observed baryon asymmetry. GUT-scale baryogenesis models rely on superheavy particles (with GUT-scale masses) decaying into baryons through $C$ and $CP$ violating reactions. As discussed in Section \ref{sec:InstPreh}, such superheavy particles can be produced non-perturbatively after inflation, out of thermal equilibrium, making (p)reheating the ideal setting for these models. Leptogenesis, the generation of a number asymmetry, $n_{l}=n_{L}-n_{\bar{L}}$, between leptons and antileptons, can also account for the observed baryon-to-entropy ratio. The lepton number, $l$, can be converted into the baryon number via Standard Model sphalerons -- transitions between degenerate topologically different $SU(2)$ electroweak gauge field configurations -- they become suppressed at temperatures $<300\,\rm{GeV}$. Essentially, for sphaleron transitions (you can think of them as reactions) $b-l=\rm{const}$, but $b\neq\rm{const}$ and $l\neq\rm{const}$. Thus, if one starts with $n_b=0$, by the end of the transition it is converted into $n_{b}\sim n_{b-l}$. This is an example of electroweak baryogenesis %. The non-thermal particle productions in preheating after low scale inflation (not too far from the electroweak scale) can enhance electroweak baryogenesis by delaying thermalization, when leptogenesis is inefficient. 
and preheating can provide a way for generating the initial lepton asymmetry. Another possibility is the Affleck-Dine baryogenesis mechanism. It involves a complex scalar field, $X$, that carries a baryon number and whose non-equilibrium dynamics does not conserve $b$. The field normally starts in a spatially homogeneous configuration in which invariance under $C$ and $CP$ may or may not be spontaneously broken. Then it evolves into a non-thermal configuration, in which the $C$ and $CP$ symmetries are spontaneously broken, with a final non-zero $b$, which is eventually converted into baryons. A version of this model where the inflaton plays the role of the scalar field is the subject of \cite{LozAmin}. There we show that the non-linear dynamics of reheating can play an important role for the prediction of $\eta$.

\subsubsection{Magnetic fields}
\label{sec:MagFieldsIntro}

Magnetic fields are abundant in our universe \cite{Kandus:2010nw}. They have been observed in galaxies $B_{10-10^2\,\rm{kpc}}\sim10^{-5}\,\rm{G}$ and galaxy clusters $B_{0.1-1\,\rm{Mpc}}\sim10^{-6}\,\rm{G}$. There is a (conservative) lower bound on the strength of magnetic fields with cosmic scale correlation lengths $B_{>1\,\rm{Mpc}}>10^{-17}\,\rm{G}$. While galactic fields can be accounted for by the amplification of seed fields via the dynamo mechanism \cite{Durrer:2013pga}, the origin of those seeds, as well as the large-correlation-length fields that are unaffected by magnetohydrodynamic processes remains an open problem. It can be explained by a primordial magnetic field component. CMB observations have put upper bounds on it $B_{1\,\rm{Mpc}}^{\rm{prim}}<10^{-9}\,\rm{G}$ \cite{Ade:2015cva}, whereas the seed amplitude needed for the dynamo mechanism is model and scale-dependent.

It is difficult to connect the causal non-linear dynamics of reheating with the large scale magnetic fields. However, the linear stage of preheating can provide the perfect setting for magnetogenesis. Low momentum magnetic field modes can be resonantly amplified \cite{Finelli:2000sh} or undergo tachyonic instabilities \cite{Fujita:2016qab,Kobayashi:2014sga}. The biggest challenge is to avoid back-reaction of small-scale modes before low-momentum modes have been sufficiently amplified \cite{Lozanov:2016pac,Adshead:2016iae}. Tachyonic instability can be achieved quite easily. A conformal coupling of the form $\mathcal{L}_{\rm{Maxwell}}=-f(\tau)F_{\mu\nu}F^{\mu\nu}/4$ yields
\Beq
\mathcal{A}^{T}_{k}{}''+\left(k^2-\frac{f''}{f}\right)\mathcal{A}^{T}_{k}(\tau)=0\,,
\Eeq
where $\mathcal{A}^{T}_{k}(\tau)=a(\tau)f(\tau)A^{T}_{k}(\tau)$ are the canonically-normalized transverse (Fourier) modes. $f$ tends to $1$ at late times, but if it is $\propto \tau^{\alpha}$ earlier on, certain choices of $\alpha$ and the magnitude of $f$ could lead to a successful magnetogenesis via tachyonic preheating \cite{Markkanen:2017kmy}.

\subsubsection{Miscellaneous}

In many dark matter models, the relic abundance is determined by the self-interactions of a thermalized dark matter sector. After the inflaton resonantly excites the Standard Model and dark matter degrees of freedom during preheating, the two sectors can attain different equilibrium temperatures -- a phenomenon known as asymmetric reheating, which can be sensitive to the non-linear dynamics of reheating \cite{Hardy:2017wkr}. A detection of a temperature difference can put constraints on the inflaton mass and couplings \cite{Adshead:2016xxj}.

The non-linear dynamics of reheating can lead to the formation of stable topological defects, see Section \ref{sec:NonLinIntro}, for which there are no observational evidence \cite{Ade:2015xua}. Overproduction of such defects could overclose the universe or affect CMB anisotropies. In fact, CMB measurements provide the tightest constraints \cite{Ade:2015xua}.

\subsection{Metric fluctuations}
\label{sec:MetrFluc}

Departures from the FRW universe described by matter and metric perturbations are at the heart of modern cosmology. Within current observational limits, an adiabatic curvature perturbation (a scalar mode) with Gaussian statistics can explain the measured CMB temperature anisotropies \cite{Ade:2015xua}. Furthermore, the detection of polarization $B$-modes generated by primordial tensor fluctuations in the metric (gravitational waves) is one of the main goals of the upcoming Stage-4 CMB experiments \cite{Abazajian:2016yjj}. The linear and non-linear stages of reheating can give rise to gravitational waves, as well as entropic and non-Gaussian contributions to the curvature perturbation. Their non-detection constrains different reheating scenarios, as we discuss in the remainder of this section.

\subsubsection{Gravitational waves}

\begin{figure}[t] %  figure placement: here, top, bottom, or page
   \centering
\vspace{-0.75in}
   \includegraphics[width=4.75in]{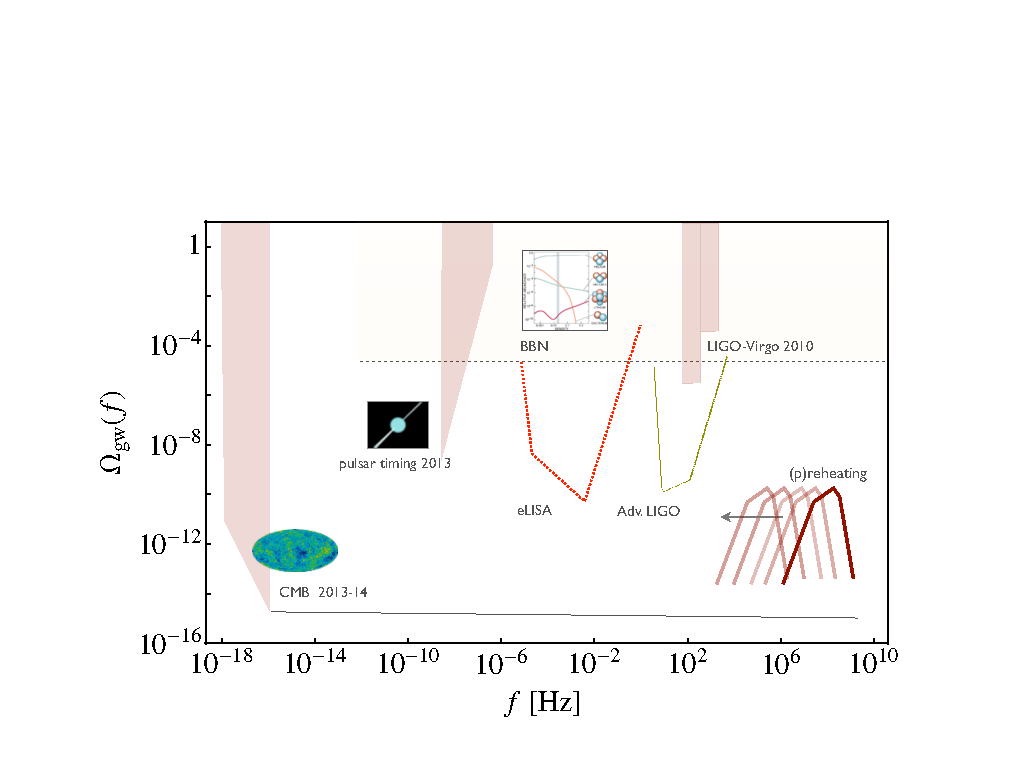} 
   \caption{The gravitational wave energy density per logarithmic interval in frequency in the current universe \cite{Amin2014}. The peaked thick curve on the right is what is expected typically from preheating soon after $10^{15}\,\rm{GeV}$-scale inflation.}
\label{fig:OmegaGWReviewReheating}
\end{figure}

Shortly after it was appreciated that non-perturbative particle production during preheating can lead to the fragmentation of the inflaton condensate, it was shown that the non-linear dynamics can give rise to a stochastic gravitational wave background \cite{Khlebnikov:1997di} in addition to the one generated during slow-roll inflation. Unlike the gravitational waves from inflation \cite{Guzzetti:2016mkm}, whose origin is quantum mechanical and power-spectrum scale-invariant, the gravitational waves from reheating are sourced by the classical evolution of inhomogeneities on sub-horizon scales and their power-spectrum is strongly peaked around a single frequency. Typically, the frequency of the peak is determined by the fragmentation lengthscale, which can be estimated from the linear analysis of preheating. Taking into account the expansion of the universe between reheating and today, one can show, see \cite{Amin2014},
\Beq
\label{eq:GWsPeakIntro}
f_0\sim \beta^{-1}\sqrt{\frac{H_{\rm{br}}}{\mpl}}\times 4\times10^{10}\,\rm{Hz}\,,\qquad \Omega_{\rm{GW},0}\sim10^{-6}\beta^2\,,
\Eeq
where $f_0$ and $\Omega_{\rm{GW},0}$ are the peak frequency and gravitational energy density per logarithmic frequency interval normalized by the critical energy density, respectively. Both quantities are evaluated today. $H_{\rm{br}}$ is the Hubble rate at back-reaction -- the time when most of the signal is generated, and $\beta H_{\rm{br}}^{-1}$ gives the physical wavelength of the excited mode causing the back-reaction on the condensate. Typically, $\beta=\mathcal{O}(10^{-2}-10^{-3})$, thus for efficient preheating after GUT-scale inflation $f_0\sim10^{10}-10^{11}\,\rm{Hz}$ and $\Omega_{\rm{GW},0}\sim 10^{-10}-10^{-12}$. These frequencies lie above the highest frequency ranges $10^3-10^{4}\,\rm{Hz}$ of planned gravitational wave detectors \cite{Moore:2014lga}, see Fig. \ref{fig:OmegaGWReviewReheating}. Decreasing the back-reaction energy scale drives the peak frequency towards the observable range, but the small amplitude of the signal is outside the reach of any of the upcoming gravitational wave observatories.

%~~~~~~~~~~~~~~~~~
%~~~~~~~~~~~~~~~~~

When the inflaton condensate fragments as a result of the resonant particle production of light scalar fields, the gravitational wave background can get tiny modulations on large scales \cite{Bethke:2013vca,Bethke:2013aba}. In addition to the prominent peak corresponding to the fragmentation (sub-horizon) lengthscale, the gravitational wave power-spectrum features a small component on low frequencies, too. The latter is a consequence of the superhorizon scale-invariant power-spectrum of light degrees of freedom developed during inflation, see eq. \eqref{eq:ChiPSInflIntro}, which can be interpreted as the light fields having non-zero vevs that vary between different Hubble patches as discussed in Section \ref{sec:MiscIntro} (recall that individual, causally disconnected, patches evolve independently of each other). The preheating and subsequent non-linear dynamics, including the amplitude of the generated gravitational waves, can be sensitive to these vevs, leading to a super-horizon modulation of the stochastic gravitational wave background. When a light scalar remains a spectator during inflation and reheating, i.e., remains decoupled from the inflaton, similar effects are observed if it is allowed to decay non-perturbatively. This was shown for the gravitational waves produced out of the resonant decay of the Standard Model Higgs into $W$ and $Z$ bosons and their subsequent non-linear evolution \cite{Figueroa:2016ojl}, assuming no coupling between the inflaton and the Standard Model sector.

Stochastic gravitational wave backgrounds from reheating with additional features can be generated as a consequences of the formation of defects \cite{Figueroa:2012kw} and non-topological solitons \cite{Zhou:2013tsa,Antusch:2016con,Kusenko:2008zm}. Even if non-linear effects never become important during reheating, different expansion histories affect the spectrum of the gravitational wave background generated during inflation \cite{Watanabe:2006qe}.

\subsubsection{Non-Gaussianities}

Scalar metric perturbations can also be generated during reheating. While the adiabatic curvature perturbations are unaffected by reheating, see eq. \eqref{eq:Adiabat} and the subsequent discussion, the generation of an entropy (or isocurvature) perturbation, $\mathcal{S}$, during reheating could modify the total curvature perturbation, $\mathcal{R}$. A significant growth of super-Hubble modes of $\mathcal{R}$ occurs if on these scales $\Delta_{\mathcal{S}}^2\gtrsim\Delta_{\mathcal{R}}^2$ \cite{Allahverdi:2010xz}. However, in models with interacting fields (prone to peheating), e.g., $V=\lambda\phi^4/4+g^2\phi^2\chi^2/2$, the super-Hubble power-spectra at the end of slow-roll single-field inflation are $\Delta_{\mathcal{S}}^2\ll\Delta_{\mathcal{R}}^2\sim10^{-9}$, where $\mathcal{S}=(H/\dot{\bar{\phi}})\chi$ \cite{Frolov:2010sz,Bassett:2005xm}. Even if the entropy perturbation is resonantly amplified during preheating, back-reaction takes place while on super-Hubble scales $\Delta_{\mathcal{S}}^2\lesssim\Delta_{\mathcal{R}}^2$ \cite{Bond:2009xx} and the observationally-relevant part of the power-spectrum of $\mathcal{R}$ is affected weakly (at most). %remains virtually unaffected. 

On the other hand, important statistical properties of the curvature perturbation, such as the bispectrum of its non-Gaussianities \cite{Chambers:2007se,Chambers:2008gu}, can be affected significantly by the resonant entropy production during preheating and the subsequent non-linear dynamics. The same mechanism responsible for large-scale modulations in the gravitational wave spectrum from preheating (see the above discussion of gravitational waves) also leads to strong non-Gaussianities in the curvature perturbation \cite{Bond:2009xx}. Essentially, extreme sensitivity is shown to the vevs of light fields within individual Hubble patches by the expansion of these patches. The latter is equivalent to the curvature perturbation, implying, e.g.,
\Beq
\mathcal{R}({\bf{x}})=\mathcal{R}_{\rm{G}}({\bf{x}})+F_{\rm{NL}}\left(\chi_{\rm{G}}({\bf{x}})\right)\,,
\Eeq
where $\mathcal{R}_{\rm{G}}$ is the standard nearly Gaussian adiabatic mode from single-field slow-roll inflation and the last term comes from the back-reaction and non-linear dynamics following the resonant amplification of the nearly Gaussian and scale-invariant (at the end of inflation) $\chi_{\rm{G}}$, see eq. \eqref{eq:ChiPSInflIntro}. As shown in \cite{Bond:2009xx}, the transfer function $F_{\rm{NL}}$ is highly non-linear and describes non-Gaussianities very different from the standard (weak) local ones
\Beq
\mathcal{R}({\bf{x}})=\mathcal{R}_{\rm{G}}({\bf{x}})+\frac{3}{5}f_{\rm{NL}}(\mathcal{R}_{\rm{G}}^2({\bf{x}})-\langle\mathcal{R}_{\rm{G}}^2({\bf{x}})\rangle)\,.
\Eeq

The $F_{\rm{NL}}$ term can in principle lead to non-Gaussian components that could be observable in the CMB \cite{Bond:2009xx} and their non-detection \cite{Ade:2015ava} constrains preheating scenarios. Other reheating scenarios that lead to potentially observable levels of non-Gaussianity include curvaton reheating (where curvature perturbations are generated by the decay of a slightly inhomogeneous curvaton field after inflation) and modulated reheating (where curvature perturbations are generated due to the dependence of the decay rate of the inflaton on the local value of a spatially varying field). Evolution of non-Gaussianity during reheating after multi-field inflation was studied in  \cite{Leung:2012ve,Leung:2013rza}.

We should point out that the modification of large-scale curvature perturbations during preheating is consistent with causality, since it involves no transfer of energy across super-Hubble scales. Entropy perturbations are simply resonantly amplified and then converted into curvature perturbations.\footnote{Conversely, if during reheating all particle species enter thermal equilibrium, having a common temperature and vanishing chemical potentials, the super-horizon curvature perturbations become purely adiabatic and no isocurvature perturbations are present at late times.}

Even when the reheating dynamics is perturbative and no strong resonances take place, the local non-Gaussianity prediction in single-field inflationary scenarios \cite{Maldacena:2002vr}
\Beq
f_{\rm{NL}}\sim\frac{n_{\rm{s}}-1}{4}\,,
\Eeq
depends on the expansion history of reheating, see eqs. \eqref{eq:NstarReh} and \eqref{eq:nsstrstInro}.

\newpage

\section{Afterword}

By examining the dynamics in realistic models of reheating, we can tie together the well-understood and well-tested high-energy physics from laboratory experiments and the more speculative physics of inflation. We can determine not only how the Universe was populated with ordinary matter, but also the origin (and perhaps the nature and the fundamental properties) of cosmic relics such as dark matter, the baryon asymmetry, stochastic gravitational wave backgrounds, etc. The study of ever-more realistic models of reheating has been successful in recent years, and this looks set to continue. With the upcoming Stage-4 CMB experiments set to provide superb data to constrain further inflationary observables  \cite{Abazajian:2016yjj}, understanding reheating-related uncertainties will become increasingly important for narrowing the range of viable models of inflation. We hope and expect that reheating will continue to be at the forefront of research in theoretical cosmology in the coming years.

%\appendix
%\section{Appendix: Potential terms in the minimal composite Higgs model }

%%%%%%%%%%%%%%%%%%%%%%%%%%%%%%%%%%%%%%
%\bibliography{eft}

%\addcontentsline{toc}{part}{Bibliography}

\newpage

\bibliography{mybib}
\bibliographystyle{utphys}

%\providecommand{\href}[2]{#2}\begingroup\raggedright\begin{thebibliography}{10}

%\end{thebibliography}\endgroup
%%%%%%%%%%%%%%%%%%%%%%%%%%%%%%%%%%%%%%
\end{document}